\documentclass[iop,rpp,reprint,twocolumn,longbibliography]{revtex4-1}[2010/07/25]
\usepackage[utf8]{inputenc} 

\usepackage{cmap} 
\usepackage[T1]{fontenc}
\usepackage{amsmath,amssymb,amsfonts}
\usepackage{txfonts}

\usepackage[warn]{textcomp}

\usepackage{tabularx}
\usepackage{dcolumn}
\usepackage{graphicx}
\graphicspath{{figures/}}

\usepackage{url,hyperref}
\usepackage[usenames,dvipsnames]{xcolor}
\hypersetup{colorlinks=true, linkcolor=BrickRed, urlcolor=blue!50!black, citecolor=blue!50!black}

\usepackage[capitalise]{cleveref}
\crefname{equations}{Eqs.}{Eqs.}
\Crefname{equations}{Equations}{Equations}
\creflabelformat{equations}{\textup{(#2#1#3)}}

\newcommand\allowhyphenation{\rightskip=0pt plus .2\hsize}

\renewcommand\epsilon{\varepsilon}
\renewcommand\phi{\varphi}
\renewcommand\theta{\vartheta}
\renewcommand\vec[1]{\boldsymbol{\mathrm{#1}}}
\newcommand\dotprod{\boldsymbol{\cdot}}
\newcommand\expect[1]{\left\langle\vphantom{\big(}#1\right\rangle}
\newcommand\diff{\text{d}}
\newcommand\e{\text{e}}
\renewcommand\i{\text{i}}
\DeclareMathOperator\Real{Re}
\DeclareMathOperator\Imag{Im}
\DeclareMathOperator\Pf{Pf}

\makeatletter
\DeclareRobustCommand*\textsubscript[1]{%
  \@textsubscript{\selectfont#1}}
\def\@textsubscript#1{%
  {\m@th\ensuremath{_{\mbox{\fontsize\sf@size\z@#1}}}}}
\makeatother

\newcommand\dw{{d_\text{w}}}
\newcommand\df{{d_\text{f}}}
\newcommand\kB{{k_\text{B}}}
\newcommand\T{\mathcal{T}}

\begin{document}

\author{Felix Höf\/ling}
\affiliation{Max-Planck-Institut für Intelligente Systeme, Heisenbergstraße 3,
70569 Stuttgart, and Institut für Theoretische Physik IV,
Universität Stuttgart, Pfaffenwaldring 57, 70569 Stuttgart, Germany}

\author{Thomas Franosch}
\affiliation{Institut für Theoretische Physik, Friedrich-Alexander-Universität
Erlangen--Nürnberg, Staudtstraße~7, 91058 Erlangen, Germany}

\title{Anomalous transport in the crowded world of biological cells}

\begin{abstract}
A ubiquitous observation in cell biology is that the diffusive motion of
macromolecules and organelles is anomalous, and a description simply based on
the conventional diffusion equation with diffusion constants measured in dilute
solution fails.
This is commonly attributed to macromolecular crowding in the interior of cells
and in cellular membranes, summarising their densely packed and heterogeneous
structures.
The most familiar phenomenon is a sublinear, power-law increase of the
mean-square displacement as function of the lag time, but there are other
manifestations like strongly reduced and time-dependent diffusion coefficients,
persistent correlations in time, non-gaussian distributions of spatial
displacements,  heterogeneous diffusion, and a fraction of immobile particles.
After a general introduction to the statistical description of slow, anomalous
transport, we summarise some widely used theoretical models: gaussian models
like fractional Brownian motion and Langevin equations for visco-elastic media,
the continuous-time random walk (CTRW) model, and the Lorentz model describing
obstructed transport in a heterogeneous environment.
Particular emphasis is put on the spatio-temporal properties of the transport
in terms of two-point correlation functions, dynamic scaling behaviour, and how
the models are distinguished by their propagators even if the mean-square
displacements are identical.
Then, we review the theory underlying commonly applied experimental techniques
in the presence of anomalous transport like single-particle tracking,
fluorescence correlation spectroscopy (FCS), and fluorescence recovery after
photobleaching (FRAP).
We report on the large body of recent experimental evidence for anomalous
transport in crowded biological media: in cyto- and nucleoplasm as well as in
cellular membranes, complemented by \emph{in vitro} experiments where a variety
of model systems mimic physiological crowding conditions.
Finally, computer simulations are discussed which play an important role in
testing the theoretical models and corroborating the experimental findings.
The review is completed by a synthesis of the theoretical and experimental
progress identifying open questions for future investigation.
\end{abstract}


\received{25 June 2012}
\revised{7 December 2012}
\accepted{19 December 2012}

\maketitle
\tableofcontents

\section{Preface}

Transport of mesoscopic particles suspended in simple solvents is governed by
Brownian motion and is one of the pillars of biological and soft condensed
matter physics.
The pioneering works of Einstein and von Smoluchowski identified the random
displacements as the essence of single-particle motion and---employing ideas of
the central-limit theorem---suggest a gaussian probability distribution with a
mean-square displacement that increases linearly in time.
This paradigm constitutes the basis of numerous applications ranging from
colloidal suspensions, emulsions, and simple polymeric solutions to many
biological systems.

From this perspective, the emergence of \emph{anomalous transport} in more
complex materials, typically characterised by a sublinear increase of the
mean-square displacement, appears as exotic.
Yet experiments from many different areas reveal that anomalous transport is
ubiquitous in nature, signalling that slow transport may be generic for complex
heterogeneous materials; examples are crowded biological media, polymeric
networks, porous materials, or size-disparate mixtures.
One prominent ingredient common to many of these systems is that they are
densely packed, known as \emph{macromolecular crowding} in the biophysics
community.
The presence of differently sized proteins, lipids, and sugars in the cell
cytoplasm, as well as the filamentous networks permeating the cell, is believed
to be at the origin of the observed slowing down of transport, rendering
diffusion constants meaningless that were measured in dilute solution.
Similarly in heterogeneous materials displaying pores of various sizes, the
agents interact strongly with surfaces of complex shape giving rise to an
entire hierarchy of time scales, which implies a power-law increase of the
mean-square displacement.

The goal of this review is to provide a pedagogical overview of the existing
theoretical framework on anomalous transport and to discuss distinguished
experiments from the different fields, thus establishing a common language for
the observations coming from various experimental techniques.
The focus here is on crowded biological systems \emph{in vivo} and \emph{in
vitro} both in the bulk systems such as the cytoplasm or biological model
solutions as well as in planar systems, for example, crowded cellular
membranes.
We also want to emphasise the similarities with complex heterogeneous materials
as they naturally occur, e.g., porous rock, to technologically relevant
materials such as molecular sieves and catalysts.

In \cref{sec:basics}, we provide some elementary background to quantify the
stochastic motion of a single particle in a complex disordered environment.
We recall the standard arguments that underlie the theory of Brownian motion
both on the level of an ensemble average (Fickian diffusion) as well as in
terms of single-particle trajectories. 
In particular, we rederive the widely used mean-square displacement of a tracer
and more generally the entire two-time conditional probability density.
Thereby, we introduce some notation that will be employed throughout the review
and exemplify the important concept of scaling.
Then we contrast complex transport from normal transport by requiring that
deviations from normal transport are persistent way beyond the natural time
scales of the system.
The most celebrated indicator is the subdiffusive increase of the mean-square
displacement.
A second example which has drawn significant interest in the statistical
physics community, although almost entirely ignored by biophysicists, is given
by the velocity autocorrelation function.
We recall that even the case of a single particle in a simple solvent encodes
non-trivial correlations and long-time anomalies due to the slow diffusion of
transverse momentum.
Further, we emphasise that anomalous transport can no longer be described by a
single diffusion \emph{constant}, rather the transport properties depend on the
considered length and time scales.
Thus, it is essential to make testable predictions for the spatio-temporal
behaviour, which we elucidate in terms of the nongaussian parameter and, more
generally, the shape of the distribution of displacements.

The theoretical description of the phenomenon of anomalous transport is
reviewed in \cref{sec:theoretical_models}.
We focus on the three most widely used concepts each of which has been
corroborated from some experiments and excluded from others.
The first one consists of relaxing the white-noise assumption in a Langevin
approach.
Then to induce a sublinear increase of the mean-square displacement a power-law
time correlation has to be imposed on the increments while the probability
distribution is still assumed to be gaussian.
In this simple model, one easily derives two-time correlation functions such as
the van Hove correlation function or its spatial Fourier transform, the
intermediate scattering function.
Simple generalisations to models where the correlation functions encode only a
power-law tail at long times then lead to the gaussian model.
A second, complementary approach is given in terms of a distribution of waiting
times characteristic of the complex medium.
The class of these models is known as the continuous-time random walk model and
allows for a simple solution after a Fourier--Laplace transform to the complex
frequency domain.
While these two descriptions are phenomenological in nature, the third concept
is a microscopic approach that also provides insight into the mechanism how
subdiffusion can emerge from simple dynamic rules in a complex environment.
In this class, which we refer to as Lorentz models, the origin of slow
transport is found in an underlying geometric percolation transition of the
void space to which the tracer is confined.
We recall how self-similar, fractal structures emerge and why transport in
these systems is naturally subdiffusive.
We compare different microscopic dynamics and discuss the crossover to
heterogeneous diffusion at long time scales, scaling behaviour, and the
appearance of an immobile fraction of particles.

The section is concluded by delimiting the scope of the review to subdiffusion
that is related to molecular crowding or the excluded volume effect.
Polymer physics and single-file diffusion, where subdiffusive motion is well
known too, will not be covered here.
In addition, while many experimental data are interpreted in terms of
subdiffusion, some of them actually display only deviations from normal
transport that are not persistent over large time windows.
Hence one has to be careful whether one speaks of subdiffusion or simple
crossover scenarios.
Typical examples where such apparent subdiffusion is observed comprise
diffusion of more than one species each moving with its own diffusion
coefficient.
In the same spirit a particle could change its conformation thus exhibiting
several internal states again characterised by several diffusive regimes.
Sometimes, the data only suggest power laws which could be interpreted as
subdiffusion, yet the correlations are due to the measurement technique and
have no relation to the underlying physical processes.

\Cref{sec:techniques,sec:crowding} review anomalous diffusion in crowded
biological systems from an experimental point of view.
We first introduce the experimental techniques that have proved themselves as
useful tools in biophysics to measure transport properties in mesoscopic
samples and on microscopic scales.
An important technique that is both intuitive and powerful is provided by
single-particle tracking.
Here the trajectory of some agent in a complex environment is recorded over
sufficiently long time, allowing for the evaluation of, in principle, all
correlation functions.
The most widely used quantity is of course the mean-square displacement, since
it is rather robust with respect to statistical fluctuations and appears easy
to interpret.
A second powerful method that has been widely applied in the biophysical
context is fluorescence correlation spectroscopy.
Here the basic idea is to label few molecules by a fluorescent dye and record
the fluctuating fluorescent light upon illuminating a small part of the sample
with a laser.
We briefly introduce the theory underlying the measurement and discuss how
anomalous transport manifests itself in the corresponding correlation function.
Complementary to these two single-particle methods is fluorescence recovery
after photobleaching, which detects the diffusion front of fluorophores after
depleting a small spot by an intense laser pulse.
The technique is apt to measure very slow transport and immobile particles, and
we outline how anomalous transport becomes manifest in the recovery curve as
function of time.

\cref{sec:crowding} is devoted to the plethora of experiments on anomalous
transport in the cell interior and related model systems and their
interpretation in terms of the three theoretical frameworks introduced before.
We have compiled and discuss the most significant experiments in the field,
focusing on the past decade.
Almost all experiments agree that transport is hindered and slowed down by
molecular crowding, manifested in a suppression of the diffusion constant.
A large subset of experiments provides clear evidence for subdiffusive motion,
but there are notable experiments which report normal diffusion, and some
findings appear to be specific to certain biological conditions.
Important progress in the understanding of anomalous transport was made by
biologically motivated model systems, where in contrast to living cells key
parameters are adjustable.
The discussion is completed by addressing computer simulations of simple and
complex model systems, which provide essential support for the interpretation
of experiments.
The presentation is divided into three-dimensional transport in cellular
fluids, e.g., in the cytoplasm of living cells, and transport in cellular
membranes, which may be approximated as a (curved) two-dimensional manifold.
Finally, the consequences of anomalous transport on reaction kinetics are
briefly sketched and some recent progress on this emerging topic is reported.

In \cref{sec:conclusion}, we first summarise what the state of the art in the
field of anomalous transport currently is and provide some general conclusions
on what has been achieved so far both theoretically and experimentally.
We point out which questions are still under debate and give suggestions where
future research may go and what crucial issues still need to be addressed.

\section{Basics of Brownian motion}
\label{sec:basics}

The naming of the observation of \emph{anomalous transport} frequently found
in complex systems already suggests that the phenomena are fundamentally
different from the standard case that therefore qualifies as normal transport.
Before reviewing how anomalous transport can be addressed theoretically and
measured experimentally, we discuss the  framework  for normal transport
connecting  macroscopic diffusion with fluctuations at small scales.

The erratic movement of a mesosized particle suspended in a simple solvent is
referred to as \emph{Brownian motion,} after the Scottish botanist Robert
Brown, who observed the continuously agitated motion of minute particles
ejected from certain pollen grains under a light microscope.
The first theoretical description has been achieved independently by
Einstein~\cite{Einstein:1905} and Smoluchowski~\cite{vonSmoluchowski:1906} in
terms of a probabilistic approach.
These ideas have been rephrased shortly after by Langevin~\cite{Langevin:1908,
Lemons:1997} in terms of stochastic differential equations by separating the
force balance into a deterministic and a random part.
The characterisation of the random forces is largely due to
Ornstein~\cite{Ornstein:1917} thereby laying the foundations of the modern
calculus of Langevin equations.

The experimental demonstration of the probabilistic route to a macroscopic law
has been achieved by Perrin~\cite{Perrin:1909} and his students around the same
time by meticulously analysing single trajectories  of colloidal particles
observed under a microscope.
His contribution was awarded with the Nobel prize in 1926 as a breakthrough
in proving the physical reality of molecules.

The impact of  Einstein's theory on Brownian motion, i.e., the normal case, can
hardly be underestimated and constitutes one of the milestones in physics.
Recently, on the occasion of 100\textsuperscript{th} anniversary of
Einstein's \emph{annus mirabilis} 1905, a series of reviews have been published
highlighting the new concepts and future directions in the field of Brownian
motion~\cite{Haenggi:2005, Frey:2005, Sokolov:2005, Renn:2005}.

\subsection{Simple diffusion}
\label{sec:simple_diffusion}

In the molecular kinetic approach advocated by Einstein and Smoluchowski the
suspended particle experiences rapid collisions with the solvent  molecules.
These events occur at the time scale of the liquid dynamics, typically in the
picosecond regime, and at each encounter a tiny amount of momentum is
exchanged.
Simultaneously, the collisions are responsible for the macroscopic friction
force counteracting the random kicks.
At the time scales where the suspended particle moves significantly, the
increments $\Delta \vec{R}(t) = \vec{R}(t+t') -\vec{R}(t')$ after an elapsed
time $t$ are considered as random variables that are identically and
independently distributed.
By the central-limit theorem, the total displacement, being the sum of many
independent tiny increments, then is governed by a gaussian distribution
\begin{equation}
  P(\vec{r},t) = \bigl[2 \pi \delta r^2(t)/d\bigr]^{-d/2}
  \exp \left(-\vec{r}^2 d/2 \delta r^2(t) \right) \,,
\end{equation}
where $d$ is the dimension of the embedding space.
The probability distribution $P(\vec{r},t)$ is known in general as the
\emph{propagator} or the \emph{van Hove self-correlation
function}~\cite{Hansen:SimpleLiquids, BoonYip:1980}.
For independent increments the variance
$\delta r^2(t) := \expect{[\vec{R}(t)-\vec{R}(0)]^2}$ grows linearly with the
number of steps, implying a linear increase of the mean-square displacement,
$\delta r^2(t) = 2d D t$.
The only transport coefficient characterising the Brownian motion is then the
diffusion constant $D$, completely specifying the propagator,
\begin{equation}
 \label{eq:DiffusionPropagator}
 P(\vec{r},t) = \frac{1}{(4 \pi  D t )^{d/2}} \exp\left(\frac{-\vec{r}^2}{4  D t}\right)\, .
\end{equation}
The self-similarity of free Brownian motion becomes evident by writing the
propagator in a scale-free form,
\begin{equation}
  P(\vec{r},t) = r^{-d} \mathcal{P}_\text{gauss}(\hat r) \,,
  \qquad \hat r \propto rt^{-1/2} \,,
  \label{eq:diffusion_scaling}
\end{equation}
introducing a dimensionless scaling variable, $\hat r := (2  D t)^{-1/2} r$,
$r=|\vec r|$, and a scaling function,
\begin{equation}
  \mathcal{P}_\text{gauss}(\hat{r}) = (2\pi)^{-d/2} \hat{r}^d \exp(-\hat{r}^2/2) \,.
  \label{eq:Pscaling_gauss}
\end{equation}
Other scaling forms will be encountered in the course of this review.

The connection to the macroscopic description arises when considering many
particles performing  Brownian motion such that the probability cloud
$P(\vec{r},t)$ displays the same space-time dynamics as the macroscopic
concentration.
In particular, one verifies that the gaussian propagator fulfils the diffusion
equation
\begin{equation}
 \partial_t P(\vec{r},t) = D \nabla^2 P(\vec{r},t),
\end{equation}
and satisfies the initial condition of a spatially localised distribution,
$P(\vec{r},t) = \delta(\vec{r})$.

For  applications, a representation in terms of spatial Fourier modes is
advantageous, $P(\vec{k},t) = \int \!\diff^d r\, \e^{- \i  \vec{k}
\dotprod \vec{r}} \, P(\vec{r}, t)$, and $P(\vec{k},t)$ is known as the
(self-)intermediate scattering function~\cite{Hansen:SimpleLiquids,
BoonYip:1980}.
It can be measured directly by neutron scattering employing the spin-echo
technique~\cite{Lovesey:Neutron_Scattering} or, on larger length scales, by
photon correlation spectroscopy~\cite{BernePecora:DynamicLightScattering}.
The momentum transferred from the sample to the photon or neutron is then
simply $\hbar \vec{k}$.
For the diffusion propagator, one readily calculates
\begin{equation}
 P(\vec{k},t) = \exp(- Dk^2 t) \,,
\end{equation}
implying that density modulations decay with a rate $ 1/\tau_k = Dk^2$.
Then, long-wavelength perturbations are long-lived since by particle
conservation the relaxation involves transport of particles over large
distances.

For future reference, we also provide the van Hove correlation function in the
complex frequency domain,
\begin{equation}
  P(\vec{r},\omega) = \int_0^\infty\! \e^{\i \omega t}\, P(\vec{r}, t)\, \diff t\/ ,
 \qquad \Imag[\omega] \geq 0.
\end{equation}
The one-sided Fourier transform reduces to the standard Laplace transform,
provided one identifies $s = -\i \omega$; the advantage of introducing
complex frequencies as above is that the transform can be readily inverted
numerically by evaluating the integral
\begin{equation}
 P(\vec r, t) = \frac{2}{\pi} \int_0^\infty \!
 \Real[P(\vec r,\omega)] \cos(\omega t) \, \diff \omega \,.
 \label{eq:temporal_backtransform}
\end{equation}
For the diffusive propagator one finds
\begin{equation}
  P(\vec{r},\omega) = \frac{1}{(2\pi D)^{d/2}}
  \left(\frac{\sqrt{- \i \omega D}}{r} \right)^{d/2-1}
  K_{d/2-1}\left( r \sqrt{\frac{- \i \omega}{D}} \,\right) \,,
  \label{eq:propfreq_general}
\end{equation}
where $K_\nu(\cdot)$ denotes the modified Bessel function of the second kind.
The expression simplifies for the dimensions of interest,
\begin{subequations}
  \label[equations]{eq:propfreq}
  \begin{align}
  P(\vec{r},\omega)
  &= \frac{1}{2 \sqrt{-\i\omega D}} \exp\left(-r\sqrt{\frac{-\i\omega}{D}} \, \right)
  & (d=1), \\
  P(\vec{r},\omega)
  &= \frac{1}{2 \pi D} K_0\left(r \sqrt{\frac{-\i\omega}{D}} \, \right)
  &(d=2), \\
  P(\vec{r},\omega)
  &= \frac{1}{4 \pi D r} \exp\left(-r \sqrt{\frac{-\i\omega}{D}} \,\right)
  &(d=3).
  \end{align}
\end{subequations}

In scattering techniques, where the energy transfer to the sample is also
recorded, the central quantity is the frequency- and wavenumber-dependent
scattering function
$P(\vec{k},\omega) = \int_0^\infty\! \e^{\i  \omega t}\, P(\vec{k},t)\, \diff t$.
The scattering cross section corresponding to an energy transfer
$\hbar \omega$ and a momentum transfer $\hbar \vec{k}$ is then essentially
given by $\Real[P(\vec{k},\omega)]$, known as the (incoherent) dynamic
structure factor~\cite{Hansen:SimpleLiquids}.
In the case of simple diffusion, the dynamics is represented by a simple pole
on the negative imaginary axis,
\begin{equation}
 P(\vec{k},\omega) = \frac{1}{-\i  \omega + D k^2}\, .
\end{equation}


An equivalent way to characterise the dynamics of a tracer is to give a
prescription on how individual trajectories are generated as a result of the
stochastic fluctuations in the medium~\cite{Kubo:StatisticalMechanics_II,
vanKampen:StochasticProcesses}.
The propagator $P(\vec{r},t)$ is then the result of a suitable  average over
the possible individual realisations of the jittery motion for the tracer.
The equations of motion naturally become stochastic differential equations
referred to as \emph{Langevin equations}, which incorporate the randomness of
the kicks by the medium as ``noise''.
The modern formulation in terms of a Langevin equation is mostly due to
Ornstein, who shaped the notion of what is now known as random gaussian white
noise. A mathematical rigorous introduction to the stochastic differential
equations  and Brownian motion can be found in the excellent textbook by
{\O}ksendal~\cite{Oksendal:Stochastic_Differential}.

For overdamped motion, the displacements $\vec{R}(t)$ are assumed to obey the
stochastic differential equation
\begin{equation}
 \label{eq:langevin}
 \partial_t \vec{R}(t) = \vec{\eta}(t),
\end{equation}
with noise terms
$\vec{\eta}(t) = \boldsymbol (\eta_1(t), \dots, \eta_d(t) \boldsymbol)$
that are considered as independent, random quantities on sufficiently
coarse-grained time scales.
In fact, these $\eta_i(t), \quad i=1,\dots,d$, represent already averages over
many independent processes occurring on even shorter time scales such that the
central-limit theorem applies.
The probability distribution then corresponds to a multivariate gaussian,
symbolically written as
$P[\vec{\eta}(t)] \propto \mathcal{D}[\vec{\eta}(t)]
\exp\left( - \int\! \diff t\, \vec{\eta}(t)^2/4 D \right) $,
which is characterised completely by the only non-vanishing
cumulant~\cite{Kubo:StatisticalMechanics_II},
\begin{equation}
  \expect{ \eta_i(t) \, \eta_j(t') } = 2 D \delta_{ij} \delta(t-t') \, , \qquad
  i,j=1,\ldots, d \,.
\end{equation}
Such a noise displays only short-time correlations and corresponds to a power
spectral density that is flat at the frequencies of interest, commonly referred
to as \emph{white} noise.
We have imposed that different Cartesian directions $\eta_i(t)$ are
uncorrelated and, invoking isotropy, the strength of the noise, $2D$, is
identical for all directions.
The idea of coarse-graining and the seemingly innocent assumption of
independence then necessarily leads to gaussian white noise as the universal
law for the statistics of the displacements at small times.
Any deviation from this law indicates the existence of non-trivial persistent
correlations in the system.

The displacement after a finite lag time follows from formally integrating the
Langevin equation,
$\Delta \vec{R}(t) = \vec{R}(t)-\vec{R}(0) = \int_{0}^{t}\! \diff t' \, \vec{\eta}(t')$,
and being a sum of gaussian variables, it obeys again a gaussian distribution.
Thus it suffices to calculate the first two cumulants.
Since the mean of the noise vanishes, one infers
$\expect{\Delta \vec{R}(t)} = 0$,
and the correlation function of the displacements follows from the
delta-correlated noise as
\begin{equation}
 \expect{\Delta R_i(t) \, \Delta R_j(t)} = 2 D t \delta_{ij} \,.
\end{equation}
In particular, one recovers $\delta r^2(t) =
\expect{\Delta \vec{R}(t)^2} = 2 d D t$, and the probability distribution is
determined by the diffusion propagator, \cref{eq:DiffusionPropagator}.

\subsection{Anomalous and complex transport}
\label{sec:complex_transport}

The probabilistic reasoning presented in the previous subsection suggests that
normal diffusion emerges as a statistical law essentially by the central-limit
theorem.
In particular, the mean-square displacement is expected to increase linearly in
time for time scales much larger than microscopic ones.
In simple systems such as normal liquids~\cite{Hansen:SimpleLiquids,
BoonYip:1980} one observes diffusion already at time scales exceeding the
picosecond scale.
The phenomena of anomalous or complex transport deal with dynamics where this
diffusive regime is not visible even on time scales that are by many orders of
magnitude larger than picoseconds.
Conventionally, a non-linear growth of the mean-square displacement $\delta
r^2(t)$ is taken as indicator of such unusual behaviour.
Typically, the mean-square displacement is proportional to a power law,
$\delta r^2(t) \propto t^\alpha$, with an exponent $0< \alpha < 1$.
Hence the mean-square displacement increases slower than for normal diffusion,
formally the diffusion coefficient becomes zero, nevertheless the tracer is not
localised.
This kind of behaviour is referred to as \emph{subdiffusion} or \emph{anomalous
transport}\/\footnote{In different contexts one finds also superdiffusive
transport corresponding to $\alpha>1$, which is beyond the scope of this
review.}.
Theoretically, the phenomenon then calls for reasons why the central-limit
theorem does not apply at the time scales of interest.
Rephrasing the argument in terms of increments reveals that persistent
correlations are hidden in the dynamics on meso- or macroscopic time scales.

\begin{figure}
  \includegraphics[width=\linewidth]{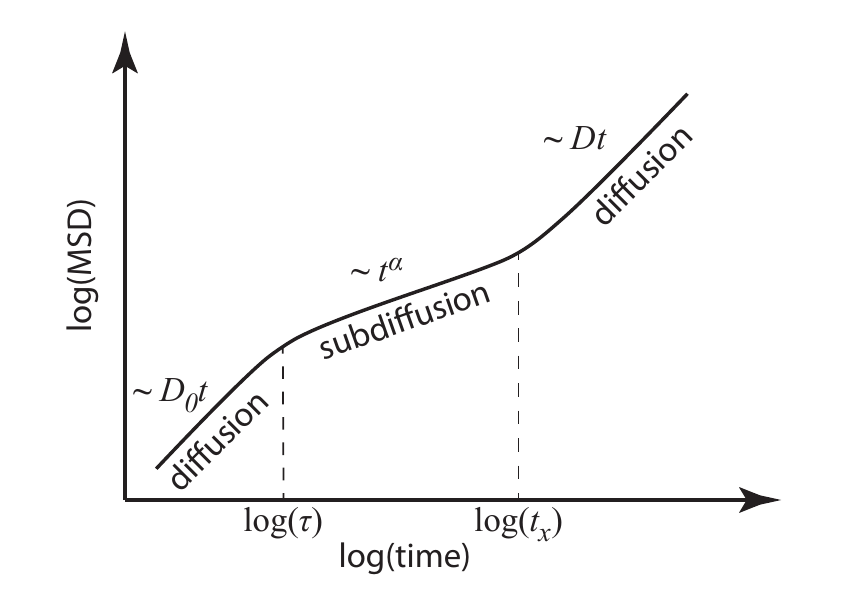}
  \caption{Schematic mean-square displacement (MSD) for intermediate
  subdiffusive transport. Free diffusion at microscopic scales is followed by
  subdiffusive transport at intermediate time scales. In a physical system, the
  subdiffusive growth ends typically at a second crossover, where the MSD grows
  linearly again with reduced diffusion constant, $D\ll D_0$, or where it
  saturates, e.g., due to boundaries like the cell membrane.}
  \label{fig:msd_sketch}
\end{figure}

We would like to make a distinction between a simple violation of the
central-limit theorem in some intermediate time window and mechanisms leading
to subdiffusive behaviour that can in principle persist forever.
In the first case some dynamic processes are unusually slow that spoil the
central-limit theorem on these scales, yet ultimately normal transport sets in.
This scenario of complex transport occurs generically by having constituents of
the medium of different sizes or soft interactions, e.g., polymers.
Then the mean-square displacement displays only a crossover from some
short-time motion to long-time diffusion.
Since the crossover can extend over several decades (due to a series of slow
processes occurring in the medium), fits by power laws are often a satisfactory
description.
In the second case, the correlations in the increments decay slowly and upon
tuning suitable control parameters the window of subdiffusion can become
arbitrarily long.
Hence in a well-defined limit the subdiffusion persists forever and the
central-limit limit theorem never applies.
We reserve the term \emph{anomalous transport} for the latter scenario.
Typically, the mean-square displacement is expected to display two crossover
time scales, see \cref{fig:msd_sketch} and Ref.~\onlinecite{Saxton:2007}, which is
also found in experiments~\cite{Kusumi:2005}.
The long-time diffusion coefficient is then strongly suppressed compared to the
microscopic motion at short times, for example, subdiffusion with $\alpha=0.6$
over 4~decades in time yields a reduction of $D$ over $D_0$ by a factor of
$(t_x/\tau)^{1-\alpha}=10^{4\times 0.4}\approx 40$.
We postpone the discussion what physical mechanisms can lead to such drastic
changes and focus here on the measurable quantities suited to reveal complex
and anomalous transport.

\subsubsection{Van Hove self-correlation function}

The basic observable is the fluctuating single-particle density,
$\rho(\vec{r},t) = \delta\boldsymbol(\vec{r}-\vec{R}(t)\boldsymbol)$, and the
corresponding correlation function,
$P(\vec{r}-\vec{r}',t-t') = V \expect{ \rho(\vec{r},t) \, \rho(\vec{r}',t') }$,
is referred to as van Hove (self-)correlation function~\cite{Hansen:SimpleLiquids}.
Here $V$ is the volume of the container and the thermodynamic limit
$V\to\infty$ is anticipated.
Furthermore by translational invariance and for a stationary stochastic
process, $P$ depends only on the elapsed time $t-t'$ and the accumulated
displacement
$\vec{r}-\vec{r}'$.
Then the van-Hove correlation function can be cast into the form $P(\vec{r},t)
= \expect{ \delta(\vec{r}-[\vec{R}(t)-\vec{R}(0)] ) }$ which is
interpreted  directly  as probability density for an observed displacement
$\vec{r}$ after a lag time $t$.
Furthermore for isotropic systems, to which we restrict the discussion, only
the magnitude $r = |\vec{r}|$ enters the van Hove function.

The  first generalisation of simple diffusion consists of assuming that the
increments $\Delta \vec{R}(t)$ follow a gaussian probability distribution with
zero mean,
\begin{equation}
 \label{eq:gaussian_propagator}
 P_\text{gauss}(\vec r, t) = \frac{1}{\left[ 2 \pi \delta r^2(t)/d \right]^{d/2}}
 \exp\left(\frac{-r^2}{2 \delta r^2(t)/d} \right) \, .
\end{equation}
Here the mean-square displacement $\delta r^2(t)$ characterises the width of
the distribution; for ordinary diffusion, it holds $\delta r^2(t) = 2 d D t$,
of course.
The higher moments,
\begin{equation}
 \delta r^n(t) := \expect{| \Delta \vec{R}(t)|^n} =
 \int\! \diff^d r\, |\vec{r}|^n\, P(\vec{r},t) \,,
\end{equation}
are obtained  by performing gaussian integrals, for example, one finds for the
mean-quartic displacement,
$\delta r^4(t) = [(d+2)/d] \bigl[\delta r^2(t)\bigr]^2$.

Equivalently to the van Hove function, one can study the single-particle
dynamics by monitoring the decay of density fluctuations in the wavenumber
representation $\rho(\vec{k},t) = \exp( \i  \vec{k} \dotprod \vec{R}(t) )$.
The corresponding correlation function,
$P(k,t) = \expect{ \rho(\vec{k},t)^*\, \rho(\vec{k},0) }$,
is the (self-)intermediate scattering function and merely the spatial Fourier
transform of the van Hove function.
Again by isotropy, $P(k,t)$ depends only on the magnitude $k=|\vec{k}|$ of the
wavenumber.
The explicit representation
$P(k,t) = \expect{\!\exp[ -\i  \vec{k} \dotprod \Delta \vec{R}(t) ] }$
permits an interpretation of $P(k,t)$ as the characteristic function of the
random variable $\Delta \vec{R}(t)$, such that the lowest order moments can be
obtained from a series expansion for small wavenumbers, $k\to 0$,
\begin{equation}
 \label{eq:Pmoments}
 P(k,t) = 1 - \frac{k^2}{2 d}\,\delta r^2(t) + \frac{k^4}{8 d(d+2)} \, \delta r^4(t)
 + \mathcal{O}\bigl(k^6\bigr) \,.
\end{equation}
Here we used that the orientational average over a $d$-dimensional sphere
yields%
\footnote{%
  Let $n_i$ be the Cartesian components of a unit vector.
  Then by symmetry one argues that
  \begin{align*}
  \overline{ n_i n_j} &= A \delta_{ij} \nonumber \\
  \overline{ n_i n_j n_k n_l } &= B (\delta_{ij} \delta_{kl}
    + \delta_{ik} \delta_{jl} + \delta_{il} \delta_{jk} ),
  \end{align*}
  where $\overline{\ldots}$ indicates a spherical average.
  Contracting the indices $i$ and $j$ in the first relation reveals $A=1/d$,
  contracting in the second shows $B=1/d(d+2)$. Thus $ \overline{ n_z^4} =
  3/d(d+2)$.
}
$\overline{ \cos^2 \vartheta} = 1/d$, $\overline{ \cos^4 \vartheta } = 3/d(d+2)$.

The logarithm of the characteristic function generates the cumulants,
\begin{multline}
 \ln P(k,t) = - \frac{k^2}{2d} \delta r^2(t)
 + \frac{k^4 \bigl[\delta r^2(t)\bigr]^2}{8 d^2}
 \left( \frac{d}{d+2}\frac{\delta r^4(t)}{\bigl[\delta r^2(t)\bigr]^2} - 1 \right)  \\
+ \mathcal{O}(k^6)  , \qquad k\to 0.
\end{multline}
In the case of gaussian transport, the Fourier transform of
\cref{eq:gaussian_propagator} yields
$P_\text{gauss}(k,t) = \exp \bigl(- k^2 \delta r^2(t)/2 d \bigr)$,
and all cumulants apart from the second one, $\delta r^2(t)$, vanish
identically.
Thus a simple, dimensionless indicator for transport beyond the gaussian
approximation is the non-gaussian parameter~\cite{Hansen:SimpleLiquids,
BoonYip:1980},
\begin{equation}
 \label{eq:nongaussian}
 \alpha_2(t) :=  \frac{d}{d+2}\frac{\delta r^4(t)}{\bigl[\delta r^2(t)\bigr]^2} - 1\, .
\end{equation}
The subscript indicates that there is an entire series of similarly defined
quantities involving higher moments of the displacements.
The inequality
$\bigl\langle{X^2}\bigr\rangle \geq \expect{X}^2$
for the random variable $X=|\Delta \vec R(t)|^2$
implies a lower bound on the non-gaussian parameter,
$\alpha_2(t) \geq -2/(d+2)$.

The fact that probability theory imposes certain constraints on  $P(k,t)$ as a
function of wavenumber $k$ naturally poses the question if additional
conditions apply if $P(k,t)$ is considered as a function of time $t$.
More generally, what class of functions are permissible for correlation
functions?
Decompose the fluctuation density into Fourier modes
$\rho_T(\vec{k},\omega) = \int_{-T/2}^{T/2} \diff t\, \e^{-\i \omega t}
\rho(\vec{k},t)$ for real frequencies $\omega$ and a long but finite
observation time $T>0$.
Then the corresponding \emph{power spectral density} is obtained via the
Wiener--Khinchin theorem~\cite{Kubo:StatisticalMechanics_II, vanKampen:StochasticProcesses},
\begin{equation}
  \lim_{T\to\infty}
  \frac{1}{T} \expect{ \bigl|\rho_T(\vec{k},\omega)\bigr|^2 } = 2 \Real[P(k,\omega)] \/.
\end{equation}
Thus $\Real[P(k,\omega)] \geq 0$ and  inversion of the one-sided Fourier
transform yields
\begin{equation}
  P(k,t) = \frac{2}{\pi} \int_0^\infty \! \Real[P(\vec{k},\omega)]
  \, \cos(\omega t)\, \diff \omega \/,
\end{equation}
i.e., the propagator is the cosine transform of a non-negative function.
Transforming again to complex frequencies by one-sided Fourier transform one
derives relations of the Kramers--Kronig
type~\cite{Forster:Hydrodynamic_Fluctuations, Goetze:MCT}; in particular, one
observes that $\Real[ P(k,\omega)] \geq 0$ not only for real but
for all complex frequencies in the upper half-plane, $\Imag[\omega]>0$.

Since the particle is to be found somewhere a particle conservation law holds,
and the intermediate scattering function approaches unity in the
long-wavelength limit, $\lim_{\vec{k}\to 0}P(\vec{k},t) = 1$.
For the one-sided Fourier transform, the particle conservation law suggests the
representation $P(k,\omega) = 1/[ -\i \omega + k^2 D(k,\omega)]$, where
$D(k,\omega)$ is known as the frequency- and wavenumber-dependent diffusion
kernel~\cite{Hansen:SimpleLiquids, BoonYip:1980}.
From $\Real[P(k,\omega)] \geq 0$ for $\Imag[\omega] >0$ the same property
is inherited for the diffusion kernel, $\Real[D(k,\omega)] \geq 0$.

Of particular interest is the long-wavelength limit $Z(\omega) = D(k\to
0,\omega)$, which encodes the spatial second moment of  the tracer motion.
Note again $\Real[Z(\omega)] \geq 0$ for $\Imag[\omega] >0$.
Expanding for small wavenumber,
\begin{equation}
  P(k,\omega) = \frac{1}{-\i \omega + k^2 D(k,\omega)}
  = \frac{1}{-\i \omega} - \frac{k^2 Z(\omega)}{(-\i \omega)^2} + {\cal O}(k)^4 \,,
  \label{eq:P_k_omega_expansion}
\end{equation}
and comparing to \cref{eq:Pmoments}, one finds
$Z(\omega) = - (\omega^2/2 d) \int_0^\infty \!\diff t\, \e^{\i \omega t} \delta r^2 (t)$\/.
Integration by parts reveals that $Z(\omega)$ is the one-sided Fourier
transform of  the velocity autocorrelation function (VACF),
\begin{equation}
 Z(t) = \frac{1}{d} \expect{ \vec{v}(t) \dotprod \vec{v}(0) }
 = \frac{1}{2d} \frac{\diff^2}{\diff t^2} \delta r^2(t).
\end{equation}
Reversely, the mean-square displacement is obtained by integration,
\begin{equation}
 \delta r^2(t) = 2 d \int_0^t \!\diff t'\, (t-t') \, Z(t') \,.
\end{equation}
For stochastic processes where the derivative of the increments $\Delta
\vec{R}(t)$ does not exist, e.g., for a Brownian particle, the VACF may be
defined via the mean-square displacement and can be shown to be a negative and
completely monotone function~\cite{Lorentz_2D:2010}.

In the case of ordinary diffusion, $\delta r^2 (t) = 2 d D t$, the diffusion
kernel simply assumes a constant, $D(k,\omega) \mapsto Z(\omega) =D$, and the
velocity decorrelates instantaneously, $Z(t) \mapsto D\, \delta(t-0)$.
Furthermore, the non-gaussian parameter vanishes identically, $\alpha_2(t)
\equiv 0$, as do all higher cumulants.

\subsubsection{Distribution of squared displacements}
\label{sec:squared_displacements}

The measurements of the squared displacements $\Delta \vec{R}(t)^2$ for a
single particle along its trajectory often does not represent well the
\emph{mean}-square displacement $\delta r^2(t)$, rather a significant
scattering of the data is observed.
This observation suggests to introduce the distribution function for the
squared displacements~$u$,
\begin{equation}
 p(u,t) := \expect{\delta\left(u - \Delta \vec{R}(t)^2\right)} \, .
 \label{eq:squared_prob_definition}
\end{equation}
The probability distribution is obviously properly normalised,
$\int_0^\infty\! p(u,t)\, \diff u = 1$,
its mean reproduces the mean-square displacement, and higher moments
yield the even displacement moments,
\begin{equation}
  \delta r^{2k}(t) = \expect{ |\Delta \vec{R}(t)|^{2k} }
  =  \int_0^\infty \! u^k p(u,t) \,\diff u \, .
\end{equation}
The fluctuations in the squared displacements are given by the second cumulant
and are via \cref{eq:nongaussian} already encoded in the non-gaussian parameter,
\begin{align}
 \mathrm{Var}\left[\Delta \vec{R}^2(t)\right]
  &= \expect{ |\Delta \vec{R}(t)|^4 } - \expect{ \Delta \vec{R}(t)^2 }^2 \\
  &=  \left[ \frac{d+2}{d} \, \alpha_2(t) + \frac{2}{d} \right] \bigl[\delta r^2(t)\bigr]^2 \,.
\end{align}

Rather than dealing with the probability distribution it is often favourable to
work with a  moment-generating function, which is the Laplace transform
of the probability distribution,
\begin{equation}
 M(w,t) := \int_0^\infty \! \diff u \, \e^{-  u/w^2} p(u,t)
 = \expect{\!\exp\left(-  \Delta \vec{R}(t)^2/w^2 \right) } \,,
 \label{eq:squared_moment_generating}
\end{equation}
where the second representation follows from \cref{eq:squared_prob_definition}.
The convention is chosen such that $1/w^2$ is the variable conjugate to $\Delta
\vec{R}(t)^2$, and $w$ carries the dimension of a  length.
Since the exponential is approximated by unity for small displacements,
$\Delta \vec{R}(t)^2 \ll w^2$, and rapidly approaches zero for large ones,
$\Delta \vec{R}(t)^2 \gg w^2$, $M(w,t)$ essentially constitutes the
probability for the particle to be still or again within a distance
$w$ of its initial position.
In \cref{sec:fcs}, it will be shown how this quantity can be directly
measured by fluorescence correlation spectroscopy (FCS), where $w$ corresponds
to the beam waist of the illuminating laser~\cite{FCS_scaling:2011}.

The van Hove correlation $P(\vec{r},t)$ is the probability distribution of all
vector displacements $\vec{r}$, and the probability distribution of the squared
displacements $p(u,t)$ follows by marginalising,
\begin{equation}
 p(u,t) = \int\! \diff^d r \, \delta\bigl(u-\vec{r}^2\bigr) \,P(\vec{r},t)  \, .
\end{equation}
For the important case of statistically isotropic samples, $P(\vec{r},t)$
depends only on the magnitude of the displacement $r=|\vec{r}|$ and the
integral can be evaluated.
In spherical coordinates, the angular integration yields the surface area of
the $d$-dimensional unit sphere, $\Omega_d$, as a factor.
The radial integral collapses due to the Dirac delta function,
$\delta\bigr(u-r^2\bigl) = \delta\bigr(\!\sqrt{u}-r\bigl)/2 u$, and one obtains
\begin{equation}
 p(u,t) = \frac{\Omega_d}{2} u^{(d-2)/2} P\bigl(r=\sqrt{u}, t\bigr) \, .
\end{equation}
For gaussian transport, solely characterised by the time-dependent mean-square
displacement, $\delta r^2(t)$, \cref{eq:gaussian_propagator} yields
\begin{equation}
  p(u,t) = \frac{1}{\Gamma(d/2)}
  \frac{u^{d/2-1}}{\bigl[2 \delta r^2(t) / d\bigr]^{d/2}}
  \exp \left(\frac{-u}{2 \delta r^2(t) / d} \right) \, ;
\end{equation}
the gamma function evaluates to $\Gamma(d/2)=\sqrt{\pi}, 1, \sqrt{\pi} / 2$ for
$d=1,2,3$.
From these expressions one readily calculates also the cumulative distribution
functions, $\int_0^U \! p(u,t) \, \diff u$.

The van Hove function $P(\vec{r},t)$ and its spatial Fourier transform, the
intermediate scattering function, $P(\vec{k},t)$, the distribution of the
squared displacements $p(u,t)$, and the corresponding moment generating
function $M(w,t)$ all encode spatio-temporal information on the motion of the
tracer.
For isotropic systems they are all equivalent in principle, in practice they
are sensitive to different aspects of transport occurring on different length
scales.
Correlation functions that involve more than two times provide even more
information on the dynamics, and may hold the key to distinguish different
theoretical models that yield the same two-time correlation functions.

\section{Theoretical models}
\label{sec:theoretical_models}

The paradigm of anomalous transport is tantamount with a violation of the
central-limit theorem on arbitrarily long time scales.
Modelling such processes requires including persistent correlations that
manifest themselves as self-similar dynamics in the mean-square displacements.
Different models and theoretical approaches have been pursued that generically
lead to subdiffusion.
Here we focus on the three most widely used frameworks. The perhaps simplest
approach is based on stochastic differential equations where the noise term
displays persistent correlations which then transfer to the increments.
Since usually the statistics of the noise is still assumed to be gaussian, they
differ essentially only in the form of the temporal correlations they
incorporate and we summarise them as gaussian models.
The second category, the continuous-time random walk (CTRW), consists of jump
models where particles undergo a series of displacements according to a
distribution with large tails.
Here the central-limit theorem does not apply since the  mean waiting time for
the next jump event to occur becomes infinite.
The last class of Lorentz models relies on spatially disordered environments
where the tracer explores fractal-like structures that induce anomalous
dynamics.

\subsection{Gaussian models}
\label{sec:gaussian_models}

Here we collect properties of a class of models that give a phenomenological
description of complex and anomalous dynamics which all result in gaussian
propagators.

\subsubsection{Fractional Brownian motion}
\label{sec:fbm}

A simple model for subdiffusion is fractional Brownian motion, introduced
rigorously by Mandelbrot and van Ness~\cite{Mandelbrot:1968} as superposition
of Brownian processes with power-law memory.
Here we follow a heuristic approach~\cite{Kou:2004, Goychuk:2007, Goychuk:2009}
that  summarises the essence of fractional Brownian motion.
Assuming the stochastic differential equation \eqref{eq:langevin},
$\partial_t \vec{R}(t) = \vec{\eta}(t)$, we have already seen that if the noise
$\eta_i(t)$ is delta-correlated in time, the mean-square displacements increase
linearly.
If additionally the noise  obeys a gaussian statistics, this property is
inherited also for the displacements $\Delta \vec{R}(t)$ and transport is
completely characterised by the  diffusion propagator $P(\vec{r},t)$,
\cref{eq:DiffusionPropagator}.
The idea is now to incorporate persistent correlations in the noise such that
transport is drastically slowed down with respect to normal diffusion.
Since the noise plays the role of a fluctuating velocity, we use the same
notation as in \cref{sec:complex_transport}, to express the noise
correlator in terms of the velocity autocorrelation function,
\begin{equation}
 \label{eq:NoiseCorrelator}
 \expect{\eta_i(t)\, \eta_j(t')} =  d \,Z(| t-t'|) \, \delta_{ij}\, , \qquad i,j=1,\dots ,d.
\end{equation}
Here the different  Cartesian components are taken as uncorrelated, which
certainly holds for an isotropic system.
In the Fourier domain, this implies
\begin{equation}
 \expect{\eta_i(\omega)^*\,\eta_j(\omega')}
  = 4 \pi d \,\delta_{ij} \,\delta (\omega-\omega') \Real[Z(\omega)] \,,
\end{equation}
where $Z(\omega) = \int_0^\infty\! \e^{\i \omega t} Z(t)\, \diff t$ is again the
one-sided Fourier transform of the velocity autocorrelation function.
For ordinary diffusion, $Z(\omega) = D$ is constant and the noise corresponds
to white noise.

The  case of subdiffusion, $\delta r^2(t) = 2 d K_\alpha t^\alpha$ with an
exponent $0 < \alpha < 1$ and a generalised diffusion coefficient $K_\alpha>0$,
then yields a spectral density
$Z(\omega) = (-\i \omega )^{1-\alpha} K_\alpha \Gamma(1+\alpha)$~\cite{Mandelbrot:1968}.
Hence the strength of the noise approaches zero as the frequencies become
smaller, which explains that transport slows down with increasing correlation
time.
In the temporal domain, $Z(t)$ is represented by a pseudofunction%
\footnote{The noise correlator corresponds to a distribution, and
 pseudofunction means that integrals with test functions $\phi(t)$ extract only
 Hadamard's finite part~\cite{Zemanian:Distribution_Theory},
 $\int \! \phi(t) \Pf | t|^{\alpha-2} \,\diff t
  := \int [\phi(t)-\phi(0)] |t|^{\alpha-2} \,\diff t$.
 In particular, one easily verifies that the one-sided Fourier transform of
 $Z(t)$ yields $Z(\omega)$.
},
\begin{equation}\label{eq:FractionalNoise}
 Z(t) = \alpha (\alpha-1) K_\alpha \Pf | t|^{\alpha-2}.
\end{equation}
Up to here only the mathematical frame has been set and no assumptions on the
nature of the stochastic process has been made.

In fractional Brownian motion, the statistics of the noise correlator is
assumed to be characterised  by the only non-vanishing cumulant $Z(t)$,
\cref{eq:NoiseCorrelator,eq:FractionalNoise}, i.e., a stationary
gaussian process although not white noise.
Then the statistics of the increments $\Delta \vec{R}(t)$ is again gaussian,
and the propagator reduces to $P_\text{gauss}(\vec{r},t)$,
\cref{eq:gaussian_propagator}.
Its scaling form corresponds to that of simple diffusion,
\cref{eq:diffusion_scaling},
\begin{equation}
  P_\text{FBM}(r, t) = r^{-d} \mathcal{P}_\text{gauss}(\hat r) \,,
  \qquad \hat r \propto r t^{-\alpha/2},
  \label{eq:fbm_scaling}
\end{equation}
sharing the gaussian scaling function, \cref{eq:Pscaling_gauss}, but not the scaling
variable, $\hat r$.
In particular, the non-gaussian parameter $\alpha_2(t)\equiv 0$ vanishes by the
construction of fractional Brownian motion.

In contrast to simple diffusion, fractional Brownian motion is not a Markov
process; in particular, the van Hove correlation function is not sufficient to
characterise the statistical properties completely.
Multiple-time correlation functions encode non-Markovian behaviour, for which
fractional Brownian motion makes detailed predictions.
As an example,
\begin{multline}
 \expect{ [\vec{R}(t)-\vec{R}(0)]^2 [ \vec{R}(t+T)-\vec{R}(T)]^2 }  \\
  = 4 d^2 K_\alpha^2  t^{2\alpha}+ 2 d K_\alpha^2
  \bigl(|t+T|^{\alpha} + | t-T|^{\alpha} - 2 T^\alpha \bigr)^2 \,,
\end{multline}
which has been derived recently to study the ergodic properties of fractional
Brownian motion~\cite{Deng:2009}.
For $T=0$, this expression reproduces the quartic moment, $\delta r^4(t)$, and
is compatible with a vanishing non-gaussian parameter, \cref{eq:nongaussian}.

\subsubsection{Langevin equations for visco-elastic media}
\label{sec:generalised_langevin}

The erratic motion of a spherical particle immersed in a complex medium can be
described quite generally by a Langevin equation.
Rather than directly addressing the displacements one may base the description
on the velocity $\vec{v}(t) = \dot{\vec{R}}(t)$ and formulate a force balance
equation.
The paradigm has been given by \textcite{Langevin:1908} himself,
\begin{equation}
 \label{eq:underdamped_langevin}
 m \dot{\vec{v}}(t)  = - \zeta \vec{v}(t) + \vec{f}(t) \,,
\end{equation}
where $m$ denotes the mass of the particle, the deterministic friction force
$-\zeta \vec{v}(t)$ is merely the Stokes drag, and $\vec{f}(t)$ is a
fluctuating force with zero mean, $\expect{ f_i(t) } =0$.
The friction constant constant, $\zeta = 6 \pi \eta a$, is directly connected to
the solvent viscosity $\eta$ and the particle radius $a$.
The statistics of the random forces $\vec{f}(t)$ is characterised completely by
the only non-vanishing cumulant~\cite{Ornstein:1917},
\begin{equation}
 \expect{f_i(t) \, f_j(t')} = 2 \kB \T \zeta \, \delta_{ij} \, \delta(t-t') \,,
\end{equation}
where $\T$ is the temperature of the environment and $\kB$ denotes Boltzmann's
constant.  Thus the Cartesian components $f_i(t)$ of the forces are gaussian
distributed and independent for different times.
The variance at equal times is again dictated by the fluctuation--dissipation
theorem, see, e.g., Ref.~\onlinecite{Kubo:StatisticalMechanics_II} for details.
The delta-correlation in the temporal domain for the forces translates to white
noise for the corresponding power spectral density.
The velocity autocorrelation then decays exponentially~\cite{Langevin:1908},
\begin{equation}
\expect{v_i(t)\, v_j(t')} = (\kB \T/m) \, \delta_{ij} \, \exp(-|t-t'|/\tau_p) \,,
\end{equation}
where $\tau_p = m/\zeta = m/6 \pi \eta a$ is the momentum relaxation time.
Similarly, the mean-square displacement of the $d$-dimensional motion is
calculated to
\begin{equation}
 \delta r^2(t) = 2 d\, D \left[ t + \tau_p \left( \text{e}^{-t/\tau_p} -1 \right) \right],
\end{equation}
where $D = \kB \T /\zeta$ is the diffusion constant according to the
Stokes--Einstein relation.

The description can be easily generalised for the case of visco-elastic
media~\cite{Mason:1995}.
Here the response of the complex solvent to shear is encoded in the complex
frequency-dependent viscosity, $\eta(\omega)$.
In the conventions employed in this review, $\Real[\eta(\omega)]\geq 0$
corresponds to the dissipative part and $\Imag[\eta(\omega)]$ encodes the
reactive part.
Equivalently, one may employ the complex shear modulus
$G(\omega) := - \i \omega \eta(\omega)$.
For example, in the Maxwell model,
$G(\omega) = -\i \omega \tau_M G_\infty / (1-\i \omega \tau_M)$,
the modulus is characterised by a high-frequency elastic response $G_\infty$ and
a crossover time scale $\tau_M$.
The Stokes drag in a visco-elastic medium then depends on the frequency and the
Langevin equation is discussed conveniently in the Fourier
domain~\cite{Grimm:2011},
\begin{equation}
 -\i \omega m  \vec{v}(\omega) = - \zeta(\omega) \vec{v}(\omega) + \vec{f}(\omega),
\end{equation}
where $\zeta(\omega) = 6\pi \eta(\omega) a$ replaces the Stokes friction
coefficient.
By the fluctuation--dissipation theorem, the force correlator has to be
modified accordingly,
\begin{equation}
 \label{eq:FDT}
 \expect{f_i(\omega)^* \, f_j(\omega')} =
 4 \pi \kB \T \Real[\zeta(\omega)] \, \delta_{ij} \, \delta (\omega-\omega') \,.
\end{equation}
Since the fluctuations arise in the surrounding solvent as a sum over
uncorrelated regions, the forces are again assumed to be gaussian.
Then it is clear that the van Hove correlation function for the particle
corresponds to a gaussian propagator and the dynamics is specified entirely by
the mean-square displacement, $\delta r^2(t)$.
Rather than solving for $\delta r^2(t)$, we solve for the one-sided Fourier
transform of the velocity autocorrelation function~\cite{Mason:1995,
Grimm:2011},
\begin{equation}
 \label{eq:VACFzeta}
 Z(\omega) = \frac{\kB \T}{-\i m\omega + \zeta(\omega)}.
\end{equation}
Relying on the relation $\zeta(\omega) = 6\pi \eta(\omega) a$, the local
visco-elastic response of a complex medium is inferred from the motion of
tracer particles in microrheology experiments~\cite{Mason:1995}.

Subdiffusion at long times is obtained if $Z(\omega)  = (-\i \omega)^{1-\alpha}
K_\alpha \Gamma(1+\alpha)$ for $\omega\to 0$~\cite{Kou:2004}, i.e., the elastic
modulus displays power-law behaviour, $G(\omega) \sim (-\i \omega)^\alpha$,
which appears to be generic in many biological materials and soft matter
systems for intermediate frequencies.
This empirical observation is formulated in the \emph{soft glassy rheology
model}~\cite{Sollich:1997}.
The microscopic mechanism remains in general unspecified, yet for the case of a
solution of semiflexible polymer networks, the bending rigidity of a single
filament suggests an elastic power-law response,
$G(\omega) \sim (-\i \omega)^{3/4}$~\cite{Amblard:1996, Gittes:1997,
Gittes:1998, Konderink:2006}.
Similarly, by coupling to the elastic degrees of freedom of a membrane,
effective fractional friction kernels can be generated in the same way with
various exponents depending on the level of description of the
membrane~\cite{Taloni:2010, Taloni:2010a}

\subsubsection{Long-time anomalies}
\label{sec:long_time_anomalies}

The assumption of an instantaneous friction term in the Langevin equation
\eqref{eq:underdamped_langevin} is in fact incorrect even at long times, as has been
pointed out already by Hendrik Antoon Lorentz.
The reason is that the Stokes formula applies for steady motion of the particle
only and the friction is accompanied by a long-ranged vortex pattern in the
velocity field of the entrained fluid.
For unsteady motion, the particle excites incessantly new vortices diffusing
slowly through the fluid.
As a consequence the friction force depends on the entire history of the
particle's trajectory, an effect known as \emph{hydrodynamic memory}.
The theoretical description is achieved most conveniently in the frequency
domain.
The drag force for a sphere performing small-amplitude oscillations of angular
frequency $\omega$ has already been calculated by Stokes~\cite{Stokes:1851} and
leads to a frequency-dependent friction
coefficient~\cite{Kubo:StatisticalMechanics_II},
\begin{equation}
  \zeta(\omega) = 6 \pi \eta a \left(1 + \sqrt{-\i \omega \tau_\text{f}} \right)
  - \i \omega m_\text{f}/2.
\end{equation}
For steady motion, $\omega=0$, the formula reduces to the Stokes drag.
The last term appears as an acceleration force for half of the displaced fluid
mass, $m_\text{f} = 4 \pi \rho_\text{f} a^3/3$, and it is natural to absorb
this contribution by introducing an effective mass for the particle,
$m_\text{eff} = m + m_\text{f}/2$.
The second modification is a  non-analytic contribution due to the slow vortex
diffusion in the liquid as the particle performs unsteady motion.
The characteristic time scale, $\tau_\text{f} = \rho_\text{f} a^2 /\eta$, is
the time needed for a vortex to diffuse over the distance of the radius of the
particle.
By the fluctuation--dissipation theorem, \cref{eq:FDT}, the spectrum of the random
forces is no longer white but displays a coloured component that increases as a
square root with frequency.
Recently, the power spectral density of the thermal noise has been measured
experimentally for a single bead by ultra-sensitive high-bandwidth optical
trapping ~\cite{Resonances_Nature:2011} in excellent agreement with theoretical
predictions.

The velocity autocorrelation function in the frequency domain,
\cref{eq:VACFzeta},  acts as an admittance or frequency-dependent mobility and
displays a non-analytic low-frequency expansion,
$Z(\omega) = D \left[1- \sqrt{-\i \omega \tau_\text{f} } + \mathcal{O}(\omega)\right]$.
An explicit expression in the temporal domain is achieved in terms of error
functions~\cite{Hinch:1975}, here we focus on the long-time anomaly,
\begin{equation}
 Z(t) \simeq \frac{D}{2} \sqrt{\frac{\tau_\text{f}}{\pi}} \, t^{-3/2}\/,
 \qquad t\to \infty,
\end{equation}
which is a direct consequence of the non-analytic terms in $\zeta(\omega)$.
The most striking feature is that $Z(t)$ encodes persistent correlations
manifested by a self-similar tail in strong contrast to the exponential decay
of Langevin's original theory.
These long-time tails have been discovered first in computer simulations for
fluids~\cite{Alder:1967, Alder:1970} and, only recently, have directly been
observed for colloidal particles in suspension~\cite{Lukic:2005,
Atakhorrami:2005, Jeney:2008, Franosch:2009}.
The mean-square displacement follows directly by integration,
\begin{equation}
 \delta r^2(t) = 6 D t \left[ 1 - 2 \sqrt{\tau_\text{f}/\pi t}
  + \mathcal{O}\bigl(t^{-1}\bigr) \right] .
\end{equation}
The algebraic tail in the velocity autocorrelation manifests itself in a slow
approach to normal diffusive transport.

The persistent correlations in the mean-square displacement, buried under the
leading linear increase, show that the assumption of independent increments is
not satisfied and that the regime of truly overdamped motion is never reached
due to the hydrodynamic memory, even at long time scales.
Nevertheless the central-limit theorem remains valid, although the convergence
is slow due to the persistent power-law correlations induced by vortex
diffusion.

\subsection{Continuous-time random walks (CTRW)}
\label{sec:ctrw}

A different class of models that is widely discussed is the continuous-time
random walk (CTRW)~\cite{Metzler:2000, Bouchaud:1990,
benAvraham:DiffusionInFractals, Hughes:Random_Walks}, originally introduced by
\textcite{Montroll:1965} for hopping transport on a disordered lattice.
The particles spend most of the time bound to a trap with an escape time that
depends sensitively on the depth of the trap.
Rather than dealing explicitly with the quenched disorder on the lattice, the
medium is treated as homogeneous with the new ingredient of a waiting-time
distribution for the next hopping event to occur.
Anomalous transport can be generated within this framework by assuming waiting
time distributions such that the mean waiting time becomes infinite.
The central-limit theorem does not apply since longer and longer waiting times
are sampled. It turns out that the CTRW as mean-field approximation to hopping
in the quenched trap model gives the same result in dimensions 2 and
higher~\cite{Bouchaud:1990}. The trapping in biology seems rather natural, due
to  chemical attachments of molecules in the cell, such that binding on broadly
distributed time scales may lead to CTRW dynamics.
First we provide the general description of the CTRW model, then we show how
subdiffusion can emerge and discuss scaling properties of the propagator.

\subsubsection{Model definition}

In the CTRW model, the particle is assumed to traverse the $d$-dimensional
space by a series of jumps such that the displacement, $\ell$, and the waiting
time to perform the next jump, $t$, are drawn from a given distribution,
$\psi(\vec{\ell},t)$.
For simplicity, we assume that the observation starts when the process is
initialised.
Hence, a CTRW belongs to the broad class of renewal--reward processes.
In particular, the propagator $P(\vec{r},t)$ fulfils a renewal
equation~\cite{Feller:ProbabilityBd2}, which follows by conditioning on the
event that the particle has accumulated a displacement $\vec{r}$ at time $t$
within the first step.
If the first step occurs later than at time~$t$ the propagator is simply
$\delta(\vec{r})$, otherwise the process is renewed~\cite{Montroll:1965,
Metzler:2000, Scalas:2004, Gorenflo:2008},
\begin{multline}
  P(\vec{r},t) = \delta(\vec{r}) \int_t^\infty \!\! \diff t' \!\int\! \diff^d \ell\,
  \psi( \vec{\ell},t') \\
  + \int_0^t \! \diff t' \! \int \! \diff^d \ell\,
  P(\vec{r}-\vec{\ell},t-t') \, \psi( \vec{\ell},t') \,.
\end{multline}
The first term corresponds to the propagator provided that no jump occurred and
can be rewritten as
\begin{equation}
 P_0(\vec{r},t) = \delta(\vec{r}) \left[ 1 - \int_0^t \! \diff t' \, \psi(t') \right] ,
 \label{eq:ctrw_nojump}
\end{equation}
where $\psi(t) = \int \! \psi(\vec{\ell},t) \, \diff^d \ell$ is the jump
probability density irrespective of the size of the jump.
The solution of the renewal equation is most easily achieved after a spatial
Fourier transform, $P(\vec{k},t) = \int \! \diff^d r \, \e^{-\i\vec{k}
\cdot \vec{r}}\, P(\vec{r},t)$,
and a subsequent temporal one-sided Fourier transform,
$P(\vec{k},\omega) =  \int_0^\infty \! \diff t \, \e^{\i  \omega t} P(\vec{k},t)$
for $\Imag[\omega] \geq 0$.
By the convolution theorem the renewal equation simplifies to
$P(\vec{k},\omega) =P_0(\vec{k},\omega)+   \psi(\vec{k},\omega) P(\vec{k},\omega)$.
The spatio-temporal Fourier--Laplace transform of \cref{eq:ctrw_nojump} can be be
performed directly, $P_0(\vec{k},\omega) = [1- \psi(\omega)] / (-\i \omega)$.
Combining both results yields the Montroll--Weiss
relation~\cite{Weiss:Aspects_Random_Walks, Weiss:1983, Klafter:1987, Scalas:2004,
Metzler:2000} for the propagator in terms of the jump probability distribution,
\begin{equation}
  P(\vec{k},\omega) =  \frac{1}{-\i  \omega}
  \frac{1 - \psi(\omega)}{1 - \psi(\vec{k},\omega)} \,.
\end{equation}

In many applications the jump distribution is not known, and additional
assumptions are necessary to define the model.
First, we assume that the jumps exhibit no preferred direction such that $\psi(
\vec{\ell},t)$ depends only on the magnitude $| \vec{\ell}|$, Hence, the
dynamics becomes isotropic and $P(\vec{k},\omega)$ is a function of the
wavenumber, $k=|\vec{k}|$, only.
Second, the waiting-time distribution is often taken to be independent of the
jump size, and the corresponding jump distribution factorises,
$\psi( \vec{\ell},t)= \lambda(\vec{\ell})\, \psi(t)$
such that $\int \! \lambda(\vec{\ell})\,\diff^d \ell =1$.
Then the propagator assumes the simple form
\begin{equation}
 P(k,\omega) =  \frac{1}{-\i  \omega} \frac{1 - \psi(\omega)}{1 - \lambda(k) \, \psi(\omega)},
 \label{eq:ctrw_P_k_omega}
\end{equation}
which will be the starting point for the discussion.

We shall assume that the jump size distribution $\lambda(\vec \ell)$ is
well-behaved, in particular it decays rapidly for large distances.
Then the characteristic function of the jump sizes, $\lambda(k)$, encodes all moments,
\begin{equation}
 \lambda(k) = 1- \frac{k^2}{2d} \bigl\langle \ell^2 \bigr\rangle
 + \frac{k^4}{8 d(d+2)} \bigl\langle \ell^4 \bigr\rangle + \mathcal{O}\bigl(k^6\bigr) \,.
 \label{eq:ctrw_lambda_k}
\end{equation}
For the mean-square displacement, only the second moment of the spatial
distribution is relevant,
$\bigl\langle \ell^2 \bigr\rangle / d = \diff^2 \lambda(k) / \diff k^2 |_{k=0}$,
and, by \cref{eq:P_k_omega_expansion}, the velocity autocorrelation function
is computed to
\begin{equation}
  Z(\omega) = \frac{-\i  \omega \, \psi(\omega)}{1-\psi(\omega)}
  \frac{\bigl\langle \ell^2 \bigr\rangle}{2 d}\, .
  \label{eq:ctrw_Z_omega}
\end{equation}
By normalisation of the waiting-time distribution, it holds
$\psi(\omega \to 0) = 1$, and for a well-behaved distribution the low-frequency
expansion provides the moments of the waiting time,
$ \expect{\tau} := \int_0^\infty t \, \psi(t) \, \diff t$,
\begin{equation}
 \psi(\omega) = 1 + \i \omega \expect{\tau} + \mathcal{O}\bigl(\omega^2\bigr) \,.
\end{equation}
In this case, the diffusion coefficient $D = Z(\omega\to 0)$ is finite with
value $D= \bigl\langle \ell^2 \bigr\rangle / 2 d \expect{\tau}$.

\subsubsection{Anomalous transport}

Anomalous transport is obtained if the jump rate distribution is non-analytic
at zero frequency, e.g.,
\begin{equation}
  \psi(\omega) = 1 - (- \i  \omega \tau )^\alpha + \mathit{h.o.t.},
  \qquad \omega \to 0\, ,
\end{equation}
with non-integer exponent, $0 < \alpha < 1$, some time scale $\tau>0$,
and neglecting higher order terms.
Then the diffusion coefficient vanishes, $D = Z(\omega\to 0)=0$, and the VACF
inherits a leading non-analytic contribution for $\omega \to 0$,
\begin{equation}
  Z(\omega) \simeq (-\i  \omega)^{1-\alpha} K_\alpha \Gamma(1+\alpha) \,,
\end{equation}
with $K_\alpha \Gamma(1+\alpha) = \bigl\langle \ell^2 \bigr\rangle
\tau^{-\alpha} /2d$.
By means of a Tauber theorem~\cite{Karamata:1931, Feller:ProbabilityBd2}, the
non-analyticity at zero frequency corresponds to a negative long-time tail in
the VACF,
\begin{equation}
  Z(t) \simeq -\alpha (1-\alpha) K_\alpha t^{\alpha-2}\/,
  \qquad t \to \infty \, .
\end{equation}
Thus, persistent anticorrelations become manifest in the VACF.
Integration over time shows that the time-dependent diffusion coefficient,
$D(t)$, approaches zero at long times only algebraically,
$D(t) \simeq \alpha K_\alpha t^{\alpha-1}$ for $t\to \infty$.
Eventually, one finds for the mean-square displacement,
\begin{equation}
  \delta r^2(t) \simeq 2 d K_\alpha t^\alpha\/,
  \qquad t \to \infty \,,
\end{equation}
i.e., a subdiffusive increase that persists for arbitrarily long lag times.
Reversely, the low-frequency singularity in $\psi(\omega)$ is connected to a
power-law tail in the waiting-time distribution,
\begin{equation}
  \psi(t) \simeq \frac{\alpha  \tau^\alpha}{\Gamma(1-\alpha)} t^{-1-\alpha} \,,
  \qquad t \to \infty \, ,
\end{equation}
such that even the mean waiting time is infinite.

Let us briefly discuss the next higher moment, the mean-quartic displacement.
The expansion of the incoherent dynamic structure factor, \cref{eq:ctrw_P_k_omega},
in powers of the wavenumber yields the one-sided Fourier transform of the
mean-quartic displacement, cf.\ \cref{eq:Pmoments,eq:ctrw_lambda_k},
\begin{equation}
  \int_0^\infty \! \diff t \, \e^{\i \omega t} \, \delta r^4(t)
  = 8d (d+2) \frac{Z(\omega)^2}{(-\i\omega)^3}
  + 2d \frac{\bigl\langle \ell^4 \bigr\rangle}{\bigl\langle \ell^2 \bigr\rangle}
    \frac{Z(\omega)}{(-\i\omega)^2} \,.
\end{equation}
In particular, the leading low-frequency singularity is again completely
governed by the tail of the waiting-time distribution,
\begin{equation}
\int_0^\infty\!\diff t\, \e^{\i \omega t} \, \delta r^4(t) \simeq
8 d(d+2) \, K_\alpha^2 \, \Gamma(1+\alpha)^2 (-\i  \omega)^{-1-2\alpha} \,;
\end{equation}
in particular, the 4th moment of the jump size,
$\bigl\langle \ell^4 \bigr\rangle$, does not enter the leading order.
Application of a Tauber theorem~\cite{Karamata:1931, Feller:ProbabilityBd2}
yields the asymptotic long-time behaviour,
\begin{equation}
  \delta r^4(t) \simeq 8 d (d+2) \frac{\Gamma(1+\alpha)^2}{ \Gamma(1+2\alpha)}
  K_\alpha^2 t^{2 \alpha}, \qquad t\to \infty \,.
\end{equation}
One concludes that for a CTRW with power-law distributed waiting times, the
non-gaussian parameter approaches a constant value,
\begin{equation}
 \alpha_2(t\to\infty) = \frac{2 \Gamma(1+\alpha)^2}{\Gamma(1+2\alpha)}-1\, ,
\end{equation}
irrespective of the jump size distribution $\lambda(\vec{\ell})$.
The method can be easily extended to higher moments, and one can show that
\emph{all} non-gaussian parameters assume non-vanishing long-time limits, with
values that depend only on the exponent of subdiffusion, $\alpha$.

\Cref{eq:ctrw_P_k_omega,eq:ctrw_lambda_k,eq:ctrw_Z_omega}
imply a small-wavenumber approximation of the
propagator,
\begin{equation}
  P(k,\omega) \simeq \frac{1}{-\i  \omega + k^2 Z(\omega)}\/, \qquad k \to 0 \,.
  \label{eq:generalised_hydrodynamics}
\end{equation}
Propagators of this form will be referred to as \emph{generalised hydrodynamics
approximation.}
It allows for an explicit solution of the spatial inverse Fourier transform;
actually, all that needs to be done is to replace the diffusion coefficient $D$
in \cref{eq:propfreq} by its frequency-dependent generalisation, $Z(\omega)$.
For example in three dimensions, one finds
\begin{equation}
 P(\vec{r},\omega) = \frac{1}{4 \pi Z(\omega) r}
 \exp\left( -r \sqrt{\frac{- \i  \omega}{Z(\omega)}} \,\right) \qquad (d=3) \,.
\end{equation}

\subsubsection{Scaling limit and the fractional Fokker--Planck equation}

Let us discuss a simple, possibly the simplest  distribution for
the waiting time that displays a tail, the Cole--Cole distribution,%
\footnote{%
Such a frequency-dependence was introduced empirically and has since been widely
used to describe the stretched dielectric response of polymeric liquids and
glassforming materials, see Kenneth S.\ Cole and Robert H.\ Cole,  J.\ Chem.\
Phys.\ \textbf{9}, 341 (1941).}
\begin{equation}
 \psi(\omega) = \frac{1}{1 + (-\i  \omega \tau)^{\alpha}} \,.
\end{equation}
In the temporal domain, this corresponds to the waiting-time
distribution~\cite{Hilfer:1995, Gorenflo:2008}
\begin{equation}
 \psi(t) = - \frac{\diff}{\diff t} E_\alpha\bigl(- (t/\tau)^\alpha \bigr) \/,
\end{equation}
where $E_\alpha(\cdot)$ denotes the Mittag--Leffler function.
Inserting in \cref{eq:ctrw_Z_omega}, the velocity autocorrelation function assumes a
power law for all frequencies,
\begin{equation}
  Z(\omega) = \frac{\bigl\langle \ell^2 \bigr\rangle}{2 d \tau} (-\i  \omega \tau)^{1-\alpha} \,,
  \label{eq:Z_omega_cole_cole}
\end{equation}
and the corresponding mean-square displacement is subdiffusive for all times,
$\delta r^2(t) = 2 d K_\alpha t^\alpha$.
The non-gaussian parameter can be evaluated exactly,
\begin{equation}
 \alpha_2(t) = \frac{2 \Gamma(1+\alpha)^2}{\Gamma(1+2\alpha)}- 1
  + \frac{d}{d+2} \frac{\bigl\langle \ell^4 \bigr\rangle}{\bigl\langle \ell^2 \bigr\rangle^2}\,
  \Gamma(1+\alpha) \left(\frac{t}{\tau}\right)^{-\alpha} .
\end{equation}

\begin{figure}
  \includegraphics[width=\linewidth]{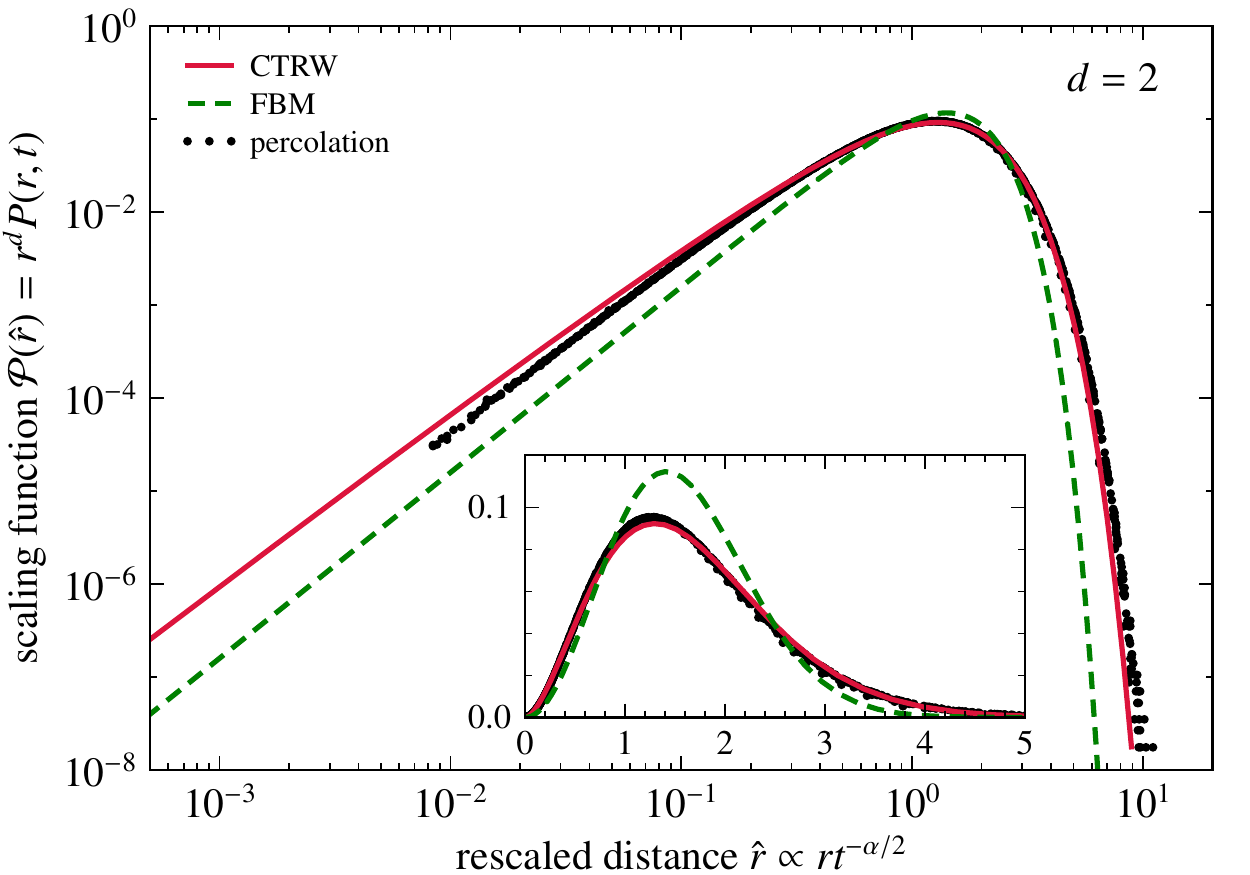}
  \caption{Scaling functions of the propagators for CTRW, fractional Brownian
  motion (FBM), and lattice percolation on the incipient infinite cluster in $d=2$
  dimensions for subdiffusive motion with $\alpha=0.695$ on double-logarithmic
  (main panel) and linear (inset) scales.
  Scaling functions and scaling variables are normalised such that
  $\Omega_d \int_0^\infty \!\diff x \, \mathcal{P}(x) / x = 1$ and
  $\Omega_d \int_0^\infty \!\diff x \, x \,\mathcal{P}(x) = d$.
  }
  \label{fig:propagators2d}
\end{figure}

Within the generalised hydrodynamics approximation, the Fourier--Laplace
transform of the propagator, $P(k, \omega)$, is given by
\cref{eq:generalised_hydrodynamics} for long wavelengths, $k\to 0$.
Together with $Z(\omega)$ given by \cref{eq:Z_omega_cole_cole}, this form
corresponds to the scaling limit~\cite{Gorenflo:2008} after coarse-graining on
large length scales and long times.
The corresponding propagator in real space and time then is the solution of the
so-called fractional Fokker--Planck equation~\cite{Metzler:2000}, which is the
scale-free limit of CTRW~\cite{Hilfer:1995, Gorenflo:2008}.
The approach presented here devoids the introduction of fractional derivatives
which naturally emerge if the equations of motion are introduced in the
temporal domain.
We refer the mathematically inclined reader to the excellent review by
\textcite{Metzler:2000}.

\begin{figure}
  \includegraphics[width=\linewidth]{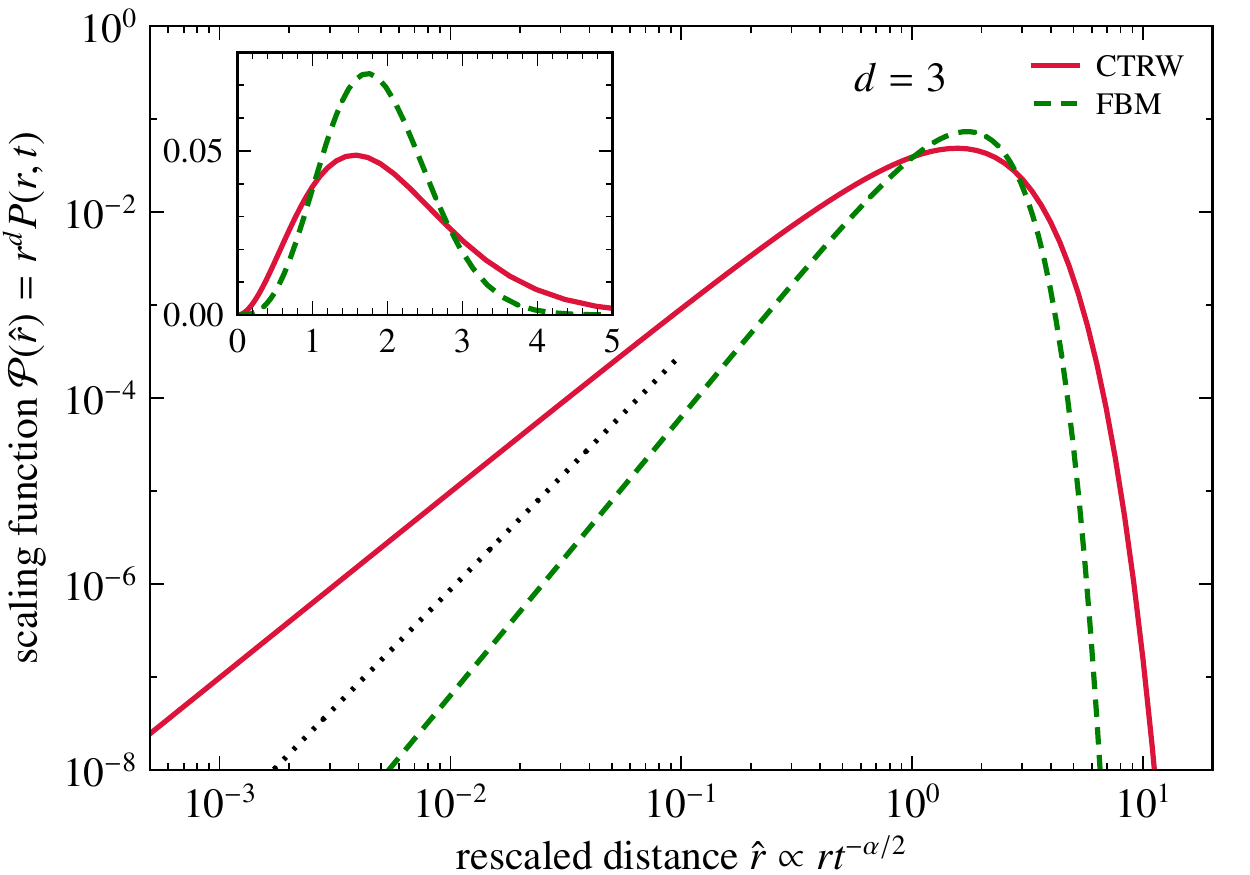}
  \caption{Scaling functions of the propagators for CTRW and fractional
  Brownian motion in $d=3$ dimensions for subdiffusive motion with
  $\alpha=0.515$ on double logarithmic (main panel) and linear (inset) scales.
  The dotted, black line indicates the anticipated asymptote for transport on
  the percolating cluster, $\mathcal{P}_\infty(\hat r) \sim \hat r^{\df}$.
  }
  \label{fig:propagators3d}
\end{figure}

The spatial backtransform is obtained from \cref{eq:propfreq_general} upon
replacing $D$ by the above result for $Z(\omega)$.
It is convenient to introduce a dimensionless, rescaled frequency,
$\hat\omega := \bigl(\bigl\langle \ell^2 \bigr\rangle/2d\bigr)^{-1/\alpha} \tau \,
 r^{2/\alpha} \omega$.
Then,
\begin{equation}
  P(\vec{r},\omega)
  =  \frac{(2 \pi)^{-d/2}}{-\i\omega  \, r^d} \,
  {(-\i \hat \omega)}^{\alpha(d+2)/4} \, K_{d/2-1}\left((-\i\hat\omega)^{\alpha/2}\right) \,.
  \label{eq:ctrw_P_r_omega}
\end{equation}
The evaluation of the complex modified Bessel function of the second kind,
$K_\nu(\cdot)$, as well as the temporal inverse Fourier transform,
\cref{eq:temporal_backtransform}, can be achieved numerically, e.g., with
\textsc{Mathematica}.
In the fractional Fokker--Planck limit, the time- and space-dependent
propagator exhibits a scaling property that holds for all times and distances,
\begin{equation}
  P(r, t) = r^{-d} \mathcal{P}_\alpha(\hat{r}) \, , \qquad \hat{r}
  \propto  r t^{-\alpha/2} \,;
  \label{eq:ctrw_scaling}
\end{equation}
the omitted prefactor,
$\bigl\langle \ell^2 \bigr\rangle^{-1/2} \tau^{\alpha / 2}$,
renders $\hat r$ a dimensionless scaling variable.
Although the scaling property is of the same form as for fractional Brownian
motion, \cref{eq:fbm_scaling}, the scaling functions are different.
Here, $\mathcal{P}_\alpha$ depends on the space dimension $d$ and  on the
exponent $\alpha$.
The scaling functions are displayed in \cref{fig:propagators2d,fig:propagators3d}
for $d=2,3$ and are compared to the scaling functions of fractional Brownian
motion and obstructed transport on the percolating cluster for the same
subdiffusion exponent.
Note that \cref{eq:ctrw_P_r_omega} was derived previously on a
lattice~\cite[Eq.~(20)]{Weissman:1989}, the subsequent inverse Laplace transform
to the time domain, however, relied on asymptotic approximations of the Bessel
function, yielding a scaling function,
$\mathcal{P}_\alpha(x)=x^d \exp\bigl(-x^{2/(2-\alpha)}/2\bigr)$
\cite{Weissman:1989, benAvraham:DiffusionInFractals}, which strongly disagrees
with our numerical findings at small arguments.

Since CTRW is not a Markov process, the propagator $P(\vec{r},t)$ does not
characterise the dynamics completely.
Higher correlation functions involving multiple times have been
calculated~\cite{Barkai:2007} and the procedure has been generalised to
arbitrary order~\cite{Niemann:2008}.
In the fractional Fokker--Planck limit, the dynamics can be written as a
subordinated Brownian motion, i.e., the individual trajectories have the very
same shape as for Brownian motion, yet the time the particle needs to follow
these trajectories is provided by a different clock~\cite{Meerschaert:2002}.
Hence to distinguish the CTRW model, respectively the fractional Fokker--Planck
description, from fractional Brownian motion, one can use the statistical
properties of the individual trajectories and derive suitable measures
highlighting the difference, see e.g.\ Refs.~\onlinecite{Magdziarz:2009,
Condamin:2008, Ernst:2012}.

The propagator  describes the probability for a tracer to have moved a distance
$\vec{r}$ after a lag time $t$.
If the stochastic process is in equilibrium this ensemble average can be
obtained also as a moving-time average, i.e., by following a single trajectory
of a tracer.
For CTRW  such that the mean waiting is strictly infinite, the process never
reaches equilibrium, no matter how remote the process has been initiated in the
past~\cite{Bel:2005, Burov:2010, Sokolov:2012, Feller:ProbabilityBd2}.
Then, time correlation functions exhibit \emph{ageing,} i.e., they
explicitly depend on the points in time of the measurements and not just
the time lag inbetween. A closely related phenomenon is referred to as weak
ergodicity breaking and was  introduced by Bouchaud~\cite{Bouchaud:1992} in the
context of spin glasses.
The reason is at the very heart of the concept of CTRW with power-law
distributed waiting times, since as time proceeds, longer and longer waiting
times are drawn from the distribution before the particle actually moves.
In consequence, individual trajectories can differ strongly from the average,
or in other words, the average is not dominated by the \emph{typical}
realisations of the trajectories.
Hence, one has to be careful when comparing predictions for individual
trajectories to ensemble averages and one should calculate probabilities for
the time averages of observable quantities~\cite{He:2008, Lubelski:2008,
Rebenshtok:2008, Sokolov:2008, Sokolov:2009}.

\subsection{Obstructed motion: Lorentz models}
\label{sec:Lorentz}

The models discussed so far implicitly assume that the particle motion occurs
in a homogeneous medium; for example, the waiting time in a CTRW does not
depend on the position.
In contrast, the Lorentz model paradigmatically describes transport in a
spatially heterogeneous medium and displays many facets of anomalous transport
like subdiffusion, crossover phenomena, immobilised particles, and long-time
tails.
In the original version, H.~A.\  Lorentz~\cite{Lorentz:1905} discussed the
motion of a ballistic particle which is elastically scattered off randomly
placed, hard spherical obstacles to lay a microscopic basis for Drude's electric
conductivity of metals.
The medium is reminiscent of a Swiss cheese: it consists of a homogeneous
phase supporting free particle transport, punched by the possibly overlapping
obstacles.
Thus at high obstacle density, a tracer faces a heterogeneous environment
characterised by a significant excluded volume and a highly ramified remaining
space.
Variants of the model include Brownian tracer particles~\cite{Kim:1992,
Lorentz_JCP:2008, Lorentz_VACF:2010, Lorentz_2D:2010} or correlated obstacles,
e.g., by simply forbidding obstacle overlaps~\cite{James:1987, Park:1989,
Kim:1992, Sung:2006, Sung:2008a, Adib:2008, Spanner:thesis}.
\textcite{Saxton:1994} and later \textcite{Sung:2006, Sung:2008a} recognised
the biophysical relevance of such obstacle models for protein transport in
cellular membranes.

The model is described by a number of parameters.
The obstacle radius $\sigma$ fixes the unit of length, and the unit of time
$\tau$ is specified in terms of the velocity $v$ of a ballistic tracer or the
microscopic diffusion constant $D_0$ in case of a Brownian tracer.
Then, the dimensionless obstacle density, $n^*=n\sigma^3$, remains as the only
interesting control parameter; $n$ is the number of obstacles per volume.
Alternatively, one may use the porosity or fraction of unoccupied, void volume,
e.g., $\phi = \exp(-n \upsilon)$ for independently inserted obstacles of
volume~$\upsilon$.

\subsubsection{Microscopic theories}

The ballistic motion of the tracer is randomised by the scattering and becomes
diffusive at long times.
Assuming uncorrelated collisions, \textcite{Lorentz:1905} computed the
diffusion coefficient about a century ago to $D_0=v \sigma/3\pi n^*$ for $d=3$;
for $d=2$, Lorentz--Boltzmann theory yields $D_0=3 v \sigma/16 n^*$
\cite{vanLeeuwen:1967}.
These results are valid only in the dilute limit, $n^*\to 0$.
Actually, subsequent collisions with the obstacles are persistently correlated due
to the frozen disorder, and a proper treatment involves the resummation of
so-called ring collisions.
A systematic low-density expansion of $1/D$ within kinetic
theory~\cite{vanLeeuwen:1967, Weijland:1968} revealed non-analytic terms at the
leading ($d=2$) or next-to-leading ($d=3$) order, supported by pioneering
computer simulations~\cite{Bruin:1972, Bruin:1974}.
For a Brownian tracer, the short-time motion is diffusive already by
construction and the long-time tail of the VACF is obtained at leading order in
a low-density expansion~\cite{Lorentz_VACF:2010}, employing a scattering
formalism for the Smoluchowski operator.

The microscopic theoretical treatment beyond the regime of low densities could
be extended using self-consistent approximations.
\textcite{Goetze:1981a, Goetze:1981b, Goetze:1982} closed the Zwanzig--Mori
equations for the incoherent dynamic structure factor by introducing a set of
approximations for the frequency-dependent memory kernel; an approach that
formed the basis of the successful microscopic mode-coupling theory for the
description of the glass transition~\cite{Goetze:MCT}.
For the Lorentz model, the theory predicts a localisation transition, i.e., a
critical obstacle density $n^*_c$, where long-range transport ceases,
$D(n^*) = 0$ for $n^* \geq n^*_c$.
As the critical density is approached, a temporal window emerges with a
subdiffusive increase of the mean-square displacement; concomitantly, the
long-time diffusion constant is predicted to vanish according to a power law,
$D(n^*) \sim |n^* - n^*_c|^\mu$ for $n^* \uparrow n^*_c$.
A similar picture was obtained within a self-consistent repeated-ring kinetic
theory~\cite{Masters:1982}, although both theories have deficiencies and differ
in their detailed predictions.

Recently, transport of a dense fluid in a porous host structure has attracted
new interest.
Significant progress in this direction has been achieved by
\textcite{Krakoviack:2005, Krakoviack:2007, Krakoviack:2009} by generalising
the mode-coupling theory of the glass transition.
There, a series of non-equilibrium phase transitions was predicted, which
roughly speaking correspond to glass-glass transitions in size-disparate
mixtures~\cite{Voigtmann:2011}.
Some of the predicted phenomena have already been observed in computer
simulations~\cite{Kurzidim:2009, Kurzidim:2011, Kim:2009, Kim:2010a}.
In the limit of small fluid density, the model degenerates to the Lorentz model
and the physics should be dominated by long-wavelength
phenomena~\cite{Lorentz_MCT:2011}; however, these features appear not to be
correctly reproduced by the current theories.
A theoretical framework combining all aspects of glassy dynamics and
localisation is yet to be developed.

\subsubsection{Percolation and random-resistor networks}

A phenomenological approach to the transport properties of the Lorentz model
relies on a mapping to random-resistor networks.
The medium is thought of as a set of voids connected by channels, and transport
occurs via hopping from void to void.
The geometry of the channels, formed by close by obstacles, imposes a
distribution of transition probabilities between the voids.
For many voids, although being neighbours in space, a direct connection is
blocked by obstacles and the transition probability is zero.
The anticipated link to random-resistor networks is provided by
representing the voids as the nodes of a regular lattice and interpreting the
transition probabilities as electric conductances between lattice nodes.
The macroscopic diffusion constant is then identified as the total conductance
of the network.
For a given obstacle configuration, the network may be constructed rigorously
from a Voronoi tessellation~\cite{Kerstein:1983, Elam:1984, Sung:2006,
Sung:2008a}.
Increasing the obstacle densities corresponds to diluting the conductive bonds
of the network, precisely as in the bond percolation
problem~\cite{Stauffer:Percolation, benAvraham:DiffusionInFractals}.
If the bonds are sufficiently diluted, the network falls apart into differently
sized clusters, and above a critical threshold and in the thermodynamic limit,
no spanning cluster exists that supports macroscopic transport.

Let us briefly summarise the essence of percolation theory; we refer the reader
to the excellent textbooks by \textcite{Stauffer:Percolation,
benAvraham:DiffusionInFractals} for a profound introduction.
Directly at the percolation transition, the incipient infinite cluster is a
fractal in a statistical sense.
It is of inextensive weight and occupies a volume $s_\infty(L) \sim L^{\df}$
within a ball of radius~$L$, defining the fractal dimension, $\df<d$.
A self-similar hierarchy of finite clusters coexists with the infinite cluster,
whereby the distribution of cluster sizes $s$ follows a power law, $s^{-1-d/\df}$.
Away from the transition, the medium is no longer scale-free and self-similarity
holds only on length scales below the correlation length, $\xi$.
In particular, the probability to encounter finite clusters of linear extent
larger than $\xi$ is at least exponentially suppressed.
On the percolating side of the transition, the infinite cluster looks
homogeneous at scales larger than $\xi$.
The correlation length exhibits a non-analytic singularity at the transition in
form of a power-law divergence, $\xi \sim |n^*-n^*_c|^{-\nu}$, which introduces
a second critical exponent,~$\nu$.
The macroscopic conductivity $\Sigma$ of a percolating random-resistor network
vanishes at the percolation threshold with a power-law singularity as well,
$\Sigma \sim |n^*-n^*_c|^\mu$.
The conductivity exponent $\mu$ constitutes a third critical exponent, which
describes transport and dynamic phenomena and genuinely extends the set of
geometric exponents $(\df, \nu)$; in general, it does not follow by means of a
scaling relation from the other two.

The percolation transition shares many aspects of a continuous thermodynamic
phase transition and is tractable by renormalisation group
methods~\cite{Stauffer:Percolation, Cardy:Scaling}.
An important lesson is that the critical exponents $\df$, $\nu$, and $\mu$ are
determined by a non-trivial fixed point of the renormalisation group flow, i.e.,
a systematic and consistent coarse-graining of the medium.
As a consequence, the microscopic details of a specific model become irrelevant
for the leading singular behaviour near the transition, and in this sense, the
critical exponents are universal numbers.
The percolation transitions of many different lattice types belong to a single
universality class and are characterised by the same set of exponents,
which depends only on the dimension of space.
A compilation of recent values for the critical exponents of random percolation
is given in \cref{tab:exponents}.

\begin{table} \centering
\newcolumntype{d}{D{.}{.}{1,6}} 
\newcolumntype{s}{D{/}{/}{2,3}} 
\setlength\tabcolsep{1em}
\begin{tabular}{cdd|d}
\hline \hline 
$d$ &
\multicolumn{1}{c}{2} &
\multicolumn{2}{c}{3} \\ \hline 
$\df$ & \multicolumn{1}{s}{91/48^\text{a}} & \multicolumn{2}{d}{2.530(4)^\text{c}} \\
$\nu$ & \multicolumn{1}{s}{4/3^\text{a}} & \multicolumn{2}{d}{0.875(1)^\text{d}} \\
$\Omega$ & \multicolumn{1}{s}{72/91^\text{b}} & \multicolumn{2}{d}{0.64(2)^\text{d}} \\ \hline 
& &
\multicolumn{1}{c|}{lattice} & \multicolumn{1}{c}{continuum} \\ \hline 
$\dw$ & 2.878(1)^\text{a} & 3.88(3)^\text{a} & 4.81(2)^\text{e} \\
$z$ & 3.036(1) & 5.07(6) & 6.30(3) \\
$y$ &
\multicolumn{1}{d}{0.5212(2)} &
\multicolumn{1}{d|}{0.42(2)} &
\multicolumn{1}{d}{0.34(2)} \\
\hline \hline 
\end{tabular}
 \caption{Static and dynamic percolation exponents for the leading and
 sub-leading critical behaviour. Numbers in parentheses indicate the uncertainty
 in the last digit. Sources: (a)~Ref.~\onlinecite{Grassberger:1999},
 (b)~Refs.~\onlinecite{Asikainen:2003, Ziff:2011},
 (c)~from exponent $\tau=1+d/\df=2.186(2)$~\cite{Jan:1998},
 (d)~Ref.~\onlinecite{Lorenz:1998},
 (e)~continuum percolation theory yields $\dw=\df+2/\nu$ for a ballistic
 tracer~\cite{Machta:1985, Lorentz_PRL:2006, Lorentz_JCP:2008}.  The values for
 $y$ and $z$ were calculated from exponent relations,
 $y \dw = \Omega \df$~\cite{Percolation_EPL:2008} and
 $z=2\dw/(2+\df-d)$~\cite{benAvraham:DiffusionInFractals}.
 The dynamic universality class does not split for $d=2$~\cite{Machta:1985,
 Halperin:1985, Lorentz_LTT:2007}.
 }
\label{tab:exponents}
\end{table}

For transport on continuum percolation clusters, a peculiarity arises: the
dynamic universality class may be different from that of lattice models.
In the context of random-resistor networks it was shown that a sufficiently
singular power-law distribution of weak bond conductances,
$\Pi_\sigma(\sigma_\text{bond})\sim \sigma_\text{bond}^{-a}$
for $\sigma_\text{bond} \to 0$,
can dominate the renormalisation flow such that the conductivity exponent $\mu$
deviates from its universal value on lattices,
$\mu=\max \bigl\{ \mu^\text{lat}, \nu (d-2)+ 1 / (1-a) \bigr\}$
\cite{Straley:1982, Machta:1986, Stenull:2001}.
In continuum percolation, the weak conductances arise from narrow gaps or
channels connecting the voids.
For uncorrelated spheres or discs, indeed a power-law distribution was derived,
which is singular enough in three dimensions to modify the exponent; the chain of
arguments has been summarised in Ref.~\onlinecite{Lorentz_JCP:2008}.
\textcite{Halperin:1985} predicted $\mu=\nu+3/2$ for the conductivity problem,
and \textcite{Machta:1986} found $\mu=\nu+2$ for the diffusion constant of
ballistic tracers.

The structural aspects of the Lorentz model with overlapping obstacles have
been studied extensively in the context of continuum percolation.
It shares the phenomenology of lattice percolation, and simulation results are
consistently described by the same universality class~\cite{Elam:1984,
Lorentz_JCP:2008}, even for obstacle mixtures with two different
radii~\cite{Rintoul:2000, vdMarck:1996}.
In the light of renormalisation group theory, this is no surprise and
corroborates the mapping to random-resistor networks.
Percolation thresholds were obtained with high precision yielding a critical
void porosity $\phi_c = 0.0301(3)$ for spheres~\cite{Elam:1984, vdMarck:1996,
Rintoul:2000} and $\phi_c = 0.323\,652\,5(6)$ for discs~\cite{Quintanilla:2007,
Quintanilla:2000}; numbers in parentheses indicate the uncertainty in the last
digit.
Simulations for the Lorentz model~\cite{Lorentz_PRL:2006, Lorentz_JCP:2008,
Lorentz_LTT:2007, Lorentz_2D:2010, Lorentz_space:2011,
Lorentz_percolating:2011} confirm the picture that the localisation transition
is indeed driven by the percolation transition of the medium.
For point tracers, the critical density of the localisation transition is
defined by the percolation threshold.
The critical porosity for localisation of a tracer with finite radius
$\sigma_t$ follows trivially in the case of hard obstacles as
$\phi_c(\sigma_t) =\exp\bigl(-n_c^* (\sigma_t + \sigma)^{-d} \upsilon\bigr)$,
e.g., the localisation transition occurs at a void porosity of
$\phi_c(\sigma_t=\sigma)=\phi_c{}^{1/8} \approx 65\%$ for tracer and obstacles
being spheres of the same size. A similarly high sensitivity of the percolation threshold to the tracer radius
was found numerically for obstacles modelled by soft repulsive
discs~\cite{Saxton:2010}.

\subsubsection{The ant in the labyrinth}

Consider a random walker (``the ant'') on percolation clusters that has to
explore the ramified and self-similar structure of the clusters (``the
labyrinth'').
The problem was posed by \textcite{deGennes:1976} and is amenable to scaling
arguments~\cite{deGennes:1976, Mitescu:1976, Straley:1980, Gefen:1983}
corroborated by Monte-Carlo simulations~\cite{Pandey:1984}.
If the walker is restricted to the incipient infinite cluster at criticality,
one expects subdiffusive motion, $\delta r_\infty^2(t) \sim t^{2/\dw}$, at all
time scales beyond the microscopic regime, $t \gg \tau:=\sigma^2/D_0$.
The exponent $\dw$ is known as walk dimension and one may prefer to consider it
the fundamental critical exponent for the dynamics rather than the conductivity
exponent, $\mu$.
Off criticality, where the infinite cluster becomes homogeneous on distances
much larger than $\xi$, tracer transport crosses over to normal diffusion,
$\delta r_\infty^2(t) \simeq 2d\,D_\infty t$, at long times, $t\gg t_x$.
The crossover time scale may be defined via $\delta r_\infty^2(t_x) = \xi^2$ and
diverges as $t_x \sim \xi^{\dw}$.
Crossover matching yields $t_x^{2/\dw} \sim D_\infty t_x$, which implies that
the macroscopic diffusion constant vanishes at the critical density in a
singular way, $D_\infty(n^*) \sim |n^*-n^*_c|^{\mu_\infty}$, where $\mu_\infty
= \nu (\dw - 2)$.
Transport on a finite cluster is subdiffusive as well with the same exponent
$2/\dw$ as long as the tracer has not fully explored the
cluster~\cite{Gefen:1983, Percolation_EPL:2008}.

If tracers on all clusters are included, the dynamics is non-ergodic since the
time average over a single trajectory differs from an ensemble average.
The average over a self-similar
hierarchy of differently sized clusters reduces the exponent of subdiffusion,
$\delta r^2(t) \sim t^{2/z}$ for $\tau \ll t \ll t_x$, with the dynamic
exponent~$z$ given by $z=\dw / [1-(d-\df)/2] > \dw$.
In the absence of a percolating cluster, the mean-square displacement saturates
for long times, $\delta r^2(t) \simeq \ell^2$ for $t\gg t_x$, and measures the
mean-cluster size, $\ell \sim t_x^{1/z} \sim \xi^{1-(d-\df)/2}$, which is
distinct from the size of the largest finite clusters,~$\xi$.
At the percolating side of the transition, the mean-square displacements for an
all-cluster average increase linearly for long times, $\delta r^2(t) \simeq
2d\,D t $ for $t \gg t_x$.
We shall refer to this behaviour also as \emph{heterogeneous diffusion}
although only the motion on the infinite cluster is diffusive, the all-cluster
averaged propagator being different from a gaussian even for long times.
Since the finite clusters do not contribute, the diffusion constant is further
suppressed by the small weight of the infinite cluster yielding $D(n^*) \sim
|n^*-n^*_c|^\mu$ with $\mu = \nu (\dw - 2 + d - \df)$ for the
all-cluster-averaged motion.
By virtue of an Einstein relation, $\Sigma \sim D$, the exponent $\mu$
corresponds to the conductivity exponent.
The previous relation connecting $\mu$ and $\dw$ emphasises that the leading
singularities of transport-related observables can be described by a single
exponent, e.g., the walk dimension,~$\dw$.

The dynamic scaling hypothesis~\cite{Hohenberg:1977} suggests that the full
time-dependence in the scaling limit is encoded in universal scaling functions,
which extends the notion of universality for the critical exponents.
More specifically, a scaling form of the mean-square displacement is expected
to hold at sufficiently long times, $t\gg \tau$,
\begin{equation}
  \delta r^2(t; n^*) = t^{2/z} \delta \mathcal{R}^2_\pm( \hat t)\/,
  \qquad \hat t := t / t_x \sim t \ell^{-z} \,;
  \label{eq:msd_scaling}
\end{equation}
$\delta \mathcal{R}^2_-$ and $\delta \mathcal{R}^2_+$ denote scaling functions
which describe the crossovers from subdiffusion to heterogeneous diffusion or
localisation, respectively.
They approach the same constant for small arguments, and the correct long-time
behaviour of $\delta r^2(t)$ is reproduced by $\delta \mathcal{R}^2_-(\hat
t)\sim \hat t^{1-2/z}$ and $\delta \mathcal{R}^2_+(\hat t)\sim \hat t^{-2/z}$
for $\hat t\to\infty$.
Numerical results for these functions can be found in
Refs.~\onlinecite{Percolation_EPL:2008, Lorentz_2D:2010} for $d=2$ and in
Ref.~\onlinecite{Lorentz_PRL:2006} for the Lorentz model in $d=3$.
The crossover from intermediate subdiffusion to the long-time behaviour
occurs slowly and extends over several decades in time.

\begin{figure}
  \includegraphics[width=\linewidth]{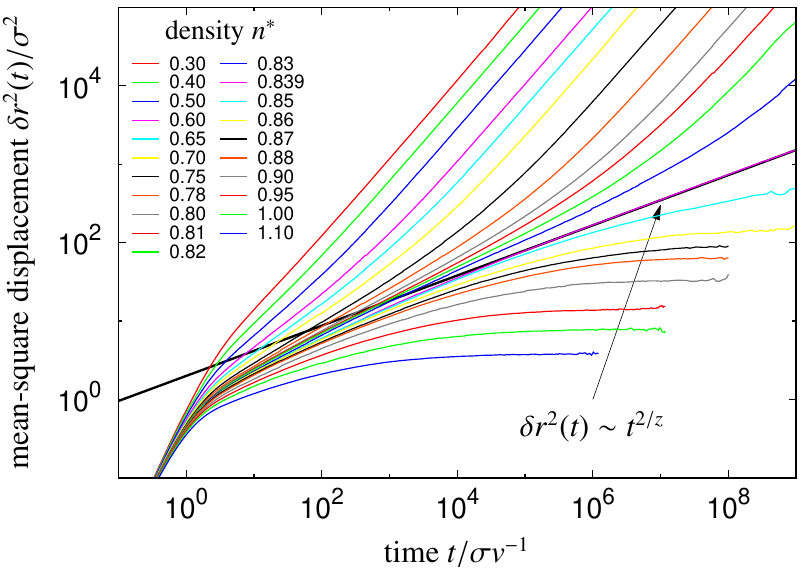}
  \caption{Ensemble-averaged mean-square displacements for different obstacle densities in the
  three-dimensional Lorentz model with a ballistic point tracer. Data taken from
  Ref.~\onlinecite{Lorentz_PRL:2006}.}
  \label{fig:lorentz_msd3d}
\end{figure}

\begin{figure}
  \includegraphics[width=.97\linewidth]{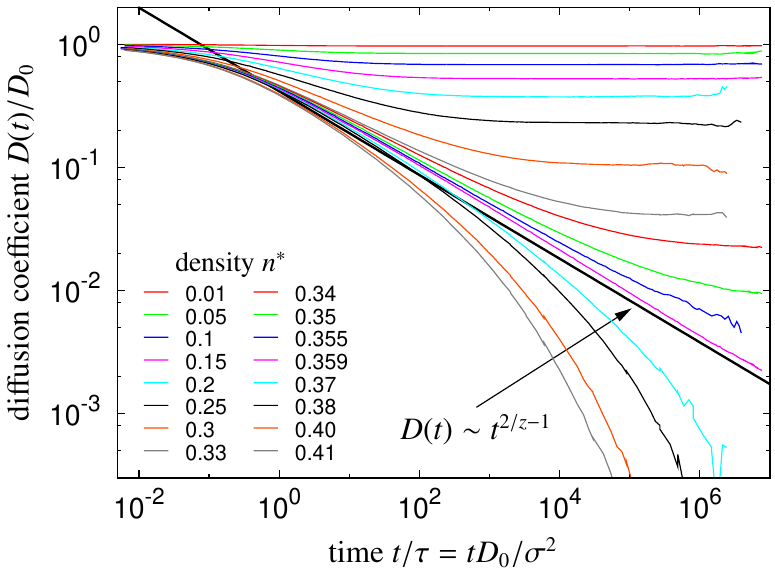}
  \caption{Time-dependent, ensemble-averaged diffusion coefficients for different obstacle
  densities in the two-dimensional Lorentz model with a Brownian point tracer.
  Data taken from Ref.~\onlinecite{Lorentz_2D:2010}.}
  \label{fig:lorentz_diff2d}
\end{figure}

\subsubsection{Simulation results}
\label{sec:Lorentz_simulations}

The ensemble-averaged mean-square displacements of a ballistic tracer in the three-dimensional
Lorentz model were determined by computer simulations over the full range of
obstacle densities~\cite{Lorentz_PRL:2006}, reproduced in
\cref{fig:lorentz_msd3d}.
A temporal window of subdiffusive motion emerges as the critical obstacle
density is approached, and at the critical density, subdiffusion is observed
over 6 decades in time.
The subdiffusion exponent has been determined as $\alpha=2/z \approx 0.32$ and is
consistent with the predictions for continuum percolation, $\mu=\nu + 2 \approx
2.88$~\cite{Machta:1986}.
Likewise, a tracer on the incipient infinite cluster obeys subdiffusion with
$2/\dw\approx 0.42$~\cite{Lorentz_percolating:2011}.
Away from criticality, the subdiffusive motion crosses over to either
heterogeneous diffusion  or localisation beyond the time scale $t_x$.
If particles on all clusters are considered, the non-gaussian parameter slowly
diverges in time, $\alpha_2(t)\sim t^{(d-\df)/\dw}$, reflecting the distinct
length scales $\ell$ and $\xi$~\cite{Lorentz_PRL:2006, Lorentz_JCP:2008}; it
approaches a constant for tracers only on the infinite
cluster~\cite{Lorentz_percolating:2011}.

The same phenomenology was found within simulations for the two-dimensional
Lorentz model with a Brownian point tracer~\cite{Lorentz_2D:2010}
and earlier for an obstructed random walker on a percolating square
lattice~\cite{Saxton:1994}.
The time-dependent diffusion coefficient, $D(t) = \partial_t \delta r^2(t) / 4$,
in \cref{fig:lorentz_diff2d} nicely exhibits a double-crossover scenario from
free diffusion at microscopic time scales to a growing subdiffusive window,
$\tau \ll t \ll t_x$, and back to heterogeneous diffusion with a suppressed
diffusion constant for long times,
\begin{equation}
  D(t) \simeq \begin{cases}
    4 D_0 t \,, & t \ll \tau\,, \\
    \alpha K_\alpha t^{\alpha-1} \,, & \tau \ll t \ll t_x \,, \\
    4 D t \,, & t \gg t_x \,; \\
  \end{cases}
\end{equation}
the data are consistent with the exponent $\alpha=2/z\approx 0.659$ from
lattice percolation in two dimensions.

In the context of ion conductors, the electric conductivity and susceptibility
are relevant material parameters.
Simulation data for the Lorentz model reveal an anomalous frequency-dependence
inherited from the time-dependent diffusion coefficient, $D(t)$, by a
Fourier--Laplace transform~\cite{Gefen:1983, Lorentz_space:2011,
Lorentz_percolating:2011}.

Closer inspection of \cref{fig:lorentz_msd3d} suggests an apparent density
dependence of the subdiffusive motion, which contrasts the notion of one
universal exponent governing the subdiffusive regime and which seems to violate
scaling, \cref{eq:msd_scaling}.
The discrepancy is resolved by noting that the scaling property only holds
asymptotically and that correction terms arise if the time scales under
considerations are not large enough.
It was argued that the leading correction term is non-analytic and well
approximated by a power law,
$\delta r^2(t; n^*) = t^{2/z} \delta \mathcal{R}^2_\pm( \hat t) \bigl(1+C t^{-y}\bigr)$,
where $C$ is a non-universal constant~\cite{Lorentz_PRL:2006}.
The correction exponent $y$, on the other hand, is universal and related to the
correction of the cluster-size distribution at criticality,
$P(s; n^*_c) \sim s^{-1-d/\df} \bigl(1+\tilde C s^{-\Omega} \bigr)$ for
$s\to \infty$.
The exponent relation $y \dw = \Omega \df$ was derived within a
cluster-resolved scaling theory for transport on percolation
clusters~\cite{Percolation_EPL:2008}.
If corrections of this form are taken into account, the scaling property,
\cref{eq:msd_scaling}, is corroborated by extensive simulation data for two- and
three-dimensional Lorentz models~\cite{Lorentz_PRL:2006, Lorentz_JCP:2008,
Lorentz_percolating:2011, Lorentz_2D:2010}.
Both, significant corrections to scaling and slow crossovers imply that a
power-law fit to experimental data over a limited time window is likely to
produce apparent subdiffusion exponents that deviate from the universal value.
Further, deviations from the idealisation of hard, immobile obstacles
potentially wash out the sharp localisation transition.

The mean-square displacement obtained in the computer simulations are ensemble
averages where the initial position of the tracers have been chosen
according to their equilibrium distriubtion, i.e., anywhere in the void
space.
Furthermore, an average over different realisations of the disordered
environment is performed to ensure that all sample-to-sample fluctuations are
averaged out.
Nevertheless, it would be interesting to investigate the convergence of
the averaging procedure.
If the tracer is confined to the infinite cluster only, one should measure how
long single trajectories have to be in order that the time-averaged mean-square
displacement reflects the ensemble-averaged counterpart.

\subsubsection{Spatially resolved transport}

The scaling form of the propagator is more involved than for the models
discussed so far due to the presence of finite clusters and the finite
correlation length.
For transport restricted to the infinite cluster, one expects~\cite{Kertesz:1983,
Percolation_EPL:2008}
\begin{equation}
  P_\infty(r, t;n^*) = \xi^{-d} \mathcal{P}_\infty \bigl(r/\xi, t \xi^{-\dw} \bigr)
\end{equation}
in the scaling limit, $r, \xi \gg \sigma$ and $t \gg \tau$; the density
dependence is encoded on the r.h.s.\ via the divergent correlation length,
$\xi \sim |n^*-n^*_c|^{-\nu}$.
At the critical density, this simplifies to the one-parameter scaling form
familiar from the preceding sections,
\begin{equation}
  P_\infty(r, t;n^*_c) = r^{-d} \widetilde{\mathcal{P}}_\infty \bigl(r t^{-1/\dw}\bigr).
  \label{eq:lorentz_inf_scaling}
\end{equation}
We have determined the scaling function $\widetilde{\mathcal{P}}_\infty(\cdot)$
for a random walker on the incipient infinite cluster of a square lattice (site
percolation) using Monte-Carlo simulations, see \cref{fig:propagators2d}.
For small arguments, the fractal dimension enters the scaling function,
$\widetilde{\mathcal{P}}_\infty(x) \sim x^{\df}$ for $x \ll 1$, with consequences
for the return probability, which is effectively probed in fluorescence
correlation spectroscopy~\cite{FCS_scaling:2011}.
The scaling of the intermediate scattering function was studied
numerically~\cite{Lorentz_ISF:2013} and it was found that certain aspects are
surprisingly well predicted by a mode-coupling approach.

For transport phenomena including all clusters, a dynamic scaling ansatz for
the van Hove function has been conjectured~\cite{Kertesz:1983,
Lorentz_PRL:2006},
\begin{equation}
  P(r, t;n^*) = \xi^{\df-2d} \mathcal{P}_\pm\bigl(r/\xi, t \xi^{-\dw} \bigr) \,;
\end{equation}
the modified prefactor follows from averaging over the cluster size
distribution~\cite{Percolation_EPL:2008} and the index $\pm$ indicates either
the localised phase ($+$) or the regime of heterogeneous diffusion ($-$).
The scaling form has implications for the mean-square displacement,
\cref{eq:msd_scaling}, and the non-gaussian parameter, both of which were
corroborated by extensive simulations~\cite{Pandey:1984, Lorentz_PRL:2006,
Lorentz_JCP:2008, Lorentz_2D:2010}.

A fraction of particles is generically trapped or ``immobile'' at long time
scales by the presence of finite clusters at all obstacle densities.
Immobile particles become manifest for all $n^*$ in a finite Lamb--Mößbauer
factor or non-ergodicity parameter~\cite{Kertesz:1983},
\begin{equation}
  f(k; n^*) := \lim_{t\to\infty} P(k, t) > 0 \,,
\label{eq:Lamb}
\end{equation}
in agreement with simulations~\cite{Hoefling:PhDthesis, Lorentz_space:2011}.
The long-wavelength limit, $\lim_{k\to 0} f(k)$, quantifies the immobile
fraction and displays a singularity at the percolation
transition~\cite{Kertesz:1983}; more generally, a scaling form has been
suggested~\cite{Lorentz_space:2011},
\begin{equation}
  1-f(k;n^*) = \xi^{\df-d} \mathcal{F}_\pm(k \xi)
\end{equation}
for small wavenumbers, $k \ll 2\pi/\sigma$, and close to the transition,
$\xi \gg \sigma$; the asymptotics of the scaling
functions are $\mathcal{F}_-(x \to 0) = \mathit{const}$,
$\mathcal{F}_+(x \to 0) = 0$,
and $\mathcal{F}_\pm(x) \sim x^{d-\df}$ for $x \gg 1$.

\subsubsection{Velocity autocorrelation function}
\label{sec:lorentz_vacf}

The obstacles impose an excluded volume on the tracer particles, which forces
the latter to reverse the direction of motion at some point.
As a result, the particle displacements carry persistent anticorrelations which
become manifest in  an algebraic long-time tail of the VACF, $-Z(t) \sim
t^{-d/2-1}$, first derived within kinetic theory for low obstacle
density~\cite{Weijland:1968}.
More generally, the power-law decay at long times emerges for diffusion in a
frozen, disordered environment~\cite{Ernst:1971a, vanBeijeren:1982}, and it is
already produced by repeated encounters of a Brownian tracer with a single
obstacle~\cite{Lorentz_VACF:2010}, or equivalently, two colloidal
particles~\cite{Ackerson:1982}.
The phenomenon is universal and occurs independently of the obstacle density
with the same exponent at sufficiently long time scales; close to the
localisation transition, the long-time tail competes with an intermediate,
critical power law inherited from the subdiffusive motion~\cite{Goetze:1981a}.
These predictions were confirmed by simulations on a
lattice~\cite{Frenkel:1992} and in the continuum~\cite{Lorentz_LTT:2007,
Lorentz_2D:2010}.
Molecular dynamics simulations for supercooled mixtures of hard
spheres~\cite{Williams:2006, Williams:2006a} and Lennard-Jones
particles~\cite{Glassy_VACF:2012}, both exhibiting slow glassy dynamics, reveal
the same power-law signature in the VACF, which indicates that the underlying
mechanism may also apply for glassforming materials.
A related long-time tail was predicted within a mode-coupling theory for the
force-force correlation function in dense colloidal
suspensions~\cite{Fuchs:2002}.

In the localised regime, $n^* > n^*_c$, where diffusive transport is absent,
\textcite{Machta:1986b} argued that another long-time tail becomes dominant if
there are cul-de-sacs with a broad distribution of exit rates.
The singular distribution of channel width in the overlapping Lorentz model
implies a long-time tail, $-Z(t) \sim t^{-[2+1/(d-1)]}$, for a ballistic
tracer~\cite{Machta:1985}, which has been detected in recent
simulations~\cite{Lorentz_LTT:2007}.

\subsection{Other sources of subdiffusion}

To conclude the section, let us mention further examples of mechanisms, which
are known to lead to  subdiffusion.
The review by \textcite{Bouchaud:1990} introduces a series of lattice models
with quenched disorder, for example random trap and random barrier models or
comb-like structures, which under certain conditions yield subdiffusive
transport.
Next, single-file diffusion describes the motion of strongly interacting
particles aligned in a tube-like structure such that the excluded volume
impedes passing.
As a consequence, the mean-square displacement grows as $\delta r^2(t) \sim
t^{1/2}$~\cite{Wei:2000, Lutz:2004, Lizana:2010}.

For flexible polymers, the chain connectivity induces strongly correlated
motion between the individual chain segments. Transport slows down as the
polymer weight increases. For intermediate time scales, a labelled monomer
displays a subdiffusive mean-square displacement until the slowest Rouse mode
has relaxed~\cite{DoiEdwards:PolymerDynamics, McLeish:2002, Kimmich:2004,
Richter:2005}.
Self-avoidance due to excluded volume and entanglement effects from topological
constraints add to the complexity of the dynamics.
The theoretical implications from the interplay of polymer physics and crowding
(excluded volume) remain largely unexplored to this date.

Apparent subdiffusion may result from incorporating several processes that
occur on different time scales, for example in multi-component mixtures,
due to a polymeric depletion layer around tracers~\cite{Ochab-Marcinek:2011,
Ochab-Marcinek:2012}, or due to internal states.
In these cases, the processes are intrinsic to the constituents or their
interactions and hence cover finite time windows only.
In particular, one can usually not manipulate these processes to generate a
self-similar distribution of time scales required for anomalous transport in
the sense of \cref{sec:complex_transport}.

\section{Experimental techniques}
\label{sec:techniques}

Before we will review the biophysical experiments addressing anomalous
transport, let us introduce some widespread experimental techniques that have
been developed during the past decades and that have proved themselves as
useful tools for the measurement of molecular transport in mesoscopic samples.

\subsection{Single-particle tracking}
\label{sec:particle_tracking}

An important technique that is both intuitive and powerful is provided by
single-particle tracking, for reviews in the context of biophysics see
Refs.~\onlinecite{Kusumi:2005} (giving an extensive historical account in the
supplement) and \onlinecite{Greenleaf:2007}.
Here, a nanoscopic reporter particle or a fluorescent dye is introduced in the
probe and followed by confocal video microscopy and digital image processing.
The trajectory is recorded over sufficiently long time, which provides direct
access to the full statistics of the spatial displacements and which, in
principle, allows for the evaluation of all correlation functions.
The most widely used quantity is, of course, the mean-square displacement as it
is rather robust with respect to experimental noise and appears easy to
interpret.

The spatial resolution of the displacements is typically on the order of a few
nanometres, while the temporal resolution is limited by the image capture rate
to about 10\,ms~\cite{Greenleaf:2007}.
High-speed tracking with a resolution of 25\,µs could be implemented by using
colloidal gold nanoparticles as tracers, which yielded a sufficiently high
signal-to-noise ratio~\cite{Kusumi:2005}.
The length of the recorded trajectories and thus the longest accessible time
scale is limited by photobleaching of the fluorophores and by residual drifts
of the sample baseline.
Single-particle tracking is well suited to study transport in cellular
membranes and to elucidate their structural details~\cite{Kusumi:2005,
Saxton:1997}.
Limitations occur for tracking inside the cytoplasm since the spatial resolution
transverse to the focal plane is naturally an order of magnitude
lower~\cite{Dix:2008}; progress was made recently using
tailored computer algorithms that address the noise and image
inhomogeneities specific to the cytoplasm~\cite{Smith:2011}.
Single-particle tracking has also become a widespread tool to probe
the microrheology of the cytoplasm, see Refs.~\onlinecite{Wirtz:2009, Yao:2009,
Squires:2010} for recent reviews and references therein.

Three-dimensional tracking at the nanoscale has seen significant advances
recently.
Truly three-dimensional trajectories in real-time at a resolution of 32\,ms are
provided by orbital tracking of a single particle, which relies on a feed-back
loop coupled to the laser-scanning microscope~\cite{Levi:2005}.
The ``blindness'' of this technique to the neighbourhood is overcome by
simultaneous wide-field imaging, which is particularly useful to track
particles with nanometre resolution in a heterogeneous
environment~\cite{Katayama:2009}.
Off-focus imaging of trapped particles provides even subnanometre resolution in
all three directions; it requires, however, information about the optical
properties of the sample~\cite{Speidel:2003}.
An alternative to strongly trapped probes is interferometric detection, where
the motion can be resolved with sub-Ångstrøm resolution at
75\,MHz~\cite{Huang:2011}, yielding significant time-correlation functions over
4 decades in time~\cite{Jeney:2008, Resonances_Nature:2011}.

Central for the analysis of single-particle tracking experiments is the
evaluation of time-correlation functions as they contain the proper statistical
description of the motion, see \cref{sec:complex_transport}.
The mere inspection of individual stochastic trajectories may easily lead to
misinterpretations~\cite{Saxton:1997, Feller:ProbabilityBd1}.
The discussion and computation of time-correlation functions and related
quantities is simplified in the case of time-translational invariance, i.e.,
for samples in equilibrium or in a steady state.
Then, the correlation functions depend on the lag time only and provide
information about the motion at a certain time scale (and not at a certain
point in time); they are naturally displayed on a logarithmic time axis.
For example, the mean-square displacement of a single trajectory $x(t)$ of
finite length $T$ given on an equidistant time grid is efficiently evaluated as
a time average employing a fast (discrete) Fourier transformation
$\mathcal{F}$,
\begin{align}
  C_{xx}(t;T)  & := \expect{x(t)\,x(0)}_T
    = \frac{1}{T} \int_0^T\!x(t'+t)\,x(t')\,\diff t'  \\
  &= \mathcal{F}^{-1}_t\left[\frac{1}{T}\,\bigl|\mathcal{F}_\omega[x(t)]\bigr|^2\right] \\
  &= \frac{1}{T}\!\sum_{n\in\mathbb{Z}} \frac{1}{T}\,\bigl|x_T(\omega_n)\bigr|^2 \cos(\omega_n t) \,,
\intertext{where $\omega_n=2\pi n/T$ and}
 x_T(\omega) & := \mathcal{F}_\omega[x(t)]
    =  \int_{-T/2}^{T/2}\!x(t)\,\e^{-\i\omega t}\,\diff t \,;
\end{align}
the mean-square displacement then follows as
$\delta x^2(t)=2[C_{xx}(0;T) - C_{xx}(t;T)]$ for $t\ll T$.
For very long trajectories, a semi-logarithmic sampling of the trajectory data
and a direct evaluation of the correlations has proved a useful
alternative~\cite{Glassy_GPU:2011}.

Automated experimental assays can measure several thousand trajectories
and thus permit the evaluation of displacement histograms, i.e., the propagator
$P(r,t)$, or of the squared displacements, see
\cref{sec:squared_displacements}.
Analysis of these distributions helps to disentangle different transport
processes, for example in mixtures of slow and fast
particles~\cite{Schuetz:1997, Deverall:2005} or when stationary and
non-stationary processes coexist~\cite{Weigel:2011}.
Second, the propagator gives insight into the spatial aspects of a transport
process that would remain undisclosed by a mere investigation of the
mean-square displacement, as has been discussed extensively in
\cref{sec:theoretical_models}.
For example, tracking experiments demonstrated that transport characterised by
a diffusive mean-square displacement may be non-gaussian in
space~\cite{Wang:2009}.
Finally, the ensemble of time-averaged mean-square displacements may display
large fluctuations, in particular if transport is anomalous.
Comparison of ensemble and time averages tests the ergodic hypothesis and may
shed light on the underlying processes in nanoporous structures \cite{Feil:2012,
Kirstein:2007} as well as in cellular membranes~\cite{Weigel:2011}.

\subsection{Fluorescence correlation spectroscopy (FCS)}
\label{sec:fcs}


A second powerful technique that has been widely applied in the biophysical
context is fluorescence correlation spectroscopy (FCS).
Here the basic idea is to label few molecules by a fluorescent dye and record
the fluctuating fluorescent light upon illuminating a small part of the sample.
We briefly introduce the theory underlying the measurement and discuss how
anomalous transport manifests itself in the corresponding correlation function,
comprehensive information about the advantages and limitations of FCS may be
found in the pertinent literature, for reviews see
Refs.~\onlinecite{RiglerElson:FCS, Krichevsky:2002, Hess:2002, Petrov:2008,
Machan:2010}.
Compared to single-particle tracking, an advantage of FCS is the high temporal
resolution in the microsecond regime and the use of tracers as small as a single fluorophore.
On the other hand, the spatial sensitivity is typically on the order of some 100\,nm
(exceptions are discussed below), orders of magnitude larger than that of high-precision tracking experiments.

A typical FCS setup consists of an illumination laser and a confocal microscope
with a photon detector.%
\footnote{In practice, the detector consists of two cross-correlated avalanche
photodiodes to reduce the detector noise.}
The laser beam illuminates the detection volume with the intensity profile
$W(\vec r)$ and excites the fluorophores in the focal volume.
The emitted fluorescent light is collected in the detector, it depends on the
fluctuating, local concentration $c(\vec r,t)$ of labelled molecules.
Thus, the detected light intensity is a spatially weighted
average~\cite[Ref.][section~6.6]{BernePecora:DynamicLightScattering},
\begin{equation}
 I(t) = \epsilon \int\!\diff^d r\, W(\vec{r})\,  c(\vec{r},t)\,,
 \label{eq:fcs_intensity}
\end{equation}
where the prefactor $\epsilon$ accounts for the total quantum efficiency of
absorption, fluorescence, and detection.
The output of the FCS experiment is the time-autocorrelation function $G(t)$ of
the intensity fluctuation $\delta I(t) = I(t) - \expect{I}$ around the mean
intensity; it is conventionally normalised as
\begin{equation}
 G(t) =  \expect{\delta I(t)\,\delta I(0)} \Big/ \expect{I}^2\,.
\end{equation}
Introducing a spatial Fourier transform,
$W(\vec k) := \int\! \diff^d r \, \e^{\i \vec{k} \dotprod \vec{r}} \,W(\vec r) \, ,$
and the intermediate scattering function,
\begin{equation}
 S(\vec{k},t) := \frac{1}{\expect{c}} \int\! \diff^d r \,
 \e^{-\i \vec{k} \dotprod \vec{r}}
 \expect{\delta c(\vec{r},t)\, \delta c(\vec{0},0)} ,
 \label{eq:S_kt}
\end{equation}
one arrives at the expression
\begin{equation}
 G(t) = \frac{1}{N} \frac{\int\! \diff^d k\, |W(\vec{k})|^2\, S(\vec{k},t)}
 {\int \diff^d k\, |W(\vec{k})|^2} \, ,
\label{eq:fcs_integrals}
\end{equation}
where $N:=\expect{I}^2 \big / \expect{\delta I^2}=\expect{c}\!V_\text{eff}$ is
interpreted as the number of fluorophores in the effective illumination volume
$V_\text{eff}$.
In the experiments, the fluorophores are highly diluted and $V_\text{eff}$
usually contains only few molecules or even less than 1, turning FCS
essentially into a single-molecule fluorescence technique~\cite{Petrov:2008}.
Thus, $S(\vec{k},t)$ reduces to the \emph{incoherent} intermediate scattering function,
\begin{equation}
 S(\vec{k},t) \simeq P(\vec{k},t) =
 \expect{\e^{-\i \vec{k} \dotprod \Delta \vec{R}(t)}} \, .
 \label{eq:inc_isf}
\end{equation}
Inserting this in \cref{eq:fcs_integrals} and interchanging the
$\vec k$-integration and the statistical average provides the fundamental
expression for the FCS correlation function~\cite{FCS_scaling:2011},
\begin{equation}
 G(t)\propto \expect{\int\! \diff^d k\, |W(\vec{k})|^2\,
 \e^{-\i \vec{k} \dotprod \Delta \vec{R}(t)}}\,.
 \label{eq:fcs_general}
\end{equation}
It depends solely on the approximation of dilute labelling, it holds for
arbitrary illumination profile and makes no assumptions on the fluorophore
transport.

For practical purposes, however, \cref{eq:fcs_general} shall be evaluated further.
A conventional laser emits a gaussian beam profile, which together with the
usual confocal setup leads to an illumination profile that is often
approximated by an elongated ellipsoid,
\begin{equation}
 W(\vec{r}) \propto \exp\left(-2 (x^2+y^2)/w^2 - 2 z^2/z_0^2 \right)
 \label{eq:gauss_beam}
\end{equation}
with beam waist~$w$ and longitudinal extension $z_0$.
This implies a gaussian filter function,
$|W(\vec{k})|^2 \propto
\exp\left( - \bigl( k_x^2 +k_y^2+\eta^2 k_z^2 \bigr) w^2/4 \right),$
introducing the anisotropy parameter $\eta=z_0/w$.
Then, the $\vec k$-integration can be carried out and gives
\begin{equation}
 G(t; w) = \frac{1}{N} \expect{\exp\left(-\frac{\Delta \vec R(t)^2}{w^2}
 +\frac{1-\eta^{-2}}{w^2}\Delta Z(t)^2\right)} \,.
 \label{eq:fcs_master}
\end{equation}
For planar motion, the displacement along the beam axis vanishes, $\Delta
Z(t)=0$, and a compact statistical expression for the FCS correlation follows.
It reveals the close relationship to the probability distribution of
the squared displacements, \cref{eq:squared_moment_generating}, if $G(t;w)$ is
interpreted as the corresponding characteristic function.
The similarity of the representations in \cref{eq:inc_isf,eq:fcs_master}
suggests that the FCS correlation encodes important spatial information
analogous to scattering methods as soon as the beam waist $w$ is considered an
experimentally adjustable parameter~\cite{FCS_scaling:2011}.
Several FCS setups with spatio-temporal resolution have been implemented
recently by introducing variable beam expanders~\cite{Masuda:2005,
Wawrezinieck:2005}, z-scan FCS~\cite{Humpolickova:2006, Gielen:2009},
sub-wavelength apertures~\cite{Rigneault:2005, Wenger:2007}, near-field
scanning optical microscopy~\cite{Vobornik:2008}, or stimulated emission
depletion (STED) permitting spot sizes as small as 20\,nm~\cite{Kastrup:2005,
Hell:2007, Eggeling:2009}.
Finally, \cref{eq:fcs_master} provides a simple scheme for the efficient evaluation
of autocorrelated FCS data in computer simulations from a given trajectory.

For the analysis of a specific FCS experiment, the statistical average needs to
be performed and some knowledge about the statistical nature of the fluorophore
displacements is required.
A common, but in the context of anomalous transport strong assumption is that
of a gaussian and isotropic distribution of the displacements $\Delta
\vec{R}(t)$ after a fixed time lag, i.e., $\expect{\Delta \vec{R}(t)} =0$ and
only the second cumulant
$\delta r^2(t) \equiv \expect{|\Delta \vec{R}(t)|^2}$
is non-zero.
It follows that
$P(\vec{k},t) = \exp\left(- k^2  \delta r^2(t)/2d\right)$
being the characteristic function of the random displacements,
and the FCS correlation in $d=3$ is given by~\cite{Shusterman:2004}
\begin{subequations}
 \label[equations]{eq:fcs_gaussian}
 \begin{equation}
  \label{eq:fcs_gaussian_3d}
  G_\text{Gauss}(t) = \frac{1}{N}
  \left[1+ \dfrac{2}{3}\dfrac{\delta r^2(t)}{w^2}\right]^{-1}
  \left[1+ \dfrac{2}{3}\dfrac{\delta r^2(t)}{z_0^2}\right]^{-1/2}\,,
 \end{equation}
 which simplifies in two dimensions to~\cite{Avidin:2010}
 \begin{equation}
  \label{eq:fcs_gaussian_2d}
  G_\text{Gauss}(t) = \frac{1}{N}
  \frac{1}{1+ \delta r^2(t)/w^2}\,.
 \end{equation}
\end{subequations}
Both expressions allow for a (numerical%
\footnote{
In three dimensions, one may alternatively consider to expand the last factor
of \cref{eq:fcs_gaussian_3d} for $\eta\gg 1$, yielding the approximation
\[
  \delta r^2(t) \simeq 3w^2 \frac{1 - N G(t)}{2 N G(t) - \eta^{-2}}
  \left\{1+\frac{3}{8}\frac{1 - N G(t)}{[N G(t)]^2}\,\eta^{-4} \right\}\,,
\]
which is useful as long as $NG(t) \gtrsim \eta^{-2}$.
}%
) inversion at fixed time $t$ and conveniently provide direct access to the
mean-square displacement from an FCS experiment~\cite{Shusterman:2008,
Winkler:2006}, without resorting to fitting algorithms.
This will be particularly useful in the generic situation where the transport
exhibits different regimes depending on the time scale, including crossovers
from subdiffusion to normal diffusion (see, e.g.,
Ref.~\onlinecite{Kusumi:2005}).
The quality of the gaussian approximation can be tested experimentally by
resolving the spatio-temporal properties of the molecular
motion~\cite{FCS_scaling:2011}, e.g., by systematic variation of the confocal
volume~\cite{Wawrezinieck:2005, Humpolickova:2006, Gielen:2009, Masuda:2005,
Rigneault:2005, Wenger:2007, Vobornik:2008, Kastrup:2005, Hell:2007,
Eggeling:2009}.
Then, the mean-square displacements obtained from \cref{eq:fcs_gaussian} for
different values of $w$ should coincide if the gaussian approximation is
applicable.

Specialising to free diffusion, it holds $P(\vec{k},t) = \exp\left(-D k^2
t\right)$, and $G(t)$ attains the well-known form
\begin{equation}
 G_\text{free}(t) = \frac{1}{N}
 \Bigl(1+ t/\tau_D\Bigr)^{-1}
 \Bigl(1+ \eta^{-2} t/\tau_D\Bigr)^{-1/2} ,
 \label{eq:fcs_free}
\end{equation}
with the dwell time $\tau_D=w^2/4D$.
Imposing subdiffusion at all times, $\delta r^2(t) \sim t^\alpha$, and assuming
gaussian spatial displacements as in fractional Brownian motion,
\cref{eq:fcs_gaussian} yield
\begin{equation}
 G_\text{FBM}(t) = \frac{1}{N}
 \Bigl[1+ (t/\tau_D)^\alpha \Bigr]^{-1}
 \Bigl[1+ \eta^{-2} (t/\tau_D)^\alpha \Bigr]^{-1/2} .
 \label{eq:fcs_fbm}
\end{equation}
This form has been widely employed in FCS-based studies of anomalous transport,
maybe because it simply supplements the usual model, \cref{eq:fcs_free}, by an
additional fitting parameter~$\alpha$.
Care must be taken that the additional fitting parameter $\alpha$ is not abused
for disclosing experimental deficiencies (e.g., drift) or apparent subdiffusion
where the physical picture would actually favour several species with different
diffusion coefficients.

For FCS experiments with variable detection volume, it is beneficial to
develop some scaling properties of the FCS correlation~\cite{FCS_scaling:2011}.
Starting from a fairly general scaling ansatz for the transport propagator,
$P(r,t) = r^{-d} \mathcal{P} \bigl(r t^{-1/\dw}\bigr)$,
\cref{eq:fcs_master} implies the existence of a scaling function
$\mathcal{G}(\cdot)$ such that
\begin{equation}
  G(t;w) / G(0; w) = \mathcal{G}\bigl(t w^{-\dw}\bigr) \,.
\end{equation}
A representation of the l.h.s.\ as function of the scaling variable $t
w^{-\dw}$ should collapse the FCS correlation data for different beam waists
$w$ onto a single master curve~\cite{FCS_scaling:2011}.
In particular, the half-value times $\tau_{1/2}$, defined via
$G(\tau_{1/2}) / G(0) = 1/2$,
are expected to obey
\begin{equation}
  \tau_{1/2}(w) \sim w^{\dw} \, .
  \label{eq:fcs_tau12_scaling}
\end{equation}

\textcite{Wawrezinieck:2005} suggested to study the phenomenological relation
$\tau_d(w)=t_0 + w^2/4D_\text{eff}$, termed \emph{FCS diffusion law}, where
$\tau_d(w)$ is obtained from fitting the simple diffusion model to the FCS data
for different beam waists.
Based on simulations, they proposed that the transport mechanism can be
inferred from the sign of the axis intercept $t_0$: negative values would
indicate transport hindered by barriers and $t_0>0$ would hint at transport in
the presence of microdomains with slower diffusion.
For a meshwork model, where the tracer diffuses freely in square domains
separated by barriers of finite probability, the relation
$\tau_{1/2}(w)=(w^2-a^2/12)/4 D$ has been derived analytically for large $w$,
with $a$ being the mesh size and $D$ the long-time diffusion
constant~\cite{Destainville:2008}.
This is consistent with the empirical findings of
Ref.~\onlinecite{Wawrezinieck:2005}.
A thorough theoretical analysis for other transport models, however, remains
still to be done~\cite{Saxton:2005}.

A general representation of a correlation function for pure relaxation
dynamics, e.g., for the incoherent intermediate scattering
function, is given by~\cite{Franosch:1999}
\begin{equation}
 P(k,t) = \int_0^\infty \! \e^{-t/\tau}\,\Pi_k(\tau)\,\diff \tau
 \label{eq:isf_distribution}
\end{equation}
with a set of positive and $k$-dependent probability distributions~$\Pi_k$ for
the relaxation times $\tau$.
From \cref{eq:inc_isf,eq:fcs_general}, one finds for the FCS correlation
function
\begin{equation}
 G(t) = \int_0^\infty \!\e^{-t/\tau} \, \Pi_W(\tau)\,\diff\tau,
 \label{eq:fcs_relaxation}
\end{equation}
where the distribution $\Pi_W$ is given by $\Pi_k$ and depends on the
illumination profile, $\Pi_W(\tau) \propto \int\!\diff^d k\, |W(k)|^2 \Pi_k(\tau)$.
An interpretation in terms of many diffusing components may be obtained by
transformation to
\begin{equation}
 G(t) = \int_0^\infty \! G_\text{free}(t;\tau_D)\, \Pi_{\tau_D}(\tau_D)\, \diff \tau_D \,,
 \label{eq:fcs_distribution}
\end{equation}
using the representation
\begin{equation}
 (1+t/\tau_D)^{-d/2} =
 \int_0^\infty \! \e^{-t/\tau}\,
 \frac{(\tau_D/\tau)^{d/2+1}\, \e^{-\tau_D/\tau}}{\tau_D \, \Gamma(d/2)}
 \, \diff \tau \,.
 \label{eq:fcs_simple_relaxation}
\end{equation}
The specialisation to two freely diffusing components is widely employed to fit
experimental data for transport in crowded media,
\begin{equation}
  \Pi_{\tau_D}(\tau_D) = f_1 \delta(\tau_D-\tau_1) + f_2 \delta(\tau_D-\tau_2)\,.
  \label{eq:fcs_two_component}
\end{equation}

For normal as well as for subdiffusive motion, $G(t)$ displays a power-law decay at long times,
$G(t)\sim t^{-\beta}$, implying a power law also for the distribution of
relaxation times, $\Pi_W(\tau)\sim \tau^{-\beta-1}$ as $\tau \to \infty$,
by means of a Tauber theorem~\cite{Karamata:1931, Feller:ProbabilityBd2}.
Iterating the application of the Tauber theorem, the distribution of residence times
displays a power-law tail, $\Pi_{\tau_D}(\tau_D) \sim \tau_D^{-\beta - 1}$
for $\tau_D\to\infty$, only in the case of subdiffusion, $\beta < d/2$.
Note that for spatially non-gaussian transport, the exponent $\beta$ and the
subdiffusion exponent may not be connected by a simple relation;
in case of the Lorentz model, it involves the fractal space dimension,
$\beta = \df/\dw$~\cite{FCS_scaling:2011}.
The numerical determination of the distributions of relaxation times from a
given data set for a slowly decaying correlation function is a formidable
task~\cite{Franosch:1999}; it amounts to performing an inverse Laplace
transform, which is a mathematically ill-posed problem~\cite{Epstein:2008, Petrov:2008}.
Imposing the additional constraint that the ``entropy'' of the distribution
shall be maximal, \textcite{Sengupta:2003} nevertheless introduced a numerical
procedure to determine the distribution of diffusion times $\Pi_{\tau_D}$ from
a given FCS correlation function $G(t)$, which was successfully applied to the
interpretation of experiments~\cite{Banks:2005, Sanabria:2007}.

For the measurement of slow, anomalous transport, it is crucial that the FCS
correlation covers several decades in time and that the data at long times are
still significant.
In particular, the choice of an appropriate fit model requires sufficient
knowledge about the physics of the sample, and physical constraints on fit
parameters or fit windows need consideration~\cite{Malchus:2010}.
The application of FCS in complex cellular environments has several limitations
which may artificially induce weakly subdiffusive behaviour~\cite{Fradin:2003,
Enderlein:2004} or modify the properties of true anomalous
transport~\cite{Malchus:2010}; some of these limitations are addressed by
advanced derivatives of the FCS technique~\cite{Bacia:2006, Petrov:2008}.
We agree that ``the physical insight gained from the empirical application of
the concept of anomalous diffusion to experimental FCS data may be quite
limited, especially if no convincing microscopic origin of the deviation from
the normal diffusion law is provided''~\cite{Petrov:2008}.
This conclusion emphasises the need for more refined theoretical models and for the
possibility of their experimental discrimination.
Long measurement times, a high signal-to-noise ratio, and the combination of
spatial and temporal information seem essential for this task.

\subsection{Fluorescence recovery after photobleaching (FRAP)}
\label{sec:frap}

Simultaneously to the FCS technique, fluorescence recovery after photobleaching
(FRAP) was developed~\cite{Peters:1974, Axelrod:1976a} and readily applied to
membrane proteins on cells~\cite{Peters:1974, Axelrod:1976, Edidin:1976,
Jacobson:1976}, see Refs.~\onlinecite{Elson:1985, Meyvis:1999,
Lippincott-Schwartz:2001, Reits:2001, Verkman:2003, Machan:2010} for reviews.
The method is based on a similar experimental setup as FCS, but initially the
fluorophores in the observation region are bleached by a brief, intense laser
pulse.
Afterwards the fluorescent light intensity emitted from this region is
monitored while it recovers due to the diffusion of unbleached fluorophores
from outside into the observation region.
The method is in some sense complementary to FCS: it uses a high fluorophore
concentration and measures the collective transport in form of a diffusion
front.
It is applicable to very slow processes (even on the scale of several
seconds~\cite{Verkman:2003}) and appears more robust than FCS if a significant
fraction of molecules is immobile~\cite{Machan:2010}.
FRAP has become a valuable tool to measure protein motion and activity in drug
delivery in pharmaceutical research~\cite{Meyvis:1999} as well as in living
cells~\cite{Reits:2001}.

A transparent presentation of the theoretical background was given by
\textcite{Elson:1985} for normal diffusion in two dimensions, which we will
briefly summarise.
The experiment measures the response of the local fluorophore concentration to
an initial quench out of equilibrium.
Similar to FCS, the detected fluorescent light intensity $I(t)$ is given by
\cref{eq:fcs_intensity} as integral over the time-dependent, local concentration
$c(\vec r, t)$ weighted with the intensity profile of the detection laser
$W(\vec r)$.
Let $c_\text{eq}(\vec r)\equiv c_0$ be the homogeneous background concentration
of fluorophores before bleaching, $c(\vec r, 0)$ the imprinted profile directly
after the bleaching pulse, and $\delta c(\vec r, t) := c(\vec r, t) - c_0$ the
partially recovered disturbance after a time $t$.
Repeated measurements of the intensity evolution $I(t)$ result in a
non-equilibrium average, $\expect{I(t)}_\text{n.e.}$; fluctuations across
different measurements are not considered.
The prebleach intensity and the recovery curve are then given by
\begin{align}
I_{-\infty} & := \expect{I(t\to\infty)}_\text{n.e.} = \epsilon c_0 \int \!\diff^d r\, W(\vec r) \,, \\
I_t & :=  \expect{I(t)}_\text{n.e.} =
I_{-\infty} + \epsilon \int \!\diff^d r\, W(\vec r) \, \expect{\delta c(\vec r, t)}_\text{n.e.} \,.
\end{align}
The postbleach intensity is defined as $I_0 := \lim_{t\to 0} I_t$, and we will
give results for the reduced fluorescence recovery function in the following,
\begin{equation}
  \mathcal{I}(t) := \frac{I_{-\infty} - I_t}{I_{-\infty} - I_0} \,.
\end{equation}

Assuming an irreversible conversion of the bleached molecules to a
non-fluorescent state, the fluorescence recovery is solely governed by
fluorophore transport.
Arguing along the lines of Onsager's regression hypothesis, the macroscopic
relaxation to the equilibrium distribution follows the same laws as the
regression of a microscopic disturbance of the local fluorophore concentration
induced by thermal fluctuations~\cite[Ref.][section~7.6]{Hansen:SimpleLiquids}.
(While the fluctuation--dissipation theorem supports this procedure for small
concentration gradients, even for complex dynamics, the validity of this
assumption beyond linear response remains to be corroborated.)
Specifically, the time evolution of a disturbance $\delta c(\vec r, 0)$ is
governed within linear response theory by the equilibrium correlation function
of the concentration fluctuations, namely the van Hove
function,
\begin{equation}
  \expect{\delta c(\vec r, t)}_\text{n.e.} =
  \int\!\diff^d r'\, S(\vec r - \vec r', t) \, \delta c(\vec r', 0) \,,
  \label{eq:conc_evolution}
\end{equation}
where
\begin{equation}
  S(\vec r - \vec r', t)=\frac{1}{c_0}\, \expect{\delta c(\vec r, t) \, \delta c(\vec r', 0)}
\end{equation}
for spatially and temporally homogeneous transport.
In Fourier space, perturbations with different wavevector decouple and their
relaxation is dictated by the intermediate scattering function, \cref{eq:S_kt},
\begin{equation}
  \expect{\delta c(\vec k, t)}_\text{n.e.} = S(\vec k, t) \, \delta c(\vec k, 0) \,.
\end{equation}
For the recovery of detected fluorescence intensity, one finds
\begin{align}
  I_{-\infty} - I_t &=  - \epsilon \int \!\diff^d r\,\diff^d r'\,
  W(\vec r) \, S(\vec r - \vec r', t) \, \delta c(\vec r', 0) \\
  &= - \epsilon \int \!\frac{\diff^d k}{(2\pi)^d}\, W(-\vec k) \, \delta c(\vec k, 0) \, S(\vec k, t) \,.
\end{align}
The reduced fluorescence recovery function follows as
\begin{equation}
  \mathcal{I}(t) = \int \!\diff^d k\, \mathcal{W}(\vec k) \, S(\vec k, t) \,,
  \label{eq:frap_general}
\end{equation}
introducing a filter function,
\begin{equation}
  \mathcal{W}(\vec k) :=
  \frac{W(-\vec k) \, \delta c(\vec k, 0)}{\int\!\diff^d k\, W(-\vec k) \, \delta c(\vec k, 0)} \,,
\end{equation}
to describe the specific experimental setup. Note that bleaching and detection
enter independently and may be implemented with different beam profiles.
$S(\vec k, t)$ is solely determined by the physics of fluorophore transport in
the investigated sample.
If the molecules diffuse freely,
\begin{equation}
  S_\text{free}(\vec k, t) = \exp(-D_c k^2 t) \,,
  \label{eq:S_kt_free}
\end{equation}
where the collective diffusion constant $D_c$ differs in general from the
diffusion constant of a single tracer molecule.

Modelling the fluorophore bleaching as an irreversible first-order reaction,
the concentration of unbleached fluorophore immediately after the bleaching
pulse is given by $c(\vec r, 0)=c_0 \e^{-K(\vec r)}$, where $K(\vec r)$ is
proportional to the beam profile of the bleaching laser and to the duration of
the pulse.
For free diffusion in two dimensions, bleaching with a gaussian beam,
\cref{eq:gauss_beam}, and detection with the same, but attenuated beam, the recovery
of fluorescence intensity can be expressed in terms of the incomplete gamma
function or as an infinite series~\cite{Axelrod:1976a},
\begin{equation}
  I_t = I_{-\infty} \sum_{n \geq 0} \frac{(-K)^n}{n!}\,\frac{1}{1+n(1+2t/\tau_d)}\,;
  \label{eq:frap_gaussian_beam}
\end{equation}
it involves the bleaching parameter $K=K(\vec 0)$ and the diffusion time
$\tau_d = w^2 / 4 D_c$. The postbleach intensity is calculated as
$I_0 = I_{-\infty} \bigl(1-\e^{-K}\bigr) / K$.

If the beam profiles of both the bleaching and the detection laser are
approximated by a step-like disc of radius~$w$, the filter function
$\mathcal{W}(\vec k)$ assumes a relatively simple form and is independent of
the bleaching parameter,
\begin{equation}
  \mathcal{W}(\vec k) = \frac{J_1(k w)^2}{\pi k^2} \,,
\end{equation}
where $J_1(\cdot)$ denotes the Bessel function of the first kind; the postbleach
intensity is $I_0 = I_{-\infty} \e^{-K}$.
In the case of free diffusion, \cref{eq:frap_general,eq:S_kt_free} yield the
fluorescence recovery curve after performing the
$\vec k$-integration~\cite{Soumpasis:1983},
\begin{equation}
  \mathcal{I}_\text{free}(t) = 1 - e^{-2\tau_d / t} [ I_0(2\tau_d / t) + I_1(2\tau_d / t)] \,;
  \label{eq:frap_disc}
\end{equation}
$I_0(\cdot)$ and $I_1(\cdot)$ denote modified Bessel functions of the first kind.

Bleaching patterns other than a spot were implemented experimentally.
In fluorescence pattern photobleaching recovery (FPPR), a fringe pattern of
periodic stripes is bleached to probe the anisotropic dynamics at finite,
non-zero wavenumber~$k$~\cite{Smith:1978, Davoust:1982, Starr:2002,
Gambin:2006}.
Using a laser-scanning confocal microscope allows for bleaching and probing an
arbitrary geometry, e.g., a line segment, with diffraction-limited
resolution~\cite{Wedekind:1996}.
Furthermore, the size of a circular bleaching spot can easily be varied, which
permits the intrinsic determination of the instrumental resolution
parameters~\cite{Smisdom:2011} and which can elucidate the spatio-temporal
properties of anomalous transport analogous to FCS experiments with variable
observation volume.

In some experiments, the initial fluorescence is only partially recovered, even
after long measurement time, which is commonly attributed to further
photobleaching, chemical reactions, or immobile
fluorophores~\cite{Meyvis:1999}.
The unrecovered fluorescence intensity is quantified by the ratio
\begin{equation}
  R = \frac{I_{-\infty} - I_\infty}{I_{-\infty} - I_0} = \mathcal{I}(t\to \infty) \,.
\end{equation}
In this case, fluorophore transport is, strictly speaking, not ergodic and one
infers a finite long-time limit of $S(\vec k, t)$ from \cref{eq:frap_general},
\begin{equation}
  F(\vec k) := \lim_{t\to\infty} S(\vec k, t) \,,
\end{equation}
determining the incomplete recovery,
\begin{equation}
  R = \int \! \diff^d k \, \mathcal{W}(\vec k) \,  F(\vec k) \,.
\end{equation}
In the context of glassy dynamics, $F(\vec k)$ is known as non-ergodicity
parameter or Debye--Waller factor~\cite{Goetze:MCT}, and it plays the same role
as the Lamb--Mößbauer factor in the case of the incoherent dynamics, c.f.\
\cref{eq:Lamb}.
It follows that the ratio $R$ for the same sample depends via
$\mathcal{W}(\vec k)$ on the geometries of the bleaching and detection spots.
In particular, FRAP experiments with variable beam waist $w$ of the laser probe
the non-ergodicity at different scales.
\textcite{Salome:1998} suggested that $R$ is an affine function of the inverse
radius, $1/w$, which is motivated by simulations and experiments on
compartmentalised membranes.
Let us assume that a fraction $R_i$ of the fluorophores is immobile and
homogeneously distributed and that the mobile part is ergodic despite the
presence of immobile molecules.
Then, $F(\vec k) = R_i$ and the unrecovered intensity is identified with the
immobile fraction, $R = R_i$, which is regularly encountered in the literature.
More generally, $R_i$, is obtained as small-wavenumber limit of the
non-ergodicity parameter,
\begin{equation}
  R_i =  \lim_{k\to 0} F(\vec k) \, .
\end{equation}

For the study of anomalous transport, \textcite{Feder:1996} generalised the
FRAP recovery curve, \cref{eq:frap_gaussian_beam}, by assuming a gaussian
subdiffusion model, which effectively amounts to replacing $t/\tau_d$ by
$(t/\tau_d)^\alpha$.
For the intermediate scattering function, this implies a gaussian ansatz in
space and a stretched exponential decay in time,
\begin{equation}
  S(\vec k,t)=\exp\bigl(- (t/\tau_k)^\alpha \bigr)\/,
  \qquad \tau_k\propto k^{-2/\alpha}.
\end{equation}
Such a stretched fluorescence recovery occurs quite slowly and long
measurements are required to distinguish the subdiffusion model from incomplete
recovery~\cite{Feder:1996}.
More general considerations on FRAP for anomalous transport are still to be
worked out; in particular, an extension to spatially non-gaussian transport
would be useful.
To this end, \cref{eq:frap_general} may serve as starting point for a similar
scaling analysis as above for the FCS correlation function.

\section{Anomalous transport in crowded biological media}
\label{sec:crowding}

Application of the discussed techniques to crowded biological media have led to
a plethora of experimental results on anomalous transport in the cell interior
and in related model systems.
In the main part of this section, we provide a compilation of the most
significant biophysical experiments addressing anomalous transport and report
on the progress made during the past decade.
Since a proper characterisation of anomalous transport requires many time and
length scales, emphasis is put on the scales investigated by the different
experiments.
The presentation is divided into three-dimensional transport in cellular fluids
(\cref{tab:crowded_cells}) and crowded model solutions
(\cref{tab:cytoplasm_in_vitro}), and quasi-two-dimensional transport in
cellular membranes (\cref{tab:cellular_membranes}) and lipid bilayer models
(\cref{tab:model_membranes}).
Each experimental subsection is supplemented by a discussion of related
computer simulations (\cref{tab:cytoplasm_simulations,tab:membrane_simulations}).
The section closes by briefly accounting for recent insights in the
implications of crowding for biochemical reactions.

The topic is rather controversially discussed in the biophysics community.
Shortly after the turn of the millennium, it was recognised that the
heterogeneous structure of intracellular environments can not be neglected in
the modelling and description of macromolecular transport and biochemical
reactions, with impact on basically all intracellular
processes~\cite{Ellis:2001, Ellis:2001a, Ellis:2003, Hall:2003}.
While a number of sound experiments on intracellular transport and related
model systems have observed pronounced subdiffusion, other experimental
findings are well explained in terms of normal diffusion~\cite{Dix:2008,
Chiantia:2009}.
The situation becomes even more delicate since ``the sub-optimal experimental
conditions often encountered in cellular measurements do not allow ruling out
simple Brownian diffusion models''~\cite{Chiantia:2009}.
Further, the microscopic mechanisms underlying the subdiffusive motion have not
been identified unambiguously until today.
We agree with \textcite{Elcock:2010} that there is clearly a need for theory
and simulation of microscopic models that can make qualitative and quantitative
predictions of the transport behaviour in crowded environments, at least
\emph{in vitro}.

\subsection{Crowded cellular fluids}

\subsubsection{Cytoplasm and nucleoplasm (\emph{in vivo})}
\label{sec:crowded_cells}

\begin{table*}
  \setlength\tabcolsep{.5em} \setlength\extrarowheight{1ex}
  \fontsize{9pt}{9pt} \selectfont 
  \begin{tabularx}{\linewidth}{%
    >{\raggedright\arraybackslash}p{.12\linewidth}%
    >{\raggedright\arraybackslash}p{.13\linewidth}%
    p{.13\linewidth}p{.18\linewidth}%
    >{\raggedright\arraybackslash}Xp{5ex}%
    >{\raggedright\arraybackslash}p{9ex}}
  \hline \hline
  cell type & probe (size) & experimental \newline technique
    & temporal and spatial \newline scales investigated
    & subdiffusion exponent $\alpha$ \newline or diffusion constant $D$ & year & reference
  \vspace{2pt} \\ \hline
  \emph{S.~pombe}  & lipid granules (300\,nm) & laser tracking, \newline video microscopy
    & 50\,µs\textminus{}1\,ms, \newline 40\,ms\textminus{}3\,min
    & 0.70\textminus{}0.74 & 2004 & ref.~\citenum{Tolic-Norrelykke:2004}
  \\
    & & laser tracking
    & 0.1\,ms\textminus{}10\,ms (PSD) & 0.81 (interphase) \newline 0.84 (cell division)
    & 2009 & ref.~\citenum{Selhuber-Unkel:2009}
  \\
    & & video microscopy & 10\,ms\textminus{}10\,s & 0.4 (time average)
    & 2010 & ref.~\citenum{Tejedor:2010}
  \\
    & & video microscopy, & > 10\,ms, & 0.8, &  2011 & ref.~\citenum{Jeon:2011}
  \\[-\extrarowheight]
    & & laser tracking & 0.1\,ms\textminus{}1\,s
    & normal for $t\lesssim 3\,\text{ms}$, \newline then $\beta=1-\alpha\approx 0.2$ &
  \\
    \emph{E.~coli} & mRNA (100\,nm) & video microscopy
    & 1\,s\textminus{}30\,s, \newline $10^{-3}$\textminus{}1\,Hz (PSD)
    & $0.70 \pm 0.07$, \newline $0.77\pm 0.03$ & 2006 & ref.~\citenum{Golding:2006}
  \\
    & & video microscopy &  1\,s\textminus{}10\textsuperscript{3}\,s
    & $0.71 \pm 0.10$ & 2010 & ref.~\citenum{Weber:2010}
  \\
    \emph{E.~coli}, \newline \emph{C.~crescentus} & chromosomal loci (GFP labelled)
    & video microscopy & 1\,s\textminus{}10\textsuperscript{3}\,s
    & $0.39 \pm 0.04$ (ensemble and time averages coincide)
    & 2010 & ref.~\citenum{Weber:2010}
  \\
  human ostero- sarcoma cells (nucleus) & telomeres \newline (GFP labelled)
    & video microscopy & 10\,ms\textminus{}1\,h
    & subdiffusion with $\alpha$ varying with lag time, \newline
        $\alpha(t)$: 0.32 \textrightarrow\ 0.51 \textrightarrow\ 1.
    & 2009 & ref.~\citenum{Bronstein:2009}
  \\
    SV-80 cells & PS beads (3\,µm) & video microscopy
    &  40\,ms\textminus{}50\,s & 0.5\textminus{}1 for $t > 10$\,s, \newline
      3/2 for $t < 3$\,s (motor proteins)
    & 2002 & ref.~\citenum{Caspi:2002}
  \\
    mammalian \& plant cells & rhodamine dye & FCS
    & $0.1\,\text{ms} < \tau_{1/2} < 1\,\text{s}$
    & 0.6, or 2 components: \newline $D_1 = D_\text{aq}/5$, $D_1/D_2 = 40$
    & 1999 & ref.~\citenum{Schwille:1999a}
  \\
    COS-7 and AT-1 cells & EGFP proteins & FCS
    & $\tau_{1/2} \approx 1\,\text{ms}$
    & 0.7\textminus{}1, or 2 components: \newline $D_1 = D_\text{aq}/5$, $D_1/D_2 = 10$
    & 2000 & ref.~\citenum{Wachsmuth:2000}
  \\
    HeLa cells & FITC-dextran \newline (1.8\textminus{}14.4\,nm) & FCS
    & $0.4\,\text{ms} < \tau_{1/2} < 16\,\text{ms}$
    & 0.71\textminus{}0.84 (depending on tracer size)
    & 2004 & ref.~\citenum{Weiss:2004}
  \\
    HeLa and liver cells & gold beads (5\,nm) & FCS
    & 10\,µs\textminus{}1\,s, $\tau_{1/2} \approx 0.3\,\text{ms}$
    & 0.53, with sucrose added: 0.66 & 2007 & ref.~\citenum{Guigas:2007}
  \\
    mammalian cells & gold beads (5\,nm) & FCS
    & $0.1\,\text{ms} < \tau_{1/2} < 0.9\,\text{ms}$
    & 0.52 (cytoplasm), \newline 0.58 (nucleoplasm) & 2007 & ref.~\citenum{Guigas:2007a}
  \\
    HeLa cells & DNA \newline (20\,bp\textminus{}4.5\,kbp) & FCS
    & $5\,\text{ms} < \tau_{D1} < 20\,\text{ms}$, \newline $\tau_{D2}$ up to 500\,ms (3\,kbp)
    & normal for small tracers, 2~components above 250\,bp:
      $D_2 = D_\text{aq} / 40$ (size dependent)
    & 2005 & ref.~\citenum{Dauty:2005}
  \\
    \emph{D. discoideum} & GFP-actin, \newline free GFP & FCS
    & 170\,µs (globular actin), \newline 240\,µs (free GFP) & 0.83 (both tracers) & 2010 & ref.~\citenum{Engelke:2010}
  \\
    3T3 fibroblasts & dextran, Ficoll \newline (3\textminus{}58\,nm) & FRAP
    & 50\,µm spot size
    & normal with incomplete recovery, $D$ depends on tracer size, $D_\text{aq}/D$ up to 30
    & 1986, 1987 & refs.~\citenum{Luby-Phelps:1986,Luby-Phelps:1987}
  \\
    3T3 fibroblasts, MDCK cells & dextran, Ficoll \newline (4\textminus{}30\,nm) & FRAP & 1\,µs\textminus{}5\,s
    & normal, $D_\text{aq}/D \approx 4$, no dependence on tracer size & 1997 & ref.~\citenum{Seksek:1997}
  \\
    myotubes & globular proteins \newline (1.3\textminus{}7.2\,nm)
    & fringe pattern \newline photobleaching
    & $\tau_{1/2} \approx 1\,\text{s}$ and 208\,s, \newline
        10\textminus{}17\,µm wide stripes
    & normal, $D_\text{aq}/D$: 3\textminus{}7, but 10\textsuperscript{4} for largest protein
    & 2000 & ref.~\citenum{Arrio-Dupont:2000}
  \\
      &  FITC-dextran \newline (2.9\textminus{}12.6\,nm)
    & fringe pattern \newline photobleaching
    & --
    & normal, $D_\text{aq}/D$: 2.5\textminus{}7.7
    & 1996 & ref.~\citenum{Arrio-Dupont:1996}
  \\
    \emph{E. coli} & GFP & FRAP & 0.1\textminus{}1\,s & normal, $D \approx D_\text{aq} / 11$
    & 1999 & ref.~\citenum{Elowitz:1999}
  \\ \hline \hline
  \end{tabularx}
  \caption{Overview of \emph{in vivo} experiments on crowded cellular fluids as
  guide to the discussion in \cref{sec:crowded_cells}. Empty fields repeat the
  entry above. PSD refers to the power spectral density in the optical trap,
  and $D_\text{aq}$ to the diffusion constant in aqueous solution.}
  \label{tab:crowded_cells}
\end{table*}

Measurement of molecular transport in living cells faces a series of
complications typical for \emph{in vivo} experiments: the cell size of 1 to
100\,µm puts a natural upper limit on the accessible length scales, cells may
be in different internal states of their cell cycle, and artificial probes may
trigger a specific cell response.
Further, transport may depend on various cellular processes like directed and
active motion by motor proteins along the microtubule network, or cytoplasmic
flows induced by the cytoskeleton pushing and pulling organelles around and
locally liquefying the cytoplasm~\cite{Arcizet:2008, Keren:2009, Otten:2012}.
For example, the transfection pathway of viruses or gene carriers, visualised
by \emph{in vivo} single-particle tracking, crucially involves an intermediate
phase of slow passive transport, although cargo transport over large distances
is mainly driven by molecular motors~\cite{deBruin:2007, Ruthardt:2010}.
Such enzymatically driven processes may be suppressed or switched off by
reduction of temperature or by depletion of ATP~\cite{Lifland:2011}.
In the following, we will focus on passive transport induced by thermal
fluctuations.

An interesting question is to which extent may the cell be considered in a
stationary state?
The cell is certainly not in an equilibrium state, rather a plethora of
biochemical processes is continuously taking place.
The cell's life time is finite, it ages, and the cellular structures and
processes change during the cell cycle.
These changes, however, occur usually orders of magnitude slower than the
typical experimental time scales of milliseconds or seconds at which
intracellular transport is studied.
We anticipate that a description of the cell as a stationary state is a useful
approximation, which of course has its limitations.
The issue is partially addressed by some of the experiments covered in this
section, a definite answer, however, requires future experimental work.



The fission yeast \emph{Schizosaccharomyces pombe} is a eukaryote with a stiff
cell wall that maintains a stable cylindrical shape, 12\,µm in length and 4\,µm
in width.
It naturally contains lipid granules with diameters of about 300\,nm, which can
serve as endogenous, nearly spherical probe
particles~\cite{Tolic-Norrelykke:2004, Selhuber-Unkel:2009, Tejedor:2010,
Jeon:2011}.
In their prominent work, \textcite{Tolic-Norrelykke:2004} combined the
advantages of 2--3\,min long trajectories from video microscopy (at 25\,Hz) and
the high temporal resolution of laser-based particle tracking (22\,kHz) to
obtain mean-square displacements (MSDs) spanning a time window of more than 4
decades.
The data sets from the two techniques can be consistently extrapolated,
bridging the gap between 1 and 40\,ms.
The data provide clear evidence of subdiffusive motion with an exponent between
0.70 and 0.74 in essentially the whole accessible time window.
The optical trapping experiments were refined five years later by the same
laboratory~\cite{Selhuber-Unkel:2009}, focusing on the different phases of the
cell cycle.
From the positional power spectrum, the subdiffusive motion was confirmed at
time scales between 0.1 and 10\,ms in all phases with a small, but significant
variation of $\alpha$ from 0.81 in the interphase to 0.85 during cell division.
The larger exponents compared to the earlier work were attributed to possibly
disrupted intracellular actin filaments there, and the difference in exponents
during the cell cycle was interpreted as a sensitivity to changes of the
cytoskeleton.

From the corresponding video microscopy data in the interphase of \emph{S.
pombe}, time-averaged MSDs were computed from a small set of 8 long
trajectories (up to 200\,s)~\cite{Tejedor:2010}.
The MSDs cover time scales from 10\,ms to 10\,s and indicate subdiffusion with
a relatively low exponent of $\alpha\approx0.4$.
The ensemble- and time-averaged MSDs support this low exponent for
$0.1\,\text{s} < t < 1\,\text{s}$, but the statistical noise permits a possible
crossover to the previously found larger exponent at shorter times.
Additional statistical quantities like higher cumulants or the mean maximal
excursion are used to test the shape of the probability distribution of the
displacements.
It was concluded that the investigated data set is neither compatible with the
CTRW model nor with a percolation scenario, but shares some features with
fractional Brownian motion.
Further improvements of the laser-tracking setup yielded very long trajectories
with 22\,kHz resolution of lipid granules during cell division of \emph{S.\
pombe}~\cite{Jeon:2011}.
The time-averaged MSDs at time scales between $10^{-4}$ to 1\,s, clearly
exhibit a crossover around 3\,ms from normal diffusion to subdiffusion with
exponent $\beta \approx 0.2$, while the short-time MSDs obtained
earlier~\cite{Tolic-Norrelykke:2004, Selhuber-Unkel:2009} display subdiffusion
with the exponent $\alpha\approx 1-\beta$.
Such a peculiar behaviour is recreated in terms of a CTRW model with
truncated power-law waiting-time distribution and has been related to the
confinement by the optical trap.
We refer the reader to the original literature for a detailed
discussion~\cite{Selhuber-Unkel:2009, Tejedor:2010, Jeon:2011}.
Tracking the granules by video microscopy at time scales above 10\,ms yielded
subdiffusive motion with $\alpha\approx 0.8$, similarly as in the earlier
experiments.

Summarising, the transport of granules in fission yeast has been studied with
great precision over almost a decade and displays a complex behaviour
varying over the range of observation time scales.
The experimental data are partially described in terms of FBM and a properly
adapted CTRW model, but a consistent phenomenological description valid at all
time scales is yet to be found.
The microscopic mechanism leading to subdiffusive motion has not yet been
resolved.


In a highly recognised experiment addressing the physical nature of bacterial
cytoplasm, \textcite{Golding:2006} used video microscopy at 1\,Hz to follow the
motion of fluorescently labelled mRNA macromolecules (radius on the order of
100\,nm) over up to 30 minutes in \emph{E.\ coli}, a rod-shaped bacterium about
2\,µm long and half a micron in diameter.
The analysis of 70 trajectories yielded subdiffusion at time scales between 1
to 30\,s with varying prefactor, but rather robust exponent
$\alpha=0.70\pm0.07$.
Concatenating all trajectories, a power spectrum was obtained over three
decades in time, which clearly indicates subdiffusive behaviour with exponent
$\alpha=0.77\pm0.03$.
These findings agree with the limiting value of $\alpha$ in artificial crowding
conditions using dextran~\cite{Banks:2005}, see
\cref{sec:cytoplasm_in_vitro}.

Subdiffusion with a significantly smaller value $\alpha=0.39\pm0.04$ was
obtained from time-lapse fluorescence microscopy tracking of GFP-labelled
chromosomal loci in \emph{E.\ coli} and \emph{Caulobacter crescentus} over
$10^3$\,s~\cite{Weber:2010}.
For RNA proteins in \emph{E.\ coli}, subdiffusion with $\alpha=0.71\pm0.10$ was
observed in agreement with Ref.~\onlinecite{Golding:2006}.
These findings were rationalised by supplementing the Rouse model for polymer
dynamics by a fractional gaussian noise term~\cite{Weber:2010a} and arguing
that the monomer dynamics at short times is then characterised by the exponent
$\alpha'=\alpha/2$.
Further, a CTRW scenario was ruled out as time- and ensemble-averaged data
coincide.
Analysis of the VACF showed pronounced anti-correlations and a description in
terms of a Langevin equation with power-law correlated, spatially gaussian noise
was favoured.

Tracking the motion of telomeres in the nucleus of human osterosarcoma
cells in a broad time range of almost 6 orders of magnitude displayed
subdiffusive motion with varying exponent depending on lag
time~\cite{Bronstein:2009}.
Between 0.1\,ms and 1\,s, a small value of $\alpha \approx 0.32\pm 0.12$ was
found, changing to $0.51 \pm 0.20$  in the intermediate regime up to 200\,s,
and approaches normal diffusion for longer times; the amplitude of the power laws
varied over an order of magnitude from one telomer to another.
An interpretion in the framework of the CTRW was found to not fully explain the
experimental results, mainly since ageing behaviour could not be corroborated.
The dynamics is, however, compatible with the dynamics of entangled polymers:
the reptation model for a Rouse chain predicts an exponent of 1/4 for the
short-time relaxation, crossing over to 1/2, and finally free diffusion.

The interaction with cytoskeletal fibres and motor proteins was probed by
engulfing 3\,µm polystyrene beads into various eukaryotic cells and monitoring
their motion by video microscopy~\cite{Caspi:2002}.
The obtained MSD was found to grow \emph{super-}diffusively with $\alpha
\approx 3/2$ at time scales less than 3\,s, while a crossover to subdiffusive
motion was observed for longer times ($0.5 < \alpha < 1$).
The latter finding, however, is subject to statistical noise and could be
followed over less than half a decade only.
The superdiffusive transport, $\delta r^2(t)\sim t^{3/2}$, was explained in
terms of a generalised Langevin equation that accounts for the stochastic
driving by motor proteins and the power-law memory due to the subdiffusive
thermal fluctuations of semiflexible polymers, $\delta r^2(t)\sim t^{3/4}$.
Further, the crossover to subdiffusion with $\alpha=1/2$ was rationalised by
the decorrelation of motor activity at long time scales.


Owing to technological advances in the 1990s, the FCS technique has quickly
become an established tool for dynamic studies at the mesoscale, permitting
also intracellular applications.
\emph{In vivo} measurements benefit from the high temporal resolution as well
as from the choice of tracers ranging from tiny fluorescent proteins to
labelled polymer coils and inert nanoparticles.
Such tracers can be incorporated into the cell with only mild physiological
side effects.
In a pioneering work, \textcite{Schwille:1999a} established the application of
FCS to the cytoplasm by studying the diffusion of tetra-methyl-rhodamine dye
with single molecule sensitivity in various mammalian and plant cells.
They found subdiffusion in the different cell types with $\alpha\approx 0.6$,
but the same data may equally well be rationalised by fitting a mixture of two
normally diffusing components, the faster one being 5-fold slower than in
aqueous solution.
The slow component showed diffusion constants up to 40 times smaller than the
fast one and comprised 35--60\% of the molecules; the slow component was
attributed to membrane-bound dye as the diffusion coefficients are of
similar magnitude.

The variation of anomalous transport with the spatial position in the cell
was addressed using genetically modified COS-7 and AT-1 cells that express
fluorescent EGFP proteins~\cite{Wachsmuth:2000}.
The obtained FCS correlations displayed anomalous diffusion depending on the
position in the cell, the largest anomalies were observed in the nuclei.
The data were analysed in terms of subdiffusion and two diffusive components.
In the first case, the exponent $\alpha$ varied between 0.7 and 1 with
half-value times around 1\,ms.
The two-component fits yielded a ratio between the diffusion constants of about
10, where the larger value was 5-fold reduced compared to free diffusion of
EGFP in aqueous solution.

\textcite{Weiss:2004} introduced differently sized FITC-labelled dextrans in
HeLa cells and characterised their motion with FCS.
For all molecular weights studied, the correlation functions displayed
anomalous transport and were fitted with the gaussian subdiffusion model,
\cref{eq:fcs_fbm}.
The obtained values for the exponent $\alpha$ varied between 0.71 and 0.84
non-monotonically depending on the size of the dextran polymers, which covered
hydrodynamic radii in buffer solution between 1.8 and 14.4\,nm.
The dwell times $\tau_D$, on the other hand, increased systematically from 0.4
to 16\,ms.
Complementing their study by \emph{in vitro} experiments with unlabelled
dextran as crowding agent, the authors found a systematic decrease of $\alpha$
with the concentration of dextran, which suggests to quantify the degree of
crowdedness in terms of the subdiffusion exponent~$\alpha$.

Probing the cytoplasm and the nucleus of living HeLa cells as well as healthy
and cancerous liver cells at the nanoscale by introducing fluorescently tagged
5\,nm-sized gold beads revealed pronounced anomalous transport over five decades
in time~\cite{Guigas:2007}.
One advantage of using gold particles instead of (branched) polymers as probe
is to reduce interactions with the cytoplasm, like polymer entanglement,
potentially modifying the tracer transport.
The FCS data were fitted with an empiric model that interpolates between
subdiffusion at short times and normal diffusion at long times, see
\cref{eq:fcs_gaussian_3d} with
$\delta r^2(t)=6 w^2 \bigl[ t/\tau_d + (t/\tau_s)^\alpha \bigl]$.
It consistently yielded $\alpha\approx 0.53$, $\tau_s\approx 0.3\,\text{ms}$,
and $\tau_d\approx 90\,\text{ms}$ across all cell types, with $\alpha$ being
slightly larger in the nucleus;
these parameters imply a crossover time scale between the two transport
regimes,
$t_x = \left(\tau_d \tau_s^{-\alpha}\right)^{1/(1-\alpha)}\approx 56\,\text{s}$,
which is beyond the scope of current experiments.
When the cells were osmotically stressed by adding high, but non-aptotic
concentrations of sucrose, raffinose, or NaCl, the subdiffusive transport
changed markedly: $\alpha$ attained values around 0.66, while the dwell time
$\tau_s$ increased by a factor of 3 in the cytoplasm and up to 2.4\,ms in the
nucleus.
Considering the cyto- or nucleoplasm a concentrated polymer solution, the
observations were rationalised by a change of solvent conditions qualitatively
modifying the visco-elastic properties of the medium.

An extension of this study~\cite{Guigas:2007a} covered a representative
collection of mammalian cells, with different origin (from Chinese hamster
ovary (CHO) cells to human osteosarcoma cells) and state of health
(immortalised Thle-3 from a healthy liver \emph{vs.}\ highly aggressive human
glioma cells).
Again, the gold nanobead displayed subdiffusive motion with $\alpha\approx
0.52$ in the cytoplasm and $\alpha\approx 0.58$ inside the nucleus, with only
small variations across cell lines; the dwell times varied between 0.1 and
0.3\,ms with some exceptions reaching up to 0.9\,ms in the nucleoplasm.
In conclusion, very different mammalian cells show a similar degree of crowding
at the nanoscale with the nucleus being somewhat less crowded than the
cytoplasm.
In particular, differences in the macroscopic visco-elastic response depending
on a cell's development and disease state appear to be not reflected in the
transport at the nanoscale.

\textcite{Dauty:2005} investigated the size-dependent transport of DNA in the
cytoplasm of living HeLa cells with DNA molecules sized between 20 to 4,500\,bp
and labelled with a single fluorophore.
The motion of DNA was followed by FCS, the fitting of correlation functions for
sizes above 250\,bp required a two-component model with a short diffusion time
of 5$-$20\,ms independent of DNA size; the diffusion time of the slow component
ranged up to $\approx 500$\,ms for 3\,kbp.
Various sources were suggested as origin of the fast component, yet it could not
be identified unambiguously.
DNA diffusion was found to be significantly reduced in comparison to the free
diffusion in saline by factors of up to 40, with a pronounced dependence on
molecular weight above 500\,bp.
This dependence, however, almost disappeared by disrupting the actin
cytoskeleton: the FCS data exhibited simple diffusion with an approximately
5-fold reduction.
The authors have corroborated their findings by \emph{in vitro} experiments
using crowded solutions, cytosol extracts, and reconstituted actin networks.
Only for the actin networks, the suppression of the DNA diffusion constant over
its free value was sensitive to the molecular weight, recreating the behaviour
in intact HeLa cells.
In all other environments, simple diffusion was observed with reduction factors
not exceeding 5.
The authors concluded that mobile obstacles can not explain the strongly
reduced mobility of DNA in living cells and that the actin cytoskeleton
presents a major restriction to cytoplasmic transport.
Further, the sensitivity of diffusion to the molecular weight may be explained
by entanglement effects with the actin mesh and reptation dynamics of the
elongated DNA molecules.

Transport of globular actin molecules, a building block of the cytoskeleton,
was investigated by FCS in living \emph{Dictyostelium discoideum}
cells~\cite{Engelke:2010} using mutants that express either GFP-labelled actin
or free GFP.
Measurements in the cytoplasm displayed anomalous transport for both probes,
fits of the gaussian subdiffusion model yielded in both cases
$\alpha\approx0.83$ with half-value times of 170\,µs for free GFP and 240\,µs
for G-actin.
In the lysate of the actin mutant, transport was even slower with
$\alpha\approx 0.72$, but with a half-value time reduced by a factor of 2.5;
here, a two-component fit was not able to detect an anticipated slow diffusive
component.
Further, a low degree of actin polymerisation was found in the cytoplasm, and
thus, the slow transport was mainly attributed to crowding effects.
Transport in the presence of an highly polymerised actin network was monitored
by means of fluorescent LIM proteins marking the actin polymerisation front in
the cell cortex.
While transport was normal in the corresponding lysate, subdiffusive motion
with $\alpha=0.79\pm0.02$ was observed in the living cell.
Given that an exponent 3/4 is expected for \emph{in vitro} actin
networks~\cite{Winkler:2006}, the observed anomalous transport may be caused by
polymer fluctuations in this case.


Luby-Phelps \emph{et al.}\ applied the FRAP technique to fluorescently labelled
dextran~\cite{Luby-Phelps:1986} and Ficoll~\cite{Luby-Phelps:1987} molecules
microinjected into the cytoplasm of living Swiss 3T3 fibroblasts.
The tracer radii ranged from 3\,nm to 25\,nm (Ficoll) and 58\,nm (dextran).
The obtained diffusion constants were substantially reduced relative to their
values in aqueous solution, depending on the tracer size; a 30-fold reduction
was observed for the largest Ficolls.
The initial fluorescence was fully recovered for molecules smaller than 14\,nm
(dextran) or about 20\,nm (Ficoll), indicating that all tracers are mobile.
For larger molecules, the fluorescence recovery was increasingly incomplete,
and it was inferred that a fraction of the molecules is trapped.
These observations are consistent with the picture of a significant excluded
volume hindering cytoplasmic transport.

Targeting the same experimental system, \textcite{Seksek:1997} made different
observations using an improved FRAP apparatus with microsecond resolution and
measurement times of several seconds.
The diffusion constants of labelled dextran and Ficoll molecules in the
cytoplasm of T3T fibroblasts as well as MDCK epithelial cells were only 4-fold
reduced relative to aqueous solution.
In particular, no size dependence of this factor could be detected for a
comparable range of tracer radii, in contrast to the results in
Refs.~\onlinecite{Luby-Phelps:1986, Luby-Phelps:1987}.
As in these works, the FRAP analysis was based on the half-value times of
fluorescence recovery rather than on fits of the full recovery curve.
Nevertheless, investigation of the recovery dynamics at long times did not
display deviations form normal diffusion.
In agreement with the earlier experiments, fluorescence was only partially
recovered for increasing molecular weight, even at the scale of seconds.
This is again consistent with the presence of spatial heterogeneity in the form
of persistent structures or obstacles that immobilise a significant fraction of
the larger macromolecules.

A derivative of the FRAP technique is modulated fringe pattern photobleaching,
which was employed by \textcite{Arrio-Dupont:2000} to investigate the mobility
of different globular proteins in cultured myotubes.
The studied myotubes are elongated cells, 10$-$40\,µm wide and up to 1\,mm in
length.
Applying fringe patterns of 10$-$17\,µm wide stripes yielded recovery time
constants of several seconds for labelled proteins with hydrodynamic radii
between 1.3 and 5.4\,nm.
These time constants translate into a 3 to 7-fold suppression of protein
mobility relative to aqueous solution.
For $\beta$-galactosidase, which has a hydrodynamic radius of $\approx$7.2\,nm,
a drastic suppression by 4 orders of magnitude compared to aqueous solution was
found from the extremely slow fluorescence recovery with a time constant of
208\,s.
The results clearly display a monotone dependence of the suppression factor on
protein size.
Similar results were obtained for a series of differently sized dextran
molecules in the same cell type~\cite{Arrio-Dupont:1996}, but the mobility is
more size-sensitive for globular proteins.
The apparent discrepancy with the findings of Ref.~\onlinecite{Seksek:1997} may
be explained by the very different time and length scales probed by both
experiments, and in this sense they complement each other.

Combining FRAP and a photoactivation technique allowed for a non-invasive
measurement of the motion of endogenously expressed GFP in the cytoplasm of
\emph{E.~coli} bacteria~\cite{Elowitz:1999}.
The apparent diffusion constant was obtained as $7.7 \pm
2.5$\,µm\textsuperscript{2}/s at the time scale of some 0.1\,s, which is about
11~times smaller than for free diffusion in water and significantly lower than in
eukaryotic cells.

\subsubsection{Crowded model solutions (\emph{in vitro})}
\label{sec:cytoplasm_in_vitro}

\begin{table*}
  \setlength\tabcolsep{.5em} \setlength\extrarowheight{1ex}
  \fontsize{9pt}{9pt} \selectfont 
  \begin{tabularx}{\linewidth}{%
    >{\raggedright\arraybackslash}p{.12\linewidth}%
    >{\raggedright\arraybackslash}p{.13\linewidth}%
    >{\raggedright\arraybackslash}p{.13\linewidth}p{.18\linewidth}%
    >{\raggedright\arraybackslash}Xp{5ex}
    >{\raggedright\arraybackslash}p{9ex}}
  \hline \hline
  crowding agent & probe (size) & experimental \newline technique
    & temporal and spatial \newline scales investigated
    & subdiffusion exponent $\alpha$ \newline or diffusion constant $D$ & year & reference
  \vspace{2pt} \\ \hline
  F-actin & colloid (0.25\,µm) & video microscopy
    & 0.1\textminus{}10\,s
    & $0 \leq \alpha \leq 1$ depending on mesh size of actin network & 2004 & ref.~\citenum{Wong:2004}
  \\
  & colloid (0.48\,µm) & diffusing wave spectroscopy
    & 1\,µs\textminus{}100\,s
    & crossover \textit{normal} \textrightarrow\ \textit{caging} near 10\,ms,
      $D(t)$ spans 10\textsuperscript{6}
    & 1999 & ref.~\citenum{Palmer:1999}
  \\
    & nanosphere (25\textminus{}250\,nm) & video microscopy
    & 50\,ms\textminus{}10\,s
    & normal MSD, but anomalous, exponential propagator
    & 2009 & ref.~\citenum{Wang:2009}
  \\
  \emph{fd} virus & protein (3.5\,nm), \newline silica spheres \newline (35\textminus{}500\,nm)
    & FCS, dynamic light scattering, video microscopy
    & sensitivity: 10\,µs\textminus{}1\,s
    & $D_\text{aq}/D$: 1 \textrightarrow\ 10 as [\emph{fd}]~\textuparrow
    & 2005 & ref.~\citenum{Kang:2005}
  \\
    dextran & FITC-dextran \newline (10\textminus{}500\,kDa) & FCS
    & $\tau_{1/2}$: 0.3 \textrightarrow\ 30\,ms
    & 1 \textrightarrow\ 0.6 monotonically \newline as [dextran]~\textuparrow
    & 2004 & ref.~\citenum{Weiss:2004}
  \\
    & apoferritin & FCS & $\tau_{1/2} \approx 100\,\text{ms}$ \newline (675\,µs in saline)
    & 0.7\textminus{}0.9 (mean 0.82), no ageing, stationary process & 2009 & ref.~\citenum{Szymanski:2009}
  \\
    & streptavidin, EGFP & FCS & 1\,ms\textminus{}1\,s
    & 1 \textrightarrow\ 0.74 monotonically \newline as [dextran]~\textuparrow & 2005 & ref.~\citenum{Banks:2005}
  \\
    dextran (30~wt\%) & beads (50\,nm) & laser tracking & 50\textminus{}500\,ms
    & subdiffusion ($\alpha=0.82$), no signs of ergodicity breaking, FBM most likely
    & 2012 & ref.~\citenum{Ernst:2012}
  \\
    Ficoll-70 & dye, proteins, polymers, beads \newline (1\textminus{}100\,nm) & FCS
    & $\tau_{1/2}$: 0.3 \textrightarrow\ 100\,ms \newline (rhodamine green)
    & normal, $D_\text{aq}/D > 100$, exponential suppression:
    $D\sim \exp(-a\,[\text{\footnotesize Ficoll}])$
    & 2004, 2008 & refs.~\citenum{Dauty:2004,Dix:2008}
  \\
    micellar solution \newline (C\textsubscript{12}E\textsubscript{6})
    & dye, proteins, nanoparticles \newline (1.7\textminus{}190\,nm) & FCS
    & --
    & normal, 2 components, \newline $D_\text{aq}/D_1$:
      1 \textrightarrow\ 10 (free tracers), $D_1/D_2 \approx 10$
    & 2006 & refs.~\citenum{Szymanski:2006a,Szymanski:2006}
  \\
    dextran (500\,kDa) & ribonuclease~A & FCS
    & --
    & normal, free and bound protein: $D_1/D_2 \approx 7$
    & 2011 & ref.~\citenum{Zustiak:2011}
  \\
    BSA protein & BSA protein (3.6\,nm) & neutron scattering
    & 0.3\textminus{}5\,ns, \newline wavelength: 0.5\textminus{}3.3\,nm
    & normal, $D_\text{aq}/D_\text{short}$: 1 \textrightarrow\ 10,
      hydrodynamic effects
    & 2010, 2011 & refs.~\citenum{Roosen-Runge:2011,Roosen-Runge:2010}
  \\
    glycerol, PVP, Ficoll, proteins, \emph{E. coli} lysate & CI2 protein (7.4\,kDa)
    & nuclear magnetic \newline resonance
    & -- & test of Stokes--Einstein relation for translational and rotational motion
    & 2010 & ref.~\citenum{Wang:2010}
  \\ \hline \hline
  \end{tabularx}
  \caption{Overview of \emph{in vitro} experiments on crowded model fluids as
  guide to the discussion in \cref{sec:cytoplasm_in_vitro}. Empty fields repeat
  the entry above, right arrows indicate a systematic change upon variation of
  some parameter, up/down arrows indicate increase/decrease of some quantity.
  Abbreviations: FBM\dots{}fractional Brownian motion,
  [X]\dots{}concentration of X.}
  \label{tab:cytoplasm_in_vitro}
\end{table*}

Important progress in the understanding of anomalous transport was made by
biologically motivated model systems, where in contrast to living cells key
parameters are adjustable.
In many cases, the crowded solutions are made from linear or branched
(bio-)polymers of different rigidity, but proteins were used as crowding agents
just as well.


A component of the cytoskeleton is filamentous actin, an elongated,
semiflexible polymer forming dense, entangled networks, e.g., in the cell
cortex.
Solutions of reconstituted F-actin offer a model system to investigate the
mechanical properties of these networks \emph{in vitro}.
The thermal motion of colloidal tracer particles (radius 0.25\,µm) in semidilute
F-actin solutions was tracked with video microscopy at 30\,Hz over
20\,min~\cite{Wong:2004}.
The obtained MSDs display subdiffusion on time scales between 0.1 and 10\,s
with an exponent varying between 0 and 1 upon varying the actin concentration
and thereby the mesh size of the networks between 0.17 and 0.75\,µm.
Normal diffusion was observed for mesh sizes larger than the tracer radius.
The results can not be interpreted in terms of the macroscopic
frequency-dependent shear viscosity.
Analysis of the trajectories suggests that the tracers are caged by the network
and the motion resembles a series of infrequent and large jumps with power-law
distributed waiting times.

Diffusing wave spectroscopy experiments~\cite{Palmer:1999} on 0.48\,µm-sized
colloids embedded in F-actin solutions explored the large temporal window
between 1\,µs and 100\,s.
The MSD shows normal diffusion at short times up to 10\,ms, where caging
effects appear.
The time-dependence of the diffusion coefficient is clearly observed and
displays a suppression by 6 orders of magnitude.

In a related experiment, \textcite{Wang:2009} tracked the motion of nanospheres
in semidilute F-actin solution.
Choosing the tracer radius (25\textminus{}250\,nm) much smaller than the mesh
size, normal diffusion was expected.
The measured MSDs indeed grow linearly in the window between 50\,ms and 10\,s.
Yet, transport is anomalous: the van Hove function $P(r,t)$ decays
exponentially with $r$ rather than following a gaussian.
Likewise, the propagator exhibits scaling upon rescaling distances as in normal
diffusion, $\hat r \propto r t^{-1/2}$, but the resulting scaling function is
anomalous.
Similar findings were made for the quasi-one-dimensional diffusion of beads
along phospholipid bilayer tubes.

Another model system for a semiflexible biopolymer is the \emph{fd} virus,
being a stiff filamentous phage of contour length 880\,nm, diameter 6.6\,nm,
and persistence length 2.2\,µm.
\textcite{Kang:2005} studied transport in semidilute suspensions of \emph{fd}
virus using differently sized tracers: a protein of radius 3.5\,nm and labelled
silica spheres with radii from 35 to 500\,nm.
Measurement of the long-time self-diffusion constant required suitable
techniques to cover the large range of particle sizes: fluorescence correlation
spectroscopy for the protein and the small colloids, dynamic light scattering
for intermediate-sized spheres, and video microscopy was found most appropriate
for the large spheres.
The obtained diffusion constant are systematically suppressed by up to an order
of magnitude when the concentration of \emph{fd} increases; the effect is more
pronounced for the larger tracers.
Above the overlap concentration, i.e., in the semidilute regime, the \emph{fd}
rods span a network and an average mesh size can be assigned.
The diffusion constants as function of the tracer radius divided by the mesh
size, however, display strong deviations from scaling---in contrast to the
findings for cross-linked networks.
This may be explained by the dynamically changing structure of the \emph{fd}
network, whose constituents themselves are subject to Brownian motion.


Aqueous solution of dextran molecules may also serve as a model system for a
crowded fluid~\cite{Weiss:2004}.
FCS measurements on a small fraction of labelled dextran molecules (10, 40, and
500\,kDa in weight) revealed a systematic decrease of the exponent $\alpha$
with a concomitant increase of the half-value time as the crowding agent is
more concentrated.
The half-value times varied over an order of magnitude as a function of
unlabelled dextran concentration, and the ``anomaly parameter'' $\alpha$
reached values below 0.6 for the 40\,kDa-dextran tracers.

A benefit of using well-defined model systems is the possibility to perform a
detailed statistical analysis of anomalous transport at a given point in
parameter space.
FCS measurements on labelled apoferritin in a concentrated dextran solution
were repeated many times at the same spot to generate histograms of the
residence times and of the exponent $\alpha$~\cite{Szymanski:2009}.
The exponent covered values between 0.7 and 0.9 with mean 0.82; a systematic
variation of $\alpha$, indicating ageing of the sample, was not observed.
These distributions were then critically compared to computer simulations of
the FCS experiment for various models of anomalous transport: CTRW, FBM, and
obstructed diffusion on a percolating cluster (see
\cref{sec:theoretical_models}),
all of them generating the same subdiffusive MSD with (apparent) exponent
$\alpha=0.82$.
While FBM and obstructed diffusion provide a good description of the
experimental data, the results from the CTRW model are in qualitative and
quantitative disagreement.
This indicates that crowding-induced subdiffusion is more appropriately
described by a stochastic motion with stationary increments than by a CTRW with
untruncated waiting-time distribution.

In the same lab, single-particle tracking experiments of fluorescent
50\,nm beads in a solution of 30\% 500\,kDa-dextran were performed with
temporal and spatial resolution of 4\,ms and 10\,nm,
respectively~\cite{Ernst:2012}.
For lag times between 50\,ms and 500\,ms, the time-averaged MSDs exhibited
subdiffusion with an average exponent $\alpha\approx 0.82$ as before.
At about 1\,s, a crossover towards normal diffusion was observed.
The analysis of 21 trajectories showed no signs of ergodicity breaking, and
comparing their geometric properties to the models of
\cref{sec:theoretical_models}, fractional Brownian motion was found to match
quantitatively.

The monotone dependence of the subdiffusion exponent $\alpha$ with the
concentration of crowding agent was corroborated by \textcite{Banks:2005}, who
used the globular proteins streptavidin and EGFP as tracers in dextran
solution.
The conventional $\chi^2$-fit of the FCS data with the gaussian subdiffusion
model was substantiated further: On a double-logarithmic representation, the
long time decay of the FCS correlation is compatible with the subdiffusion
model down to the noise level of about $5\times10^{-3}$ and clearly deviates
from the power law $t^{-3/2}$, expected for normal diffusion.
Second, an analysis assuming a large number of diffusive species,
\cref{eq:fcs_distribution}, yields a continuous distribution of diffusion times
between 0.1 and 10\,ms, rather than, e.g., a bimodal distribution.
The resulting exponents of subdiffusion decrease rapidly for increasing
concentrations of dextran obstacles and approach $\alpha=0.74 \pm 0.02$ for
high concentrations.
This value appears to be related to the exponent 3/4 characterising the
internal polymer dynamics of the dextran chains.
Dextran aggregates are ruled out as source of anomalous diffusion, and the
subdiffusive motion is remarkably robust against temperature variations, merely
the half-value times reflect the changing viscosity of water.
Further, the motion of the small fluorescein molecules as well as of the
dextran crowders itself shows simple diffusion even at high dextran
concentration.
If a globular protein is used as crowding agent, streptavidin motion is only
slightly anomalous with $\alpha\approx 0.91$ at high concentrations.

Ficoll-70 served as crowding agent as well, being an inert, highly branched
polysaccharide of approximately spherical shape with a hydrodynamic radius of
5.5\,nm.
Verkman \emph{et al.}~\cite{Dauty:2004, Dix:2008} characterised the transport
of differently sized tracers in Ficoll-crowded solution using FCS with an
illumination region of 0.5\,µm in diameter.
The tracers covered about two decades in size and ranged from rhodamine green
over labelled proteins, dextrans, and DNA fragments to fluorescent polystyrene
beads with up to 100\,nm in diameter, the latter mimicking cellular organelles.
Although the FCS correlations displayed simple diffusion in all cases, tracer
transport slows down drastically upon systematically crowding the solution.
The obtained diffusion coefficients are suppressed by two to three orders of
magnitude as Ficoll concentration increases up to 60\,wt\% and follow an
exponential law.
All tracers showed qualitatively the same behaviour, independently of their
size.
Using the smaller glycerol as crowding agent, the reduction of diffusion is
smaller, but still exponential.
For the large tracers, the slow diffusion can to a large extent be explained by
the change of the macroscopic viscosity of the crowded fluid.
This correlation is less pronounced for the small rhodamine green molecule,
which appears to sense the microviscosity of its local environment.

Another model for a crowded environment is given by aqueous micellar solution
made of the non-ionic surfactant C\textsubscript{12}E\textsubscript{6}.
\textcite{Szymanski:2006a} studied diffusion constants of TAMRA-labelled
lysozyme proteins with FCS for a wide range of
C\textsubscript{12}E\textsubscript{6} concentrations and found systematic
deviations from free diffusion for concentrations above 3\,wt\%.
A fit of the FCS correlation data, however, required a two-component diffusion
model with only 10\% of the proteins in the slow component; a possible
explanation for the slow diffusion is the formation of protein--micelle
complexes.
Slow and fast diffusion constants are both suppressed by up to an order of
magnitude at the highest concentration studied (37\,wt\%) and are separated by
another order of magnitude.
The gaussian subdiffusion model fits the data equally well yielding a
systematic decrease of the exponent $\alpha$ down to 0.88, but the
interpretation in terms of two components was favoured due to its clear
physical interpretation.
Focusing on the unbound fraction, the exponential suppression of diffusion with
concentration was consistently reproduced for a variety of fluorescent tracers
covering sizes from 1.7 to 190\,nm~\cite{Szymanski:2006}.
For the larger tracers, the diffusion constants are determined by the
macroscopic viscosity of the micellar solution via the Stokes--Einstein
relation.
In contrast, the diffusion constant of small tracers show a high sensitivity to
the tracer size: in the concentrated solution, apoferritin diffuses about
100-fold slower than lysozyme although the radii of the two proteins differ
only by a factor of 3.6.
The crossover length scale of 17\,nm was identified with the persistence length
of the micelles.

The interplay of binding and crowding was addressed by means of a model system
of negatively charged dextran molecules (500\,kDa) as crowders and the
positively charged protein ribonuclease~A as tracer~\cite{Zustiak:2011}.
The tracer protein binds reversibly and non-specifically to a dextran molecule.
Already at dilute concentrations of dextran (1\,µM), the FCS correlation
signals the presence of a large fraction of bound protein with about 7-fold
suppressed diffusion constant, reflecting the increased hydrodynamic radius of
the compound.
Protein diffusion in solution of positively charged and neutral dextrans is
unaffected at these low concentrations and reduced by merely a factor of~2 at
the highest polymer concentration studied (100\,µM), crowding effects are still
small.
The data are fully explained by the coexistence of free and bound protein, both
of them exhibiting simple diffusion.


Neutron scattering is a non-invasive technique to access protein solution
samples at high protein concentrations at nanosecond and nanometre scales.
Using quasi-elastic neutron backscattering, \textcite{Roosen-Runge:2011,
Roosen-Runge:2010} probed the self-diffusion of bovine serum albumin (BSA)
proteins in crowded aqueous solutions, where the same protein served as
crowding agent.
The protein motion was inferred from the incoherent dynamic structure factor
$P(k, \omega)$ of the hydrogen atoms, and the data are compatible with simple
diffusion for wavelengths between 0.5 and 3.3\,nm and time scales between 0.3
and 5\,ns.
Increasing the protein content up to 30\% volume fraction, a 10-fold reduction
of the translational \emph{short-time} diffusion coefficient over its value in
dilute solutions was observed.
Previous theoretical and experimental studies for hard-sphere colloids showed
that diffusion is considerably slowed down already at short times when
hydrodynamic interactions are taken into account, followed by a further slowing
down at long times for high volume fractions~\cite{Tokuyama:1994, Banchio:2008},
see also Ref.~\onlinecite{Ando:2010} discussed in
\cref{sec:cytoplasm_simulations}.
The reported BSA diffusion constants compare well with the findings for
short-time diffusion of hard spheres if an effective hydrodynamic protein
radius is used, which accounts for the hydration shell and the oblate
ellipsoidal protein shape; the effective radius was determined to 3.6\,nm from
small-angle X-ray scattering.
Noting further that at time scales of a few nanoseconds the caging by
surrounding proteins is expected to be not yet effective, this suggests that
the observed slow-down of protein diffusion is mainly caused by hydrodynamic
interactions and that protein diffusion in a crowded environment cannot be
understood merely by excluded volume and confined motion.


Nuclear magnetic resonance spectroscopy was employed to quantify both the
rotational and translational diffusion of the protein chymotrypsin inhibitor~2
(CI2) in a variety of crowded solutions as function of crowder
concentration~\cite{Wang:2010}.
The crowding agents comprised glycerol, synthetic polymers (PVP, Ficoll),
globular proteins (BSA, ovalbumin, lysozyme), and \emph{E.\ coli} cell lysates.
The macroscopic shear viscosity of the solution, $\eta$, increases with crowder
concentration.
Then the Stokes--Einstein relation already suggests a reduced diffusivity, $D
\propto 1/\eta$, which describes the measured CI2 diffusion constants in
glycerol solution and also translational diffusion in ovalbumin, BSA, and cell
lysate.
In solutions of synthetic polymers, transport is affected less than expected
from the increase of viscosity, and the translational motion of CI2 is impeded
more than its rotational motion.
Surprisingly, the opposite effect was found in protein-crowded solutions and in
the cell lysate: rotational diffusion constants were suppressed stronger than
translational diffusion and stronger than the Stokes--Einstein relation would
imply.
The findings were attributed to weak non-specific, non-covalent chemical
interactions between proteins, while synthetic polymers tend to form a loose
mesh work.
This suggests that crowded protein solutions are preferred for modelling the
effects of the intracellular environment on protein transport.

\subsubsection{Computer simulations (\emph{in silico})}
\label{sec:cytoplasm_simulations}

\begin{table*}
  \setlength\tabcolsep{.5em} \setlength\extrarowheight{1ex}
  \fontsize{9pt}{9pt} \selectfont 
  \begin{tabularx}{\linewidth}{%
    >{\raggedright\arraybackslash}p{.22\linewidth}%
    >{\raggedright\arraybackslash}p{.15\linewidth}%
    >{\raggedright\arraybackslash}p{.2\linewidth}%
    >{\raggedright\arraybackslash}Xp{5ex}l}
  \hline \hline
  crowded environment & tracer & interactions
    & subdiffusion exponent $\alpha$ \newline and temporal window & year & reference
  \vspace{2pt} \\ \hline
  10\textsuperscript{6} spherical obstacles (overlapping, equal size)
    & point (ballistic and overdamped motion) & hard core
    & 1 \textrightarrow\ 0.32 \textrightarrow\ 0, subdiffusion over up to 6~decades in time
    & 2006, 2008 & refs.~\citenum{Lorentz_PRL:2006,Lorentz_JCP:2008}
  \\
    &&& 1 \textrightarrow\ 0.42 \textrightarrow\ 0 \newline (percolating cluster only)
    & 2011 & ref.~\citenum{Lorentz_percolating:2011}
  \\
    1000 spheres (equal size), \newline fluid in obstacle matrix & fluid particles & hard core
    & 1/2 over 4 decades in time
    & 2009, 2011 & refs.~\citenum{Kurzidim:2009,Kurzidim:2011}
  \\
    1000 spheres (bidisperse, partially quenched) & mobile particles & hard core
    & 0.3 over 4 decades in time
    & 2009, 2010 & refs.~\citenum{Kim:2009,Kim:2010a}
  \\
    2000 spheres (size-disparate mixture)
    & small particles
    & soft repulsion
    & 0.63 and 0.5 over several decades
    & 2006, 2009 & refs.~\citenum{Moreno:2006,Voigtmann:2009}
  \\
    \allowhyphenation 5000 spheres with weight distributed modelling HeLa
    & small spheres \newline (2\textminus{}5\,nm)
    & hard core, soft repulsion
    & 1 \textrightarrow\ 0.55, subdiffusion over 3~decades in time
    & 2004 & ref.~\citenum{Weiss:2004}
  \\
    atomistic \emph{E. coli} model: \newline 1000 macromolecules of 50~different types
    & macromolecules \newline (10\textminus{}1000\,kDa)
    & repulsion, vdW attraction, electrostatic, hydrodynamic
    & 0.75\textminus{}0.85 @ 100\,ps\textminus{}1\,µs
      (full set of interactions)
    & 2010 & ref.~\citenum{McGuffee:2010}
  \\
    \emph{E. coli} model: 15 molecule types, 1300 molecular-shaped or spherical particles
    & macromolecules
    & repulsion, vdW attraction, hydrodynamic (spheres only)
    & $D_0/D_\text{short}$: 3\textminus{}10 (with HI),
    anomalous\textrightarrow normal at 1\,µs,
    $D_\text{short}/D \approx 2$ (with HI) or 20
    & 2010 & ref.~\citenum{Ando:2010}
  \\
    1000 spherical obstacles (non-overlapping, equal size)
    and 10\textsuperscript{6} solvent particles
    & globular polymer (400 beads)
    & repulsion, vdW attraction, hydrodynamic (MPCD)
    & 1\textminus{}0.7 depending on excluded volume and polymer size
    & 2010 & ref.~\citenum{Echeverria:2010}
  \\
    polymeric network & free monomer
    & repulsion, vdW attraction
    & 1/2 or 3/5 over up to 2 decades in time
    & 2011 & ref.~\citenum{Tabatabaei:2011}
  \\ \hline \hline
  \end{tabularx}
  \caption{Overview of computer simulations on crowded fluids as guide to the
  discussion in \cref{sec:cytoplasm_simulations}. Empty fields repeat the entry
  above, arrows indicate a systematic change upon variation of some parameter.
  Abbreviations: vdW\dots van der Waals, HI\dots hydrodynamic interactions,
  MPCD\dots multi-particle collision dynamics.}
  \label{tab:cytoplasm_simulations}
\end{table*}

Anomalous transport and subdiffusion were unambiguously identified in computer
simulations for minimalist, generic models as well as for elaborate cytoplasm
models targeting a specific cell type.
These results provide essential support for the interpretation of experiments
and test various microscopic pictures behind anomalous transport.

A paradigm of anomalous transport is provided by the motion of a small tracer
in a disordered matrix of spherical obstacles, known as Lorentz
models~\cite{Lorentz_PRL:2006, Lorentz_JCP:2008, Lorentz_percolating:2011,
Lorentz_space:2011}; see \cref{sec:Lorentz} for a thorough
discussion and its connection to continuum percolation theory.
As a localisation transition is approached by tuning the excluded volume, a
growing window of subdiffusion emerges at intermediate time scales.
The tracer dynamics generically displays a double-crossover scenario from
microscopic motion at very short time scales to subdiffusion and to normal
diffusion at large times~\cite{Lorentz_PRL:2006}.
This phenomenology is preserved, whether the microscopic tracer motion is
ballistic like in a porous medium or overdamped like in the
cytoplasm~\cite{Lorentz_JCP:2008}.
The apparent exponent of the subdiffusive regime decreases monotonically upon
approaching the transition and converges to its universal value, $\alpha
\approx 0.32$ or $0.42$, depending on whether the tracer is restricted to the
percolating space or not~\cite{Lorentz_percolating:2011}.
Subdiffusive transport over 6 decades in times at the critical obstacle density
and a suppression of the time-dependent diffusion constant by 5 orders of
magnitude was observed in the simulations~\cite{Lorentz_percolating:2011,
Lorentz_space:2011}.
Even 10\% off the critical obstacle density, the motion is subdiffusive with
exponent $\alpha\approx 0.5$ over still 2 decades in time.
Thus for the observation of anomalous transport over a finite time window, it
is not essential that the system is fine-tuned to the critical point.
The scenario of the localisation transition appears to be robust: subdiffusive
motion with exponent 1/2 over 4 decades in time was found in simulations for a
hard-sphere fluid adsorbed in a matrix of disordered, non-overlapping
hard-sphere obstacles~\cite{Kurzidim:2009, Kurzidim:2011}.
Simulations for a similar system where the tracer fluid and the obstacles are
correlated yielded subdiffusion over 4 decades in time with an exponent $0.3$
resembling the value from the Lorentz model~\cite{Kim:2009, Kim:2010a}.
Anomalous transport was even found in simulations for transport of small
particles in a glassy, slowly rearranging matrix of large
particles~\cite{Moreno:2006, Voigtmann:2009}; although the MSD does not display
unambiguous power-law behaviour, the apparent exponent $\alpha(t)$, defined by
$\alpha(t)=\diff\log(\delta r^2(t))/\diff \log(t)$, was found to be 0.5 and
less over several decades in time \cite{Voigtmann:2009}, and there is evidence
that the localisation scenarios in the binary mixture and the Lorentz model are
related~\cite{Lorentz_BSSM:2010}.
At high volume fractions of the obstacle matrix, all of the mentioned models
share non-vanishing non-ergodicity parameters due to trapped
particles~\cite{Lorentz_space:2011, Kurzidim:2009, Kurzidim:2011, Kim:2009,
Kim:2010a, Voigtmann:2009, Lorentz_BSSM:2010} that would appear as immobile
fraction in a FRAP experiment, see \cref{sec:frap}.

Aiming at a more realistic modelling, the measured distribution of protein sizes
in the cytoplasm of HeLa cells served as input for a simple cytoplasm model
consisting of 5000 spherical particles with exponentially distributed molecular
weights~\cite{Weiss:2004}.
The particles interact via a hard core plus a soft repulsive, parabolic
potential; their overdamped motion was followed by Brownian dynamics
simulations.
The obtained MSDs display subdiffusion over 3 decades in time, where the
exponent $\alpha$ decreases as function of the particle radius in a sigmoidal
fashion from unity for the smallest particles (2\,nm) to about 0.55 for the
largest radii studied (5\,nm).
Thus, the excluded volume can, to a large extent, account for the
experimentally observed subdiffusion in HeLa cells.

A step towards a realistic modelling of the \emph{E.\ coli} bacterium was taken
by \textcite{McGuffee:2010}, who developed an atomistically detailed model of
its cytoplasm including 50 different types of macromolecules at physiological
concentrations.
For systems of 1000 macromolecules in total, Brownian dynamics simulations
including electrostatic and hydrodynamic interactions yielded system
trajectories over 6\,million steps or a time span of 15\,µs.
Thereby, molecular behaviour inside a cell is illustrated by vivid movies of
educational value.
Further, anomalous transport was observed over 4 decades in time with local MSD
exponents $\alpha(t)$ between 0.75 and 0.85.
Focusing on the role of excluded volume by including only repulsive
interactions in the model, the anomaly was markedly reduced.
Hence, the effects of macromolecular crowding extend beyond those of excluded
volume, although a ramified excluded volume appears to be essential for
pronounced subdiffusive motion.

The role of hydrodynamic interactions (HI) on protein motion was elucidated by
\textcite{Ando:2010}, who performed Brownian dynamics simulations of models for
the \emph{E.\ coli} cytoplasm comprised of 15 different macromolecule types:
proteins were represented either by their molecular shape or by equivalent
spheres of the same hydrodynamic radius, the particles interact non-specifically
via a soft repulsive parabolic potential or via attractive van der Waals
forces, and hydrodynamic interactions were included for the equivalent-spheres
model.
It was found that HI yield a 3- to 10-fold reduction of the \emph{short-time}
diffusion constants, depending on protein radius and concentration.
This finding is supported by previous results for monodisperse
spheres~\cite{Tokuyama:1994, Banchio:2008} and corroborates recent
experiments~\cite{Roosen-Runge:2011, Roosen-Runge:2010}, see above.
Long-time diffusion constants were determined as function of the protein radius
at an observation time scale of 5\,µs and display a further reduction
by a factor of 2 (with HI) or up to 20 (repulsion only).
Notably, the equivalent-sphere model including HI reproduced the experimentally
observed diffusion constant of GFP \emph{in vivo} without any adjustable
parameters.
As a key result, excluded volume effects and hydrodynamic interactions are
likely the two major factors for the large reduction in diffusion of
macromolecules observed \emph{in vivo}.

\textcite{Echeverria:2010} studied the effect of crowding on the conformation
and transport of a globular polymer in an explicit, albeit mesoscopic, poor
solvent.
The crowded environment was formed by a frozen, random array of hard spherical
obstacles, and the system was propagated using the multi-particle collision
dynamics (MPCD) scheme (which intrinsically includes hydrodynamic effects) for
up to 1~million solvent particles.
The simulation results show that the equilibrium structure of long polymer
chains is significantly altered towards a chain of polymer blobs being trapped
in voids between the obstacles.
For increasing volume fraction of obstacles, subdiffusive motion of the centre
of mass emerges with exponents $\alpha$ down to 0.7, and it appears that
$\alpha$ decreases further as the polymer size increases.
These findings are in accord with earlier lattice studies for polymers in
disordered media~\cite{Baumgaertner:1987, Slater:1995, Nixon:1999}, which
showed that entropic trapping of the polymer in the voids leads to a stark
slowing down of transport and to subdiffusive motion.

Transient binding to polymeric networks, e.g., in hydrogels, can render the
transport of tracer molecules anomalous as well as demonstrated by molecular
dynamics simulations~\cite{Tabatabaei:2011}.
Strongly bound tracers sliding along a relaxed polymer strand exhibit
pronounced subdiffusive motion with exponents $\alpha=1/2$ or $3/5$, depending
on the fractal structure of the coiled strands.
For a swollen hydrogel, the polymer coils are stretched and the window of
subdiffusion was found to decrease quickly.
Further, there is a competition between sliding along the polymer and free
diffusion within the pores of the network.
The observed anomalous transport depends sensitively on the attraction strength
between the tracer and the polymer chain, which governs this competition.

\subsection{Crowded membranes}

\subsubsection{Cellular membranes (\emph{in vivo})}
\label{sec:cellular_membranes}

\begin{table*}
  \setlength\tabcolsep{.5em} \setlength\extrarowheight{1ex}
  \fontsize{9pt}{9pt} \selectfont 
  \begin{tabularx}{\linewidth}{%
    >{\raggedright\arraybackslash}p{.15\linewidth}%
    >{\raggedright\arraybackslash}p{.15\linewidth}%
    >{\raggedright\arraybackslash}p{.11\linewidth}p{.11\linewidth}%
    >{\raggedright\arraybackslash}Xp{5ex}l}
  \hline \hline
  cell type (membrane) & probe & experimental \newline technique
    & temporal and \newline spatial scales
    & observation & year & reference
  \vspace{2pt} \\ \hline
  keratinocytes & E-cadherin and transferrin receptor & SPT, FRAP & 33 ms\textminus{}30\,s
    & heterogeneous MSDs: free, confined, immobile, directed
    & 1993 & ref.~\citenum{Kusumi:1993}
  \\
  RBL cells & IgE receptor & SPT, FRAP & 1\textminus{}100\,s
    & 56\% of tracers: subdiffusion with $\alpha=0.65\pm0.45$; 27\%~immobile
    & 1996 & ref.~\citenum{Feder:1996}
  \\
  C3H 10T1/2 fibroblasts & Thy-1 protein & SPT & 30\,ms\textminus{}1\,s
    & normal diffusion, fast and slow components & 1997 & ref.~\citenum{Sheets:1997}
  \\
  erythrocytes & band-3 protein & SPT & 0.22\,ms\textminus{}1\,s
    & $D_{10\,\text{ms}} / D_{1\,\text{s}} \approx 80\!-\!90$, \newline 1/3 immobile
    & 1998 & ref.~\citenum{Tomishige:1998}
  \\
  rat kidney fibroblasts & DOPE phospholipid  & SPT & 25\,µs\textminus{}3\,s
    & 3 subdiffusion regimes (double crossover): $\alpha=0.74 \to 0.55 \to 0.79$
    & 2002 & ref.~\citenum{Fujiwara:2002}
  \\
  various cell types &&& & subdiffusion @ 0.1\textminus{}10\,ms ($\alpha=0.53$),
    short- and long-time diffusion: $D_0 / D \approx 10$ & 2004 & ref.~\citenum{Murase:2004}
  \\
  HeLa cells & MHC class~I protein & SPT & 4\textminus{}300\,s
    & subdiffusion: $\alpha=0.49\pm0.16$
    & 1999 & ref.~\citenum{Smith:1999}
  \\
  CHO cells & MHC class~II protein & SPT & 100\,ms\textminus{}3\,s & normal diffusion
    & 2002 & ref.~\citenum{Vrljic:2002}
  \\
  \emph{C. crescentus} & PleC protein & SPT & 1\textminus{}9\,s
    & normal: $D \approx 10^{-2}$\,µm\textsuperscript{2}/s & 2004 & ref.~\citenum{Deich:2004}
  \\
  COS-7 fibroblasts, MDCK cells  & AQP-1 protein & SPT
    & 10\,ms\textminus{}6\,min & heterogeneous MSDs (free, restricted, immobile),
    normal case: $D$ depends on protein content
    & 2008 & ref.~\citenum{Crane:2008}
  \\
  HEK 293 cells & Kv2.1 potassium channel & TIRFM & 0.1\textminus{}100\,s &
    subdiffusion ($\alpha\approx 0.8$), ageing, fractal space, stationary
    and non-stationary displacements & 2011 & ref.~\citenum{Weigel:2011}
  \\
  RBL cells & diI-C\textsubscript{12} lipid & FCS & $\tau_{1/2}\approx 30\,\text{ms}$
    & subdiffusion $\alpha=0.74 \pm 0.08$, or two components: $D_1/D_2 = 30$
    & 1999 & ref.~\citenum{Schwille:1999}
  \\
  oligodendroglia & MOG protein, sphingomyelin lipid & FCS
    & 0.01\textminus{}100\,ms
    & subdiffusion of MOG protein ($\alpha=0.59$), two-component diffusion of lipid
    & 2005 & ref.~\citenum{Gielen:2005}
  \\
  \allowhyphenation HeLa cells (intracellular membranes) & Golgi-resident enzymes
    & FCS & 0.3\textminus{}100\,ms
    & subdiffusion: $\alpha\approx0.5\!-\!0.8$ & 2003 & ref.~\citenum{Weiss:2003}
  \\
  yeast cells & Fus-Mid-GFP protein & scanning FCS & $\tau_{1/2}\approx 2$\,s
    & slow, normal diffusion: $D \approx 10^{-3}\,$µm\textsuperscript{2}/s & 2006 & ref.~\citenum{Ries:2006}
  \\
  COS-7 cells & \allowhyphenation transferrin receptor (TfR), sphingolipids, anchored proteins
    & variable-area FCS & beam waist: \newline 190\textminus{}390\,nm
    & $w$-dependent effective $D$: $\tau_{1/2}(w)=t_0+w^2/4 D_\text{eff}$, \newline
    $t_0 > 0$ for lipids and anchored proteins, $t_0 < 0$ for TfR
    & 2005, 2006 & refs.~\citenum{Wawrezinieck:2005,Lenne:2006}
  \\
    && variable-area FCS & beam waist: \newline 75\textminus{}350\,nm
    & 3\textminus{}5-fold reduction of $D(w)$ for small $w$
    & 2007 & ref.~\citenum{Wenger:2007} \\
    PtK2 cells & sphingomyelin, anchored protein & STED-FCS
    & beam waist: \newline 15\textminus{}125\,nm
    & anomalous $w$-dependence: \newline $\tau_{1/2}(w\to 0) \approx 10\,\text{ms}$
    & 2009 & ref.~\citenum{Eggeling:2009}
  \\
  rat myotubes & ACh receptor & FRAP & 1\,min
    & immobile fraction of ACh in dense patches
    & 1976 & ref.~\citenum{Axelrod:1976}
  \\
  human embryo fibroblasts & WGA receptor & FRAP & $\tau_{1/2}\approx 30\,\text{s}$
  & only 75\% mobile receptors & 1976 & ref.~\citenum{Jacobson:1976}
  \\
  thymocytes, lymphoma cells, fibroblasts & Thy-1 protein & FRAP
    & $\tau_{1/2} < 1\,\text{s}$ & $\approx 50\%$ immobile & 1987 & ref.~\citenum{Ishihara:1987}
  \\
  COS-7 cells & anchored, acylated, and transmembrane proteins & FRAP & bleaching stripe: 4\,µm
    & normal, $D=$ 0.1\textminus{}1.2\,µm\textsuperscript{2}/s for different proteins
    & 2004 & ref.~\citenum{Kenworthy:2004}
  \\
    & 4 membrane proteins & FRAP & $\tau_{1/2}\approx 2\,\text{s}$, spot radius: 3.6\,µm
    & normal, but $D$ 10-fold smaller than for GPI-proteins
    & 2007 & ref.~\citenum{Frick:2007}
  \\
    HEK 293T & CD4 and chemokine receptors & variable-radius FRAP
    & spot radius: \newline 1.40\textminus{}3.45\,µm
    & $D$ and recovery ratio depend on size of bleaching spot & 2007 & ref.~\citenum{Baker:2007}
  \\ \hline \hline
  \end{tabularx}
  \caption{Overview of \emph{in vivo} experiments on cellular membranes as
  guide to the discussion in \cref{sec:cellular_membranes}. Empty fields repeat the
  entry above. Abbreviations: TIRFM\dots{}total internal reflection fluorescence microscopy,
  STED\dots{}stimulated emission depletion.
}
  \label{tab:cellular_membranes}
\end{table*}

According to the ``fluid mosaic'' model by \textcite{Singer:1972}, the plasma
membrane of cells is thought of as an essentially homogeneous fluid bilayer of
phospholipids with freely diffusing protein inclusions.
Numerous in vivo measurements of protein motion in membranes during the past
two decades, however, required several revisions of this simple picture.
Today, the plasma membrane is considered a patchy lipid bilayer densely packed
with integral and peripheral proteins, where some of the lipids are organised
into microdomains, some of the proteins form clusters, and others are tethered
to the cytoskeleton~\cite{Vereb:2003, Engelman:2005, Kusumi:2005}.


First hints on the heterogeneous landscape of the plasma membrane go back to
early FRAP experiments~\cite{Axelrod:1976, Ishihara:1987}, indicating that a
significant fraction of certain membrane proteins is immobile.
These findings were corroborated by single-particle tracking techniques
emerging in the 1990's, which enabled the observation of individual
fluorescently labelled membrane proteins~\cite{Kusumi:1993, Feder:1996,
Sheets:1997, Tomishige:1998}.
The dynamics of membrane receptors in mouse keratinocytes was
found to be highly heterogeneous within the same type of protein, and the
trajectories were classified after their MSDs into four types of motion:
free diffusion, confined diffusion, immobile, and directed
motion~\cite{Kusumi:1993}.

SPT experiments on IgE receptors by \textcite{Feder:1996}
yielded another class of MSDs, which required a subdiffusive power-law
fit for 56\% out of 241 trajectories (150 frames every 1.6\,s) with an
average exponent $\alpha=0.64\pm0.45$; 27\% of the receptors were immobile.
The findings are consistent with corresponding FRAP measurements, see below.
In similar experiments, \textcite{Sheets:1997} preferred to account for the
subdiffusive fraction by refining the classification of trajectories also
according to their ``shape'' and by introducing the additional class of slow
normal diffusion.
For the Thy-1 protein, the classes of fast and slow diffusion yielded diffusion
constants as different as 0.081 and 0.0035\,µm\textsuperscript{2}/s,
respectively.

Significant enhancement of the temporal resolution to 0.22\,ms of the SPT
experiments~\cite{Tomishige:1998} provided a more detailed picture of band-3
protein transport in erythrocyte membranes: a fraction of proteins was
immobile, the remaining two thirds exhibited normal diffusion at macroscopic
time scales of a few seconds.
The diffusion constant, however, was drastically reduced by a factor of 80$-$90
compared to its microscopic value at 10\,ms.
Refined experiments suggested a hopping mechanism due to interaction with the
spectrin network, characterised by a mesh size of 110\,nm and an average hop
frequency of 2.8\,$\mathrm{s}^{-1}$.

Some years later, the same lab implemented a high-speed SPT setup with a
temporal resolution of remarkable 25\,µs~\cite{Fujiwara:2002}, which permitted
the detailed tracking of DOPE phospholipids in rat kidney fibroblasts.
The obtained MSD covers five(!) decades in time and exhibits two crossovers at
12 and 590\,ms, connecting three regimes of motion (see supplement
of Ref.~\citenum{Fujiwara:2002}).
All three regimes are compatible with subdiffusive motion with exponents
$\alpha=0.74$ (short times), $\alpha=0.55$ (intermediate times), and
$\alpha=0.79$ (long times, but regime covers less than a decade).
Extrapolating to short time scales, the MSD appears to approach the free
diffusion in a homogeneous lipid bilayer.
It was concluded that the plasma membrane is doubly compartmentalised into
750\,nm and then into 230\,nm compartments with regard to the lateral diffusion
of DOPE lipids.
Repetition of the measurements for many other cell types~\cite{Murase:2004}
corroborated these findings qualitatively with hopping rates between 1 and
59\,$\mathrm{s}^{-1}$ and compartment sizes between 30 and 230\,nm.
For FRSK cells (fetal rat skin keratinocytes), the regimes at short and long
times exhibited essentially normal diffusion with diffusion constants separated
by an order of magnitude; the intermediate regime between 0.1\,ms and $\approx
10$\,ms was found to be subdiffusive with exponent $\alpha=0.53$.

On the basis of these observations, \textcite{Kusumi:2005a, Kusumi:2005} have
proposed a revised model for the plasma membrane, which should be viewed as a
compartmentalised fluid.
The structure is provided by barriers or obstacles like the
actin-based membrane skeleton (``fences'') and anchored transmembrane
proteins (``pickets'').
The motion of proteins and lipids is then thought of as a hopping process
between differently sized membrane compartments, and transport is slowed down
by such specific barriers.

Deeper insight into the stochastic transport process is expected from the
investigation of displacement histograms for fixed lag times, i.e., the van
Hove function.
Such a route was taken by \textcite{Smith:1999}, who performed SPT experiments
using fluorescence imaging of major histocompatibility complex (MHC) class~I
molecules on HeLa cells.
Fitting the propagator of normal diffusion to the displacement histograms
yielded the diffusion coefficients $D(t)$ as function of the observation time
scale.
A marked decrease of $D(t)$ with increasing time interval was found, and the
data were best described by subdiffusive motion with $\alpha=0.49\pm0.16$ over
the whole accessible temporal window between 4\,s and 300\,s.
Unfortunately, the histograms suffered from large statistical noise prohibiting
an answer to the question whether the distribution of displacements is gaussian
or not.

The translational diffusion of {MHC} class~{II} membrane proteins in Chinese
hamster ovary cells was studied by \textcite{Vrljic:2002} to probe the plasma
membrane for barriers from putative lipid microdomains.
Using SPT at a rate of 100\,ms, the cumulative probability distribution
function of displacements was extracted at time scales up to only a few seconds
(limited by photobleaching).
In this time window, however, almost negligible deviations from simple Brownian
motion and no significant confinement could be inferred.
Most notably, the authors also studied correlations in the transport of
\emph{close pairs of proteins}, initially separated between 0.3 and 1\,µm.
The idea was to detect a possible confinement by impermeable, but diffusing
barriers.
Again, the results followed the predictions of free diffusion and no evidence
for a restriction to small, freely diffusing domains was found at the length
scales under investigation.

Normal diffusion was also observed by tracking the transmembrane histidine
kinase PleC in the bacterium \emph{Caulobacter crescentus}~\cite{Deich:2004},
which yielded the relatively low value
$D=12\times10^{-3}$\,µm\textsuperscript{2}/s.
The experimental MSDs, however, covered only a small time window between 1 and
9\,s (due to limitations of the fluorescent label) and are subject to large
statistical errors.
Hence, the question whether transport is normal or anomalous could actually not
be addressed.
Further, transport was found to be spatially uniform across all positions in
the elongated cell with exception of the cell poles and without evidence for
active transport.

The impact of putative lipid microdomains on the motion of integral membrane
protein aquaporin-1 (AQP-1) in COS-7 fibroblasts and Madin--Darby canine kidney
cells was studied by long-time SPT experiments~\cite{Crane:2008}.
Labelling with quantum dots enabled the recording of long trajectories at 1\,Hz
over 6\,min.
Trajectories were classified into free, restricted, and immobile according to
their deviation from the extrapolated short-time MSDs at 91\,Hz.
Most AQP-1 proteins diffused freely over several microns on the scale of
minutes, and diffusion was faster by a factor of 4 in protein-poor membrane
blebs.
It was concluded that AQP-1 is a largely non-interacting protein with a
macroscopic diffusion constant determined by the concentration of obstructions
in the membrane.

Only recently, \textcite{Weigel:2011, Weigel:2012} investigated the dynamics of
Kv2.1 potassium channels in the plasma membrane of human embryonic kidney
(HEK~293) cells.
Labelling with quantum dots and tracking by means of total internal reflection
fluorescence microscopy (TIRFM) permitted the collection of 1000 trajectories
each about 10\,min long at temporal and spatial resolution of 50\,ms and 8\,nm,
respectively.
Scrutinising time- and time-ensemble-averaged MSDs, ageing behaviour, the
propagator, and the distribution of waiting times indicated the coexistence of
a non-stationary transport process compatible with a CTRW and a stationary walk
in a fractal space.
The non-stationary process was attributed to binding of the potassium channels
to the actin network, while the stationary process may have its origin in the
molecular crowding of the membrane.


During the past decade, FCS has routinely been applied to cellular membranes
for \emph{in vivo} measurements of lipids and proteins.
Introducing fluorescent lipid probes in the plasma membranes of rat basophilic
leukaemia (RBL) cells, \textcite{Schwille:1999} obtained FCS correlations which
clearly deviate from normal diffusion.
The data can be described by the gaussian subdiffusion model with
$\alpha=0.74\pm 0.08$ and half-value times around 30\,ms, or alternatively
well, by the two-component diffusion model yielding diffusion times of 3\,ms
and 100\,ms.
A control experiment with giant unilamellar vesicles (GUVs) showed normal
diffusion, but inducing phase separation of the lipid mixture by adding
cholesterol rendered the lipid transport anomalous again.

For myelin oligodendrocyte glycoproteins bound to oligodendrocyte membranes,
anomalous diffusion with $\alpha=0.59$ was reported~\cite{Gielen:2005}, while a
two-component diffusion model appeared to be less likely.
The results for sphingomyelin lipids in the same membrane suggest a
two-component description with diffusion constants of 0.37 and
70\,µm\textsuperscript{2}/s; here, a possible explanation for the highly mobile
fraction are dye molecules moving freely in solution.

Intracellular membranes were addressed by \textcite{Weiss:2003}, who monitored
three Golgi-resident enzymes both in the endoplasmic reticulum and in the Golgi
apparatus of HeLa cells.
Subdiffusive motion was found over more than two decades in time with
half-value times of 10--30\,ms and with exponents $\alpha$ that depend on the
type of enzyme and range between 0.5 and 0.8; for the same enzyme, similar values
of $\alpha$ were obtained in the two different membrane structures.
The data can be equally well fitted by the two-component diffusion model, but
it was argued that such a fit seems unlikely.

A modification of the FCS technique is \emph{scanning FCS}~\cite{Ruan:2004}, where
the illumination laser repeatedly scans the probe, e.g., along a large circle.
The same spot is visited with a low frequency, giving access to the fluorophore
dynamics at very long time scales.
Combining this method with continuous wave excitation, very slow, albeit normal
diffusion was observed for Fus-Mid-GFP proteins on yeast cell
membranes~\cite{Ries:2006}, with diffusion constants on the order of
$10^{-3}$\,µm\textsuperscript{2}/s and FCS diffusion times of up
to 2~seconds.
Compared to the motion of these slow proteins, the other membrane components
rearrange quickly, effectively homogenising the originally heterogeneous
environment of the protein.
This may explain the finding of simple diffusion at long times, resolving an
apparent contradiction with the measurements quoted above.
The example further highlights the importance of taking into account the
experimental time and length scales for the interpretation of transport in
complex environments.

Complementary to the previous FCS studies, the spatial aspects of complex
molecular transport in membranes were explicitly addressed by systematic
variation of the detection area.
Using beam waists ranging from 190 to 390\,nm, \textcite{Wawrezinieck:2005}
have measured the motion of the GFP-labelled transmembrane protein TfR
(transferrin receptor) and of a sphingolipid in the plasma membrane of COS-7
cells.
The obtained two sets of diffusion times exhibit an affine dependence on $w^2$
motivating the FCS diffusion law,
$\tau_d(w)=t_0 + w^2/4D_\text{eff}$.
Extrapolation to zero beam waist yields the axis intercepts $t_0$, which were
found to be $+$25\,ms for the lipids and $-$20\,ms in case of the TfR proteins,
with uncertainty of 10\%.
The signs of $t_0$ can give a hint on the transport mechanism, and it was
concluded that the motion of sphingolipids is hindered by isolated microdomains
and those of the TfR proteins by a meshwork of barriers.
Alternatively, the protein diffusion times may be interpreted in terms of
fractal scaling law $\tau_d(w)  \sim w^{2.6}$ according to
\cref{eq:fcs_tau12_scaling}, indicating subdiffusion with
$\alpha = 2/\dw \approx 0.77$ \cite{FCS_scaling:2011}.
The findings of Ref.~\onlinecite{Wawrezinieck:2005} were substantiated by further
experiments with a variety of molecules~\cite{Lenne:2006}, providing a broad
range of intercept values: sphingolipids and GPI-anchored proteins yielded
clearly positive $t_0$, the transmembrane TfR-GFP is characterised by a
strongly negative value, and the axis intercept vanished for
glycerophospholipid analogues.
In conclusion, the slow dynamics of sphingolipids is attributed to
cholesterol and sphingomyelin levels in the membrane, and
the transferrin receptor protein is likely to be dynamically confined
by actin-based cytoskeleton barriers.

It is clear from \cref{eq:fcs_master} that the diffusion time must vanish as the
observation area approaches zero.
Thus, the extrapolation $w\to 0$ of the above diffusion law can only hold
approximately, and it is of interest to elucidate the behaviour at small $w$,
below the diffraction limit.
Observation areas with radii ranging down to 75\,nm were achieved by equipping
the microscope coverslips with circular nanometric apertures in a metallic
film~\cite{Wenger:2007}.
This approach permitted the elucidation of nanometric membrane heterogeneity
and a considerable extension of the measured diffusion laws $\tau_d(w)$ of
Ref.~\onlinecite{Lenne:2006}, revealing non-trivial crossovers.
For ganglioside lipid analogues and GPI-anchored GFP in the plasma membrane of
COS-7 cells, the apparent, $w$-dependent diffusion constants $w^2/4\tau_d(w)$
at short scales are 5- and 3-fold reduced compared to their large-scale values,
with a crossover length scale of $w\approx100$\,nm.
Even more interestingly, the diffusion law of TfR-GFP shows two transitions
around 150 and 230\,nm; the latter transition is explained by a meshwork of
cytoskeletal barriers hindering transport of the transmembrane protein.

Application of STED-FCS to the plasma membrane of living mammalian cells gives
access to nanosized detection areas down to $w=15$\,nm in
radius~\cite{Eggeling:2009}.
Phosphoethanolamine molecules, which are assumed not to form molecular
complexes with membrane components, exhibited free diffusion, $\tau_d(w)\propto
w^2$, over the full range of investigated areas.
The diffusion of sphingomyelin, however, was found to be strongly
heterogeneous: only for large radii, $w > 80$\,nm, unobstructed, normal
transport was found.
For small detection areas, $w < 40$\,nm, the FCS correlations showed clear
deviations from normal diffusion.
Fits with the subdiffusion model yielded $\alpha \lesssim 0.7$, but an
explanation of the transport in terms of two diffusive components was favoured.
The fast fraction resembled the unobstructed transport of phosphoethanolamine,
both in space and time.
The diffusion times of the slow fraction approached a minimum of about 10\,ms
for small areas.
This offset was explained by a brief trapping event of that duration by
cholesterol-mediated complexes within an area of less than 20\,nm in diameter.
Similar findings were made for GPI-anchored proteins.


The first applications of fluorescence recovery after photobleaching (FRAP) on
plasma membranes addressed the lateral motion
of fluorescently marked receptor proteins in rat myotubes~\cite{Axelrod:1976}
and human embryo fibroblasts~\cite{Jacobson:1976}.
In both experiments, only about 75\% of the initial fluorescence was recovered
indicating that a significant fraction of receptors is immobile.
Similarly, the fluorescence recovery was found to be only about 50\% for
labelled Thy-1 antigen in the plasma membrane of lymphoid cells and different
fibroblasts~\cite{Ishihara:1987}.
The obtained diffusion constant of the mobile fraction of Thy-1 proteins was
comparable to those of the lipid analogues in the bilayer, which is consistent
with a putative lipid anchoring to the plasma membrane.
\textcite{Feder:1996} studied the motion of IgE receptors in the membrane of
leukocytes using SPT and FRAP.
Motivated by the SPT measurements, they suggested a gaussian subdiffusion model
for the analysis of the FRAP curves.
The given data sets were fitted equally well by the subdiffusion model and the
diffusion model with incomplete recovery, a fit to a combined model yielded a
recovery of 90\% and a subdiffusion exponent $\alpha\approx 0.6$.
It was shown how subdiffusion could be discriminated from an immobile fraction
if the recovery curves would span a much larger time window.

In contrast, almost complete recovery of fluorescence was found in FRAP
experiments on the plasma membrane of COS-7 cells for a variety of putative
raft and non-raft proteins~\cite{Kenworthy:2004}.
The recovery curves from bleaching a large, 4\,µm wide stripe were rationalised
with normal diffusion, yielding diffusion constants between 0.1 and
1.2\,µm\textsuperscript{2}/s for the different proteins.
These values appear to be mainly determined by the type of anchorage, since
they change uniformly upon perturbing the membrane by variation of the
cholesterol level or temperature.

In another FRAP study on four different membrane proteins expressed in COS-7
cells~\cite{Frick:2007}, it was found that transport over micron-scale
distances is normal with 80$-$90\% fluorescence recovery.
The diffusion constants, obtained from fitting a single exponential to the
recovery data, are in agreement with other \emph{in vivo} measurements, but are
an order of magnitude smaller than those of GPI-linked proteins reconstituted
into liposomes~\cite{Frick:2007}.
The diffusion constants of the membrane proteins were unaffected by
manipulations of the cortical cytoskeleton, but they increased markedly (up to
3-fold) if the protein density in the membrane was reduced in semi-intact
cells.
In conclusion, the reduced mobility of membrane proteins is likely a
consequence of the crowdedness of cellular membranes leading to anomalous
transport on smaller time and length scales than accessible by a typical FRAP
setup.

\textcite{Baker:2007} employed FRAP experiments with a variable radius of the
observation spot (vrFRAP) to address the distribution and transport of CD4
proteins and chemokine receptors on the plasma membrane of living HEK~293T
cells.
The obtained ratio of incomplete fluorescence recovery and the apparent
diffusion constants depends sensitively on the radius, which was varied between
1.40 and 3.45\,µm.
Interpretation of this dependency allowed for the conclusion that proteins and
receptors are compartmentalised into domains of about 350\,nm.

\subsubsection{Model membranes (\emph{in vitro})}
\label{sec:model_membranes}

\begin{table*}
  \setlength\tabcolsep{.5em} \setlength\extrarowheight{1ex}
  \fontsize{9pt}{9pt} \selectfont 
  \begin{tabularx}{\linewidth}{%
    >{\raggedright\arraybackslash}p{.15\linewidth}%
    >{\raggedright\arraybackslash}p{.13\linewidth}%
    p{.1\linewidth}p{.15\linewidth}%
    >{\raggedright\arraybackslash}Xp{5ex}l}
  \hline \hline
  model membrane (composition) & probe (size) & experimental \newline technique
    & temporal and \newline spatial scales & observation & year & reference
  \vspace{2pt} \\ \hline
  SLB (POPC) & DHPE lipid & SPT & 20\textminus{}300\,ms
    & normal, two components: $D_1=4.4\,$µm\textsuperscript{2}/s, $D_2  = D_1/63$
    & 1997 & ref.~\citenum{Schuetz:1997}
  \\
  \allowhyphenation SLB (SOPC) + tethered lipopolymer (``obstacles'')
    & DHPE lipid, bacteriorhodopsin & SPT
    & $t_\text{lag} = 50\,\text{ms}$
    & anomalous diffusion, \newline $D$~\textdownarrow\ linearly for [lipopolymer]~\textuparrow
    & 2005 & ref.~\citenum{Deverall:2005}
  \\
  SLB (DOPC/cholesterol) & DHPE lipid & FCS & beam radius: \newline 0.25\,µm
   & $D$: 4.2 \textrightarrow\ 0.5\,µm\textsuperscript{2}/s for [cholesterol]~\textuparrow,
      deviations from simple diffusion
   & 2003 & ref.~\citenum{Benda:2003}
  \\
  multilamellar vesicles (DMPC/DSPC) & DPPE lipid & FCS
   & $\tau_{1/2}$: 108\textminus{}230\,ms
   & two-component diffusion for fluid--gel coexistence in bilayer,
    heterogeneous structures
   & 2005 & ref.~\citenum{Hac:2005}
  \\
   & C5-Bodipy-PC & variable-waist FCS & $\tau_{1/2}$: 1 \textminus\,150\,ms \newline
    $w$: 210\textminus{}300\,nm
   & $w$-dependent diffusion constant: $\tau_{1/2}(w)=t_0+w^2/4 D_\text{eff}$, \newline
   $t_0 < 0$ in gel phase
   & 2011 & ref.~\citenum{Favard:2011}
  \\
  multibilayers (DMPC/cholesterol) & phospholipid & FRAP & spot radius: 4\,µm
    & normal diffusion, $D$ 2-fold reduced for [cholesterol]~\textuparrow & 1992 & ref.~\citenum{Almeida:1992}
  \\
  GUV (DOPC/DOPG) \newline + proteins & \allowhyphenation
    integral membrane proteins \newline (0.5\textminus{}4\,nm)
    & FCS & $\tau_{1/2} \approx 2\,\text{ms}$
    & $D$~\textdownarrow\ for [protein]~\textuparrow, max. 2-fold suppression,
    subdiffusion: $\alpha \gtrsim 0.88$ & 2009 & ref.~\citenum{Ramadurai:2009}
  \\
  SLB (SOPC) \newline + avidin & DPPE-anchored avidin protein & FCS
    & $t_\text{lag}$: 0.3\textminus{}300\,ms, \newline $w\approx 0.17\,\text{µm}$
    & development of subdiffusion, \newline
    $\alpha$: 1 \textrightarrow\ 0.68 for [avidin]~\textuparrow, spatial heterogeneity, no ageing
    & 2010 & ref.~\citenum{Avidin:2010}
  \\
  SLB (SOPC) \newline + neutravidin
    & \allowhyphenation lipid, DOPE-anchored neutravidin protein & FRAP
   & $\tau_{1/2} \approx 3\,\text{s}$
   & normal diffusion, $D_\text{protein}$ up to 7-fold suppressed, $D_\text{lipid}$ almost unchanged
    as [neutravidin]~\textuparrow
   & 2009 & ref.~\citenum{Fenz:2009a}
  \\ \hline \hline
  \end{tabularx}
  \caption{Overview of \emph{in vitro} experiments on model membranes as
  guide to the discussion in \cref{sec:model_membranes}. Empty fields repeat the
  entry above, up/down arrows indicate increase/decrease of some quantity.
  Abbreviations: SLB\dots{}supported lipid bilayer, GUV\dots{}giant
  unilamellar vesicle, [X]\dots{}concentration of X.
}
  \label{tab:model_membranes}
\end{table*}

Model membranes reduce the complexity of cellular membranes and permit a
detailed study of macromolecular transport under controlled lipid composition
and crowding conditions, vital for a physical understanding of anomalous
transport.
Membranes are usually modelled by bilayers of phospholipids, either supported
by a flat substrate or forming a giant unilamellar vesicle.
The structure of the lipid bilayer can be modified by addition of another lipid
type, and crowding effects can be mimicked by covering the bilayer with
anchored proteins or by introducing protein inclusions.
It is possible to fluorescently label lipids as well as proteins and to follow
their motion individually.


\textcite{Schuetz:1997} performed SPT experiments on a supported POPC lipid
bilayer modelling a fluid membrane, where highly diluted fluorescent TRITC-DHPE
lipids were introduced in the outer leaf.
It was found that 69\% of the labelled lipids showed normal, free diffusion
with $D=4.4$\,µm\textsuperscript{2}/s, while the motion of the remaining
part was still diffusive, although suppressed by a factor of 63.
These results rely on the analysis of the cumulative probability distribution
of the square displacements for a series of large times (from 20\,ms to 300\,ms),
an approach that is well suited to identify and account for a mixture of fast
and slowly diffusing molecules.
The data for the slow fraction indicate further that motion is confined at
short times below 100\,ms, which the authors interpreted as evidence for
corrals of about 130\,nm size after correcting for the tracking uncertainty.

A systematic development of anomalous diffusion was observed in a supported
bilayer using SOPC lipids where the motion of the monitored TRITC-DHPE lipids
was hindered by essentially immobile ``obstacles''~\cite{Deverall:2005}, the
obstacles were realised by lipopolymers in the inner leaflet tethered to the
substrate.
Similarly as in Ref.~\onlinecite{Schuetz:1997}, the distribution of squared
displacements was analysed for a fixed lag time of 50\,ms.
At obstacle concentrations below 10\,mol\%, the distribution histograms at this
time scale correspond to normal diffusion and become increasingly anomalous for
larger concentrations.
The apparent diffusion constant $\expect{\Delta \vec R^2}/4t_\text{lag}$ is
linearly suppressed with increasing obstacle concentration and vanishes near
40\,mol\%.
The scenario is consistently described by random walks on two-dimensional
lattice percolation clusters, see Refs.~\cite{benAvraham:DiffusionInFractals,
Stauffer:Percolation} and \cref{sec:Lorentz}.
Similar results were obtained for the transport of diluted bacteriorhodopsin
inclusions mimicking membrane proteins.
The work demonstrates the great potential of model systems
systematically bridging between physics and biology.


The influence of cholesterol content on the transport in planar DOPC bilayers
supported by an atomically flat mica substrate was studied by FCS of
rhodamine-labelled DHPE tracers~\cite{Benda:2003}.
The diffusion constant of DHPE reduces from 4.2 to 0.5\,µm\textsuperscript{2}/s
as up to 60\% of DOPC is replaced by cholesterol.
Deviations from normal diffusion arise for a cholesterol content of 30\% and
more.
While the data are fitted by the model with two normally diffusing components,
the diffusion constant of the fast component is much larger than without
cholesterol.
Thus, the two-component model may not be suitable for DOPC/cholesterol
bilayers.

Aiming at the reduction of coupling effects with the substrate,
\textcite{Hac:2005} studied multilamellar model membranes consisting of stacks
of about 50 layers.
The bilayers are composed of a mixture of DMPC and DSPC lipids, which exhibits
phase separation and a fluid--gel transition in each component.
The motion of fluorescent TRITC-DPPE tracer lipids was followed by FCS at
different compositions and temperatures covering the full region of fluid--gel
coexistence in the phase diagram.
While transport is normal in the fluid regimes, complex transport emerges in
the phase-separated regime with its heterogeneous structure.
The FCS correlations can be fitted by a two-component model, but again, the
obtained diffusion times deviate from the ones in the pure fluid and gel
phases.
Instead of relying on a phenomenological fit model, the authors performed
extensive Monte-Carlo simulations which quantitatively describe the
experimental FCS curves; a detailed discussion of the simulations will be given
in \cref{sec:membrane_simulations}.

In a recent FCS experiment using the same DMPC/DSPC mixture~\cite{Favard:2011},
spatio-temporal information on the lipid motion was collected by variation of
the detection area.
FCS correlation functions were fitted by the model for normal diffusion
yielding the half-value times as function of the area, which were analysed in
terms of the so-called FCS diffusion law, i.e.,
$\tau_{1/2}(w)=t_0 + w^2/4D_\text{eff}$.
In the fluid phase of the mixture, the dependence on $w^2$ is linear ($t_0=0$)
as expected for simple diffusion.
Extrapolating to zero area, a negative offset, $t_0<0$, develops as the
fluid--gel coexistence is traversed by lowering the temperature.
Such a negative offset typically arises for motion in a meshwork of permeable
barriers~\cite{Destainville:2008}, but there, transport is not subdiffusive in
the time domain.
For motion in a phase-separating lipid mixture, Monte-Carlo simulations show
that the area $4 D_\text{eff} t_0$ correlates with the typical domain size of
the spatially heterogeneous environment, demonstrating that a domain size may
be extracted from FCS experiments~\cite{Favard:2011}.
The connection between the half-value times of the FCS correlation and the
detection area may alternatively be described by a power law (see supplement of
Ref.~\onlinecite{Favard:2011}), $\tau_{1/2}(w)\sim w^{2/\alpha}$, yielding
exponents $\alpha$ between 0.5 and 1; the smallest values are realised for
temperatures close to the phase boundaries.

The lateral diffusion of a phospholipid probe was measured with FRAP in
multi-bilayers of DMPC/cholesterol mixtures, exploring a large part of the
phase diagram in the composition--temperature plane~\cite{Almeida:1992}.
In all cases, the fluorescence recovery was complete and the uniform disc model
for free diffusion, \cref{eq:frap_disc}, fitted the data.
A 2-fold reduction of the diffusion constant was observed as cholesterol
content was increased from 0\% to 50\%, and the diffusivity increased up to
8-fold as the temperature was changed from 24\,°C to 58\,°C.

The lateral mobility of fluorescently labelled integral membrane proteins
reconstituted in giant unilamellar vesicles (GUVs) of a DOPC/DOPG mixture was
measured using FCS~\cite{Ramadurai:2009}.
The hydrodynamic radius of the investigated proteins ranged between 0.5 and
4\,nm.
Consistent with the Saffman--Delbrück model, the diffusion coefficient at low
protein content displays a weak, i.e., logarithmic, dependence on the radius.
The diffusion coefficient of proteins and lipids decreases linearly with
increasing the protein concentration in the membrane up to 3000 proteins per
µm\textsuperscript{2}.
Such a concentration is an order of magnitude smaller then in biological
membranes and induces only small crowding effects.
Hence, the observed reduction of mobility did not exceed about 20\% for the
lipids and was at most 2-fold for the proteins; deviations from simple
diffusion were only moderate with the exponent of subdiffusion $\alpha$ not
falling below $0.88$.

Crowding by peripheral membrane proteins was addressed by a study of avidin
proteins irreversibly bound to biotinylated DPPE lipid anchors in a supported
single SOPC lipid bilayer~\cite{Horton:2007, Avidin:2010}.
The content of lipid anchors in the bilayer directly controls its protein
coverage.
X-ray reflectivity experiments showed that the protein layer is separated from
the lipids by a distinct water layer, and continuous bleaching experiments
indicated that the fluidity of the lipid bilayer is retained even at high
protein coverage~\cite{Horton:2007}.
Upon increasing the protein coverage, FCS measurements on dilute labelled
avidin proteins revealed a systematic development of anomalous
transport~\cite{Avidin:2010}.
Converting the FCS correlation to the mean-square displacement via
\cref{eq:fcs_gaussian_2d}, subdiffusive motion was observed over a time window
spanning from 0.3 to 300\,ms.
The exponent $\alpha$ decreases gradually from 1 to 0.68 in the most crowded
regime with an excess of biotin anchors.
The onset of anomalous transport occurs already at an anchor concentration of
0.1\,mol\% or an area coverage of 3$-$5\%.
This transition regime is characterised by a pronounced and long-lived spatial
heterogeneity: different exponents $\alpha$ were reproducibly obtained at
different spots of the sample.
The distribution of exponents becomes bimodal as crowding is increased; some
spots exhibit simple diffusion of the proteins, while an increasing fraction
displays subdiffusion.
On the time scale of hours, ageing effects could not be detected.


FRAP measurements on similar SOPC bilayers crowded with
neutravidin~\cite{Fenz:2009a} corroborate normal diffusion of the lipids with
slightly reduced diffusion constants by up to 25\% at the highest protein
coverage, a fraction of about 10\% of the lipids was detected to be immobile.
Covering the bilayer with fluorescently labelled neutravidin, the fluorescent
signal from the proteins did not recover on the time scales of the experiment,
which was attributed to proteins being arrested in a kind of gel phase.
Measuring protein transport by means of continuous bleaching, a 7-fold slowing
down was observed at the highest protein coverage, which may be compatible with
the findings in Ref.~\onlinecite{Avidin:2010} taking the different time
and length scales of the experiments into consideration.

\subsubsection{Computer simulations (\emph{in silico})}
\label{sec:membrane_simulations}

\begin{table*}
  \setlength\tabcolsep{.5em} \setlength\extrarowheight{1ex}
  \fontsize{9pt}{9pt} \selectfont 
  \begin{tabularx}{\linewidth}{%
    >{\raggedright\arraybackslash}p{.07\linewidth}%
    >{\raggedright\arraybackslash}p{.19\linewidth}%
    l%
    >{\raggedright\arraybackslash}p{.16\linewidth}%
    >{\raggedright\arraybackslash}Xp{5ex}%
    >{\raggedright\arraybackslash}p{9ex}}
  \hline \hline
  space & crowded medium & tracer & interactions & observation & year & reference
  \vspace{2pt} \\ \hline
  square lattice & point obstacles \newline (mobile, but slow) & point & excluded volume
   & $D$~\textdownarrow\ for [obstacle]~\textuparrow & 1987 & ref.~\citenum{Saxton:1987}
  \\
   & point obstacles \newline (fixed, random) & &
   & subdiffusion on growing time window,
      $D$~\textdownarrow\ linearly and $\alpha$~\textdownarrow\ for [obstacle]~\textuparrow,
      at percolation transition: $D=0$, $\alpha\approx 0.68$
   & 1994 & ref.~\citenum{Saxton:1994}
  \\
   &&&
    & dynamic scaling for cluster-resolved MSDs, universality of exponents
    & 2008 & ref.~\citenum{Percolation_EPL:2008}
  \\
  2D con- tinuum & random disc obstacles & discs & hard core repulsion
    & transient subdiffusion, $\alpha$~\textdownarrow\ for [obstacle]~\textuparrow,
     highly fragmented void space
    & 2006, 2008 & refs.~\citenum{Sung:2006,Sung:2008a}
  \\
  & random disc obstacles (overlapping) & point & hard core repulsion
    & universal scaling of localisation phenomenology, subdiffusion over
        6~decades in time, $\alpha$: 1 \textrightarrow\ 0.659 \textrightarrow\ 0
    & 2010 & ref.~\citenum{Lorentz_2D:2010}
  \\
  square lattice & extended obstacles from AFM images & point & excluded volume
    & subdiffusion, $\alpha$: 1 \textrightarrow\ 0.59 as excluded area~\textuparrow,
    agreement with experiments & 2011 & ref.~\citenum{Skaug:2011}
  \\
  & mobile ``rafts'', point and line obstacles (``fences'') & point
    & excluded volume, low mobility inside rafts
    & subdiffusion ($\alpha \approx 0.7$) only for immobile rafts and raft-excluded tracers
    & 2007 & ref.~\citenum{Nicolau:2007}
  \\
  & regular network of line obstacles (``fences'') & point & potential barrier
    & high and dense barriers reduce $D(t)$ by several orders of magnitude
    & 2008 & ref.~\citenum{Niehaus:2008}
  \\
  triangular lattice & DMPC/DSPC monolayer model (2 sites per lipid) & lipid
    & nearest neighbour, slow and fast hopping, fluid/gel state flips
    & FCS correlation: subdiffusion ($\alpha \gtrsim 0.63$), agreement with experiment
    & 2005 & ref.~\citenum{Hac:2005}
  \\
  &&&& subdiffusion ($\alpha$: 1\textminus{}0.6) from first passage times,
    structural heterogeneity
    & 2005 & ref.~\citenum{Sugar:2005}
  \\
  &&&& FCS with variable beam waist: \newline $\tau_{1/2}(w)\sim w^{2/\alpha}$,
    $\alpha$: 1\textminus{}0.5 & 2011 & ref.~\citenum{Favard:2011}
  \\
  square lattice & DMPC/DSPC monolayer model (1 site per lipid) & point
    &
    & transient subdiffusion, $\alpha\approx 0.85$ close to critical point & 2011 & ref.~\citenum{Ehrig:2011}
  \\
  &&& + heterogeneous sticky network & transient subdiffusion ($\alpha\approx 0.5$),
    \newline $D_0/D$ up to 10 & 2011 & ref.~\citenum{Ehrig:2011a}
  \\
  3D con- tinuum & atomistic DOPC bilayer (288 fully hydrated lipids) & lipid & molecular force fields
    & lipid transport at 1\,ps is correlated over distances of 2.5\,nm & 2009 & ref.~\citenum{Roark:2009}
  \\ \hline \hline
  \end{tabularx}
  \caption{Overview of computer simulations on crowded membranes as
  guide to the discussion in \cref{sec:membrane_simulations}. Empty fields repeat the
  entry above, up/down arrows indicate increase/decrease of some quantity.}
  \label{tab:membrane_simulations}
\end{table*}

\paragraph{Obstructed motion}

In his seminal works, \textcite{Saxton:1987, Saxton:1994} introduced a
minimalist model for transport in cellular membranes.
The positions of lipids and proteins are restricted to a two-dimensional
lattice for computational efficiency; tracer molecules perform a random walk
which is hindered by randomly distributed obstacles occupying a single lattice
site each.
In the limiting case of immobile obstacles, the obtained diffusion constants
are almost linearly suppressed with increasing obstacle area and vanish at the
percolation threshold~\cite{Saxton:1987}.
While the sharp localisation transition disappears if the obstacles themselves
are allowed to diffuse slowly, the systematic reduction of tracer mobility with
increasing obstacle concentration is conserved.
For immobile obstacles~\cite{Saxton:1994}, the MSD displays subdiffusive motion
over a growing time window and a crossover to normal diffusion at large time
scales.
The apparent exponents~$\alpha$ obtained from the intermediate subdiffusive
regime depend on the obstacle concentration and approach their universal value
only at the percolation threshold.
The concentration-dependence of the exponent was reconciled with the dynamic
scaling hypothesis only recently by including the leading corrections to
scaling, which are again universal~\cite{Percolation_EPL:2008}; see
\cref{sec:Lorentz_simulations} for a detailed discussion.

The spatial discreteness of lattice models is overcome by continuum models,
where the obstacles are usually implemented by a simple geometric shape.
\textcite{Sung:2006, Sung:2008a} investigated the motion of tracer discs in a
frozen obstacle matrix of randomly placed, non-overlapping, impenetrable discs;
the limit of dilute tracers is known as non-overlapping Lorentz model in the
literature~\cite{Machta:1984}.
Similarly as for the lattice model, transport slows down with increasing
obstacle concentration and subdiffusive motion emerges in an intermediate time
window.
A Voronoi map of the continuous space accessible to the tracer particles
reveals that the free space is highly fragmented into voids of different sizes,
which are connected by significantly correlated channels.
Thus at high obstacle concentration, the primary mode of transport resembles a
hopping motion between neighbouring voids; similar observations were reported
from particle tracking experiments~\cite{Murase:2004}.
Transport on large scales, however, is the result of a large number of hopping
events, and subdiffusive motion relies on the presence of a spacious and
self-similar network of connected channels meandering the membrane.

The full dynamic scaling picture of anomalous transport due to a localisation
transition was established only recently for two-dimensional continuum
models~\cite{Lorentz_2D:2010}.
Extensive simulations for a Brownian tracer between uncorrelated and
overlapping circular obstacles confirmed the universal values of the dynamic
exponents at leading and next-to-leading order, e.g., for the exponent of
subdiffusion $\alpha=2/z \approx 0.659$ if also tracers in isolated regions
(``finite clusters'') are included.
The subdiffusive motion was followed over six decades in time, and the
crossover to normal diffusion away from the localisation transition was found
to extend over several decades as well.
Further, the transport dynamics was rationalised in terms of crossover time and
length scales, which diverge at the transition.
Appropriate rescaling of the time-dependent diffusion constants for different
obstacle concentrations yielded data collapse onto a universal scaling function
that describes the crossover from subdiffusion to normal diffusion.
The transition to continuum models is a step towards a more realistic
modelling, the universal phenomenology of the localisation transition, however,
is preserved as expected at sufficiently large scales.
This is in contrast to obstructed transport in three dimensions, where lattice
and continuum models belong to different (dynamic) universality classes with
different transport exponents~\cite{Machta:1985, Halperin:1985,
Lorentz_PRL:2006}.

An uncorrelated distribution of obstacles, generated by a random process, is
certainly a simplification.
Realistic configurations of the excluded area were obtained from atomic force
microscopy (AFM) imaging of phase-separating lipid bilayers~\cite{Skaug:2011}.
For mixtures of DSPC/DOPC lipids, the gel-like DSPC-rich phase effectively
provides an excluded area to the motion of DMPE tracer lipids.
The rastered AFM images then yield ``obstacles'' of irregular shape that extent
over many lattice sites, and Monte-Carlo simulations can be used to investigate
the transport of the tracer lipids performing a random walk in the remaining
space.
This approach is somewhere in between the original lattice models and the
continuum models, and if it is justified that the macroscopic transport is not
affected by the imaging resolution (cf.\ Ref.~\onlinecite{Niehaus:2008}) it can
provide a quantitative bridge to the experiments.

Apart from excluded volume, transport of membrane proteins may be hindered by
other sources.
\textcite{Nicolau:2007} studied random walks of proteins on two-dimensional
lattice models, where the proteins are hindered by either a regular lattice of
cytoskeletal fence posts or by mobile lipid rafts that are either impenetrable
for proteins or reduce their mobility.
Appreciable subdiffusion similarly as for randomly distributed, fixed obstacles
was only observed if the protein motion is hindered by almost immobile,
impenetrable rafts.
Depending on the raft mobility, subdiffusive motion was characterised by
exponents $\alpha$ between 0.65 and 0.75.
The simulations indicated further that collisions with fence lines alone
(without binding) do not explain the experimentally reported degrees of
anomalous transport.
Fences that impose very high barriers, however, can reduce protein diffusion at
long times by several orders of magnitude~\cite{Niehaus:2008}, which was
demonstrated by Monte-Carlo simulations on a lattice compartmentalised by a
regular network of fences.
It appears that here a plateau in the MSDs rather than a subdiffusive power-law
growth is the manifestation of slow transport.
Since only a single length scale characterises the regular super-lattice of
fences, the plateau can be understood as a reflection of the mesh size.

\paragraph{Mixtures of DMPC/DSPC lipids}

Experiments on lipid transport in artificial lipid bilayers have shown that
anomalous transport emerges as function of the lipid composition.
Often binary mixtures of lipids like DMPC/DSPC or DOPC/cholesterol were used,
which can phase separate into fluid- and gel-like domains.
\textcite{Sugar:1999} developed a model for DMPC/DSPC mixtures on a triangular
lattice, where each of the two acyl chains of a lipid occupy two adjacent sites
and each site is in either the gel or the fluid state.
The thermodynamics of the model is governed by a Hamiltonian with 10 relevant
parameters, which were either directly inferred from calorimetric measurements
or estimated by means of Monte Carlo simulations to reproduce the correct phase
diagram.

Relying on this model, \textcite{Hac:2005} investigated FCS correlation
functions of the lipids using Monte-Carlo simulations with a coarse-grained
dynamics.
Each lattice update allowed for the possibility of flipping the gel/fluid state
of a lipid chain (Glauber dynamics) and of exchanging adjacent molecules
mimicking diffusion (Kawasaki dynamics).
The dynamics is controlled by a rate for the state flips and two hopping rates
for fluid- and gel-like environment; the latter two were matched to the
experimental diffusion constants for pure fluid or gel phases.
The numerically obtained FCS curves show quantitative agreement with the
experimental results if the flip rate is chosen close to the hopping rate in
the fluid.
In the regime of gel--fluid coexistence, the FCS curves deviate from normal
diffusion and are fitted by the gaussian subdiffusion model.
The exponent $\alpha$ depends on the flip rate, and its smallest value of 0.63
was reported for pure Kawasaki dynamics, i.e., in the absence of artificial
state flips.
Parameters for the hopping rates may be obtained by free volume theory, and
correlating the structure and distribution of gel clusters with the lipid
dynamics provides evidence that the anomalous transport arises from molecules
exploring a heterogeneous environment~\cite{Sugar:2005}.
The model was further employed to address spatial aspects of the lipid
transport by simulating FCS correlations for a wide range of beam waists, which
allowed for the extraction of the mean domain size~\cite{Favard:2011}; a
discussion was already given in \cref{sec:model_membranes}.

The above lattice model for a lipid membrane of Refs.\ \onlinecite{Sugar:1999}
and \onlinecite{Hac:2005} was further simplified by \textcite{Ehrig:2011,
Ehrig:2011a}.
They used a square lattice with each site representing a lipid molecule; the
coarse-grained dynamics still comprises state flips of a single lipid and
diffusion by exchange of two adjacent sites.
Together with advances in computer hardware, this permitted the investigation
of comparably large systems with a membrane area of
(0.48\,µm)\textsuperscript{2} or 600\textsuperscript{2} lattice points and
processes on time scales of up to 1\,s.
The simulations demonstrate that close to the upper critical point of
fluid--gel coexistence, the lipid motion exhibits anomalous transport over
several decades in time.
A double-crossover scenario is observed from simple diffusion at
microscopically short time scales to transient subdiffusion and finally back to
normal diffusion.
The time window of subdiffusion increases as the critical point is approached,
the local exponent $\alpha(t)$ of the MSD, however, does not yet display a
plateau and its smallest reported value is around 0.85 being still relatively
close to~1.
In consequence, only a 3-fold suppression of the macroscopic diffusion constant
is observed compared to its microscopic value.
The introduction of a frozen, spatial heterogeneity in form of a sticky
network~\cite{Ehrig:2011} renders the lipid transport significantly more
anomalous in a limited time window and induces an up to 10-fold reduction of
mobility and minimal values of the local exponent $\alpha(t)$ down to~0.5.

Let us raise the question whether an asymptotic, universal value for the
subdiffusion exponent exists in a mixture as the critical point is approached
further.
The analytical work by \textcite{Kawasaki:1966}, employing a local-equilibrium
approximation, suggests that the self-diffusion constant is only mildly
affected by the critical fluctuations; in particular, it is not singular at
leading order and does not vanish.
Similarly, recent microcanonical molecular dynamics simulations for a
three-dimensional symmetric binary mixture near its consolute point revealed
almost negligible anomalies in the single-particle transport~\cite{Das:2006,
Das:2006a}.
It appears that in two dimensions, the single-particle transport is coupled
more strongly to the critical fluctuations.
But it is unlikely that subdiffusion can exist over an arbitrarily large time
window, which would imply a vanishing diffusion constant.

With constantly growing computational resources, atomistic molecular dynamics
simulations may come into reach as an alternative source for the parameters of
coarse-grained models.
A detailed view of the structure of a DMPC/DSPC bilayer was provided by a
hybrid simulation technique, where an atomistic molecular dynamics simulation
was supplemented by a semi-grandcanonical Monte-Carlo step converting a DMPC
lipid into DSPC and \emph{vice versa}~\cite{Coppock:2009}.
Using 128 lipid molecules and 3200 explicit point charge water molecules, it
was possible to generate system trajectories covering 50\,ns in time, or two
million integration steps.
By means of a pure molecular dynamics simulation for a bilayer of 288
atomistically modelled DOPC lipids, the experimental values of the short-time
diffusion constants of the lipids could be quantitatively reproduced by a
simulation~\cite{Roark:2009}.
The study further showed that the motion of the lipids is correlated over
distances of about 2.5\,nm at the time scale of 1\,ps, which is a useful piece
of information for the refinement of lattice models: it suggests that a single
lattice site should represent a correlated area of several lipids rather than a
single lipid molecule or chain.
Molecular dynamics simulations of lipid transport over large distances seem
not yet feasible as it would be hampered by the small diffusion constant of the
lipids in the gel state ($D_\text{gel}\approx
0.05$\,nm\textsuperscript{2}/µs~\cite{Hac:2005}).
This emphasises the need of controlled coarse-graining and multiscale
approaches for the study of slow, complex relaxation dynamics at the micro- and
millisecond scale.

\subsection{Reaction kinetics}

So far, we have discussed the anomalous transport of individual macromolecules
in their crowded cellular environments.
Physiological processes, however, are determined to a large extent by
biochemical reactions.
Adding to the complexity of anomalous transport, the reaction rates as well as
the availability of reactants are significantly modified by molecular crowding.
The systematic investigation of this topic is relatively young, and we will
only sketch some of the recent findings.

Minton \emph{et al.}~\cite{Hall:2003, Zhou:2008} considered the consequences of
macromolecular crowding on the equilibrium rates of biochemical reactions from
a thermodynamics point of view, focusing on changes of the excluded volume.
They found a non-specific enhancement of reaction rates if the total excluded
volume is reduced; examples for such reactions are the formation of
macromolecular complexes in solution, binding of macromolecules to surface
sites, formation of insoluble aggregates, and compaction or folding of
proteins.
For protein association reactions, crowding is generally expected to increase
the rate of slow, transition-state-limited reactions and to decrease the rate
of fast, diffusion-limited ones.
In the context of polymer systems, it was predicted that the subdiffusion
exponent controls the crossover between reaction- and diffusion-limited kinetic
regimes~\cite{OShaughnessy:1999}.
A variation of the association rates under crowding conditions by an order of
magnitude was reported in the experimental literature reviewed until 2008 in
Refs.~\onlinecite{Hall:2003, Zhou:2008}.
\textcite{Melo:2006} reviewed the kinetics of bimolecular reactions in model
bilayers and biological membranes from an experimental perspective until 2006;
their conclusion ``When going to crowded or phase-separated systems \dots\ the
absence of experimental studies is almost complete.'' emphasises the need for
systematic investigation of the reaction kinetics in crowded membranes.

The kinetics of the reaction $A+B\to \mathit{products}$ on a percolating,
two-dimensional lattice was studied using Monte-Carlo
simulations~\cite{Saxton:2002}.
Subdiffusive motion slows down the initial encounter of reactants, but
concomitantly hinders them from escaping due to the increased return
probability.
The simulations showed that obstructions can slightly increase the reaction
rate if the reaction probability on a single encounter is small.
On the other hand, the reaction rate is reduced for high reaction probability,
and taken together, subdiffusion of the reactants narrows the range of reaction
rates.
The bimolecular reaction is similar to that of the reaction of a random walker
with an immobile target.
Realising anomalous transport of the walker by a hierarchy of non-reactive
binding sites, \textcite{Saxton:2008} found from simulations that trapping may
contribute significantly to the noise in reaction rates and further that the
subdiffusive motion yields a power-law distribution of capture times.
\textcite{Schmit:2009} identified the concentration of reactants and the
crowdedness of the medium as two antagonistic parameters for the rates of
diffusion-limited bimolecular reactions.
They showed that there is an optimal reaction rate (for a fixed ratio of the
concentrations of reacting and inert particles) if the reactants are not too
diluted and if the medium is not too crowded, i.e., diffusion of the reactants
is not too slow.
Continuum models for the reaction of ligands with an immobile receptor realised
the crowded environment by a dense hard-sphere fluid~\cite{Dorsaz:2010,
Kim:2010}.
Again, a non-monotonic dependence of the effective reaction rate on the volume
fraction of the fluid particles was found for sufficiently small receptors.

For a diffusion-limited dimerisation reaction in a computer model for the
\emph{E.\ coli} cytosol~\cite{Ridgway:2008}, it was observed that the reaction
coefficients become time-dependent in the presence of anomalous transport; the
obtained reaction rate was suppressed by two orders of magnitude and obeyed an
approximate power law over 4 decades in time.
\textcite{Hellmann:2011} found such a fractal reaction kinetics with a
time-dependent, power-law reaction rate also in two-dimensional, off-lattice
simulations for the reaction $A+B \to \emptyset$ with subdiffusive motion of
the reactants.
Specifically, fractional Brownian motion was implemented by means of an
overdamped Langevin equation with a power-law-correlated noise.
Key findings were that \emph{i)} independently of the reaction probability the
diffusion-limited formation of new reactant pairs dominantly determines the
reaction rate and \emph{ii)} subdiffusive motion enhances the segregation of
reactants.
Extending this approach to describe a multi-step enzymatic reaction, the
double-phosphorylation of mitogen-activated protein kinase (MAPK), an increased
performance due to anomalous diffusion was found when dissociation rates of the
intermediate enzyme-substrate complexes are high while the irreversible
catalytic step is slow~\cite{Hellmann:2012}.

\textcite{Froemberg:2008} set up a continuum description in form of a
reaction-\emph{sub}diffusion equation for the reaction  $A + B \to \emptyset$
with subdiffusive reactant motion governed by a CTRW.
Supplying reactants of both types from both sides of the medium, stationary
profiles of reactant concentrations and of reaction zones develop.
These profiles exhibit accumulation and depletion zones close to the
boundaries, and the reaction intensity was found to depend non-monotonically on
the reactant concentration at the boundaries.
A different approach was taken by \textcite{Boon:2012} for the
n\textsuperscript{th}-order annihilation reaction, $A + A + \cdots + A \to
\emptyset$, who derived a \emph{nonlinear} reaction-diffusion equation assuming
concentration-dependent diffusion and reaction coefficients and imposing a
scaling form of the solution, i.e., demanding the concentration to satisfy
$c(r,t) = t^{-\alpha/2}\mathcal{C}\bigl(rt^{-\alpha/2}\bigr)$.
Thereby, subdiffusive motion of the reactant is incorporated, cf.\
\cref{eq:diffusion_scaling,eq:fbm_scaling,eq:ctrw_scaling,eq:lorentz_inf_scaling}.
Then, the resulting steady-state profiles exhibit a long-range, power-law decay
and fit experimental data for the Wg morphogen gradient in \emph{Drosophila}
wing disc.

\section{Conclusion}
\label{sec:conclusion}

The advancements of biophysical experimental tools in the past two decades have
revolutionised the investigation of intracellular transport of proteins,
nucleic acids, and artificial tracers as well as the motion of membrane
receptors and lipids.
The new methods allowed for a detailed view of the plethora of dynamic
processes \emph{in vivo} directly at the cellular level.
Numerous new phenomena and features of the transport have been discovered as
more and more data of ever increasing quality have been collected.
Living cells, by nature, consist of many differently sized and shaped
constituents, each with their own physiological function, leading to densely
packed, highly heterogeneous structures at the nanoscale---commonly referred to
as \emph{macromolecular crowding}.
A widespread observation is that transport in living cells behaves rather
different from the standard picture of Brownian motion, where the erratic
trajectories are described by a probabilistic law as dictated by the
central-limit theorem.
Instead, the mean-square displacement, being the simplest quantitative measure
of single-particle transport, displays complex and often also subdiffusive
behaviour.
Here, we have focused on anomalous transport, characterised by power-law
increases on time windows covering several orders of magnitude.
A unified physical picture of this spectacular anomaly remains one of the grand
challenges of biophysics.


From a theoretical perspective, a fundamental issue is to identify mechanisms
that lead to a violation of the central-limit theorem on macroscopic time
scales.
To this end, several theoretical models have been established that achieved
this goal and yield a subdiffusive increase of the mean-square displacement.
Here, we have summarised three of the most commonly used approaches based on
different microscopic processes.

Fractional Brownian motion as a representative of spatially gaussian models is
distinguished by a strictly self-similar increase of the mean-square
displacement.
Its most prominent feature is a power-law correlated, gaussian random noise.
As a consequence, the stochastic process is stationary, but not Markovian.
The transport propagator, i.e., the probability distribution of the
displacements after a given lag time, displays a scaling property similar to
simple Brownian motion.

Second, continuous-time random walk models describe a microscopic hopping
process, which asymptotically generates subdiffusive motion in the case of
waiting-time distributions with sufficiently broad power-law tails.
It contains a fractional Fokker--Planck description as limiting case where the
propagator exhibits scaling for all time and length scales.
The scaling function depends on the subdiffusion exponent $\alpha$ and is
markedly different from the gaussian case.
In particular, the non-gaussian parameter does not vanish even for long times
and approaches a characteristic value depending only on $\alpha$.
By the very construction, the process is non-stationary if the mean-waiting
time is literally infinite, which implies ageing phenomena and weak ergodicity
breaking.

The third class consists of Lorentz models, focusing on obstructed motion in
strongly heterogeneous landscapes.
Crowding agents lead to ramified structures delimiting the space accessible to
the tracer, and anomalous transport emerges generically from the meandering
motion of the tracer in such labyrinth-like structures.
Subdiffusive motion occurs on larger and larger time windows as a localisation
transition is approached, until long-range transport would eventually cease.
In the vicinity of the transition, scaling behaviour develops and can be
rationalised in terms of a dynamic critical phenomenon.
The properties upon approaching anomalous transport are controlled by three
independent, universal exponents governing the mean-square displacement,
the crossover time scale, and the fractal spatial structure.
The averaged dynamics is stationary, exhibits heterogeneous diffusion, and
marked non-gaussian spatial transport.

All three models are indistinguishable on the level of the mean-square
displacement being subdiffusive.
However, the predictions differ drastically if transport is considered on
different length scales, for example by analysing spatio-temporal information
as encoded in the propagator.
Then, the shape of the scaling functions serves, in principle, as a fingerprint
to discriminate the various theories.
Since the full scaling function is often more difficult to measure than derived
characteristics, we have exemplified the properties of the non-gaussian
parameter as a potential clue in the quest for the microscopic origin of
anomalous transport.


For this endeavour, researchers have matured the experimental techniques that
are suited to uncover the transport properties of biophysical samples at
mesoscopic scales.
Single-particle tracking, fluorescence correlation spectroscopy, fluorescence
recovery after photobleaching, among others, have widely been applied and yield
complementary aspects of complex transport.
Single-particle tracking provides access to the detailed particle trajectories
and thereby to a complete statistical characterisation, in particular to the
full propagator.
Implementations based on video microscopy are typically restricted to limited
temporal windows at a resolution of the order of 10\,ms.
Fluorescence correlation spectroscopy, on the other hand, probes shorter time
scales and yields correlation functions over many decades in time.
The shape of the correlation function serves as indicator for anomalous
transport at the spatial scale of the illumination region.
In recent developments, the beam waist has been made adjustable to probe the
dynamics on a range of length scales, which then in principle allows for the
discrimination of different propagators.
Modern setups for fluorescence recovery after photobleaching are apt to detect
slow transport on the time scale of several seconds and the length scale
imposed by the bleaching spot; noteworthy, the method is sensitive to immobile
particles as well.
It monitors collective rather than single-particle transport, and anomalous
transport becomes manifests in the recovery curve at long times.
We see a need to evaluate the current theoretical approaches for measurable
quantities of each of the above techniques.
On the other hand, the interpretation of experimental data in the context of
anomalous transport requires careful reconsideration of the underlying
assumptions, e.g., spatially gaussian behaviour.
Similarly, if the underlying process is (weakly) non-ergodic as is the case in
CTRW or Lorentz models, time and ensemble averages generally do not coincide
and caution is advised.
Single molecule experiments are a promising tool to resolve these issues in the
future.


Experiments inside the cyto- and nucleoplasm of various eukaryotic cells and
bacteria have provided ample evidence for subdiffusive motion \emph{in vivo}
over several decades in time.
The overall investigated time scales covered the window from 100\,µs to
10\,s, and the exponent $\alpha$ mostly ranged between 0.7 and 0.85 depending
on the specific experiment, but also on the size of the tracer; an even smaller
exponent of 0.53 was reported for 5-nm gold beads in many different cell types.
These findings are supported by \emph{in vitro} measurements on crowded model
solutions, yielding systematically decreasing subdiffusion exponents $\alpha$
for increasingly crowded solution.
The smallest values for $\alpha$ were obtained in dextran solution and reached
down to 0.6 for dextran tracers and to 0.7 and 0.74 for globular proteins.
A criticism on dextran as crowding agent is a possible interference of the
tracer motion with the internal dynamics of polymeric dextran branches; for
example, anomalous transport was regularly found to be less pronounced in
solutions of globular proteins.

Computer simulations have corroborated the experimental findings at least
qualitatively.
Simplified model systems focusing on the excluded volume describe a rich
phenomenology of anomalous transport including the gradual emergence of
subdiffusive motion, spatially non-gaussian transport, heterogeneous diffusion
with strongly suppressed diffusion constants, and a fraction of (on large
scales) immobile particles.
These phenomena have been related to a localisation transition, and simulation
data have been rationalised in terms of dynamic scaling.
Subdiffusive motion over many decades in time appears already for a small
tracer in a slowly rearranging matrix of equally sized particles with apparent
exponents less than 0.5; it persists for infinitely long times for randomly
distributed obstacles at a critical value of the excluded volume, e.g., a
universal value of the exponent $\alpha \approx 0.32$ was reported from
simulations following the subdiffusive transport over 6 decades in time.
More specific models for the cytoplasms of HeLa cells and \emph{E.\ coli} use a
distribution of particle sizes close to physiological conditions, yielding
subdiffusive motion over several decades in time with exponents varying between
1 and 0.55 as function of the tracer size.


A number of experiments \emph{in vivo} and \emph{in vitro}, however, display
simple, albeit slow diffusion with diffusion constants reduced by 1 to 3 orders
of magnitude.
For some experiments, data fitting required two normally diffusing components,
ideally motivated by a physical picture like compound formation between tracer
and crowding agent.
For model solutions with adjustable degree of crowding, the reduction of
diffusivity is explained to some extent by an increased macroscopic viscosity.
On the other hand, it has been observed that the Stokes--Einstein relation is
violated for proteins as probe and that the visco-elastic response on the
nanoscale strongly deviates from the macroscopic rheological behaviour.
%
%
Eventually, transport was found to be less anomalous in simulations if
hydrodynamic interactions between the particles are taken into account;
further, hydrodynamic interactions can already result in an up to 10-fold
reduction of the \emph{short-time} diffusion constant, in agreement with theory
and recent scattering experiments probing the nanosecond scale.

Summarising, anomalous transport has ubiquitously been observed in
experiments \emph{in vivo} and can be recreated in crowded model solutions and
in computer simulations, but not in all instances. The picture is not thus
simple and the appearance of anomalous transport in experiments is far from
being understood.
At least, one can say for model systems \emph{in vitro} and \emph{in silico}
that transport is slowed down upon increasing the concentration of crowding
agent or increasing the size of the tracer.
An unsolved problem is the importance of the different contributions slowing
down the motion, examples are excluded volume, polymer effects, compound
formation, and hydrodynamic interactions.


The situation seems even more controversial for the motion of membrane proteins
or phospholipids in cellular membranes.
An central goal of such \emph{in vivo} measurements is the elucidation of the
membrane structure from the transport behaviour by indirect evidence; such
biophysical conclusions are beyond the scope of this work, and we kindly refer
the reader to Refs.~\onlinecite{Chiantia:2009, Kusumi:2005, Saxton:1997} for
pertinent reviews.
In cellular membranes, essentially all kinds of transport have been reported
ranging from normal diffusion with reduced diffusion constants over
subdiffusion to confined motion and immobile tracers.
Former measurements of membrane proteins displayed a heterogeneous dynamics in
the sense that different tracers of the same type showed qualitatively
different motion.
During the past decade, advances in the experimental techniques have given
access to the transport behaviour over large time windows.
For example, high-speed tracking of phospholipids with 25\,µs
resolution probed the mean-square displacement over five decades in time and
have provided evidence for a double-crossover scenario including a regime of
subdiffusive motion with exponent $\alpha \approx 0.53$.
Similar time scales can be covered by fluorescence correlation spectroscopy,
but crossovers between different regimes have not been observed so far; the
experimental data were typically compatible with either gaussian subdiffusion
or normal diffusion of a fast and a slow component.
For several membrane proteins, subdiffusion was reported at the millisecond scale
with exponents typically between 0.5 and 0.8,
while other experiments found normal, but very slow diffusion at the scale of seconds.
Spatial aspects of transport have been addressed by FCS experiments with
variable illumination area, revealing anomalous transport in the form of
length-scale-dependent diffusion coefficients below about 100\,nm.
Thus, the diffusion coefficient may depend both on the time and length scale
under investigation, and both should cover a sufficiently large window for a
comprehensive description of macromolecular transport in membranes.

As for cytoplasmic transport, the investigation of model systems is of
paramount importance for a firm understanding of transport in membranes.
\emph{In vitro} studies of supported lipid bilayers or giant unilamellar
vesicles agree that the lipid diffusion slows down in the presence of phase
separation, which may be induced by increasing, e.g., the cholesterol content
of the lipid bilayer.
It was found that gel-like regions effectively serve as a ramified, excluded
area rendering the lipid motion anomalous.
The experimental evidence has been corroborated by computer simulations for
two-dimensional models of lipid mixtures close to fluid-gel coexistence,
yielding subdiffusive motion for molecules exploring a heterogeneous structure.
Pronounced subdiffusion over three decades in time was detected for the motion
of proteins anchored to a lipid bilayer, with the exponent $\alpha$ developing
from 1 to 0.68 as the bilayer was increasingly crowded by like proteins.
Further, the transport of membrane proteins can become anomalous if a fraction
of the membrane lipids is immobilised, e.g., by tethering to the substrate.
This resembles obstacle models with a percolation threshold, where a ramified,
heterogeneous landscape of excluded area appears generically.
As in three dimensions, anomalous transport emerges in a finite temporal window
upon approaching the percolation threshold, and subdiffusive motion with a
universal exponent $\alpha\approx 0.66$ could be followed in simulations over 6
decades in time at the localisation transition.


As elaborated in this review, significant progress both on the experimental
side as well as in theoretical modelling has been achieved in the puzzle of
anomalous transport in crowded biological media.
A lot of insight has been obtained, in particular, to establish the phenomenon
itself and to identify implications of crowding on macromolecular transport.
Yet, the knowledge acquired also raises a series of new questions challenging
the frontier of biophysical research.
Let us conclude this review by pointing at open problems and possible
directions of future work.

So far, the interpretation of experimental data was mostly restricted to the
mean-square displacement or the FCS correlation function assuming spatially
gaussian transport.
Only few studies discuss other statistical measures like distributions of
displacements or of apparent exponents or even two-particle correlations.
We have emphasised that the characterisation of anomalous transport depends on
the time and length scales probed, highlighting the importance of collecting
spatio-temporal information.
To this end, testable theoretical predictions need to be elaborated and
experimental methods to be perfected;
a first step would be to focus on the non-gaussian behaviour, for which
detailed predictions exist.
Beyond that, multiple-time correlation functions contain information on the
stochastic process underlying the transport, which would quantify the
non-Markovian behaviour in particular.
Here, a comparison between the different theoretical approaches remains to be
done.

While basic models have been established, they need to be refined to capture
specific experimental conditions.
For example, fractional Brownian motion should be generalised to account for
crossover phenomena to normal diffusion, both at microscopically short and
macroscopically long time scales.
Similarly, several experimental groups have excluded the continuous-time random
walk model as a candidate for anomalous transport since ageing scenarios
intrinsic to the model have not been observed.
This apparent deficiency can be avoided by considering large, but finite
mean-waiting times, thus allowing for a graceful exit to normal diffusion at
long times.
The Lorentz models dealing with obstructions should be extended to account for
correlated, differently sized obstacles and slowly rearranging disordered
environments.
Furthermore, the idealisation of a hard repulsive interaction should be
relaxed, thus introducing barriers of finite rather than infinite height.
Eventually, a significant ingredient to anomalous transport in the subcellular
world are boundaries as imposed by (intra-)cellular membranes; the role of a
finite space has been studied especially for the fractional Fokker--Planck
equation.

Well-controlled experimental model systems offer a great potential to
systematically investigate anomalous transport by examining adjustable crowding
parameters over wide ranges.
They provide the largely unexplored opportunity to address the same system by
complementary techniques, allowing for a comprehensive description of the
motion in crowded media.
Further, they form a bridge from \emph{in vivo} experiments to reductionist
systems amenable to theory.
Importantly, scaling behaviour could be tested here to discriminate between
apparent subdiffusion as a mere phenomenological description of data in
contrast to genuine anomalous transport over many orders of magnitude.
As a central result of pushing experiments and theory further, it should become
possible to predict the transport properties, at least qualitatively, from the
outset of an experiment.
On the other hand, it would be desirable to develop an experimental
standard for subdiffusive transport, ideally covering sufficiently large
windows in space and time, that is able to crosscalibrate the different
experimental techniques~\cite{Saxton:2012}.
Let us remark that the connection beyond the biophysical world, e.g., to
transport phenomena in gels, porous media, and random heterogeneous materials,
has so far remained largely elusive from an experimental point of view.

In the broader context of cell biology, single-particle transport is only one
aspect of macromolecular transport and one anticipates also fascinating
collective phenomena associated with crowded environments.
Similarly, living cells comprise an abundance of active, non-equilibrium
processes, and their connection to anomalous transport needs to be unravelled
in the future.
The grand challenge, of course, is to clarify the physiological implications of
anomalous transport, most importantly: does it constitutes merely a peculiarity
or has it a biological benefit?
For example, suggested scenarios on the acceleration of target search processes
and bimolecular reactions need to be substantiated experimentally.
More generally, important ingredients for systems biology like the kinetics of
biochemical reactions, the dynamics of protein folding and unfolding, and
intracellular signalling pathways may be strongly affected by macromolecular
crowding.
The development of a unified picture of anomalous transport and its
physiological consequences will entail an immense effort of interdisciplinary
research during the next decade at least.
Or let us phrase it with \textcite{Ellis:2003}: ``Join the crowd!''

\begin{acknowledgments}
We thank
C.~Bräuchle,
C.~Fradin,
E.~Frey,
I.~Goychuk,
P.~Hänggi,
D.~Heinrich,
J.~Horbach,
S.~Jeney,
J.~Kärger,
D.~C.~Lamb,
R.~Metzler,
J.~O.~Rädler,
M.~Saxton,
F.~Schreiber,
I.~M.~Sokolov,
A.~Yethiraj,
and M.~Weiss
for valuable scientific discussions
and N.\ Destainville for correspondence.
Financial support from the Deutsche Forschungsgemeinschaft (DFG) under contract
no.\ FR 850/6-1 is gratefully acknowledged.
\end{acknowledgments}

\newcommand\enquote[1]{\emph{#1}} 
\bibliography{anomalous_transport}

\end{document}